\documentclass[aps,prc,showpacs,showkeys,nofootinbib]{revtex4}

\usepackage{color} 

\usepackage{graphicx}
\usepackage{dcolumn}
\usepackage{amsthm}
\usepackage{bm}

\usepackage{isotope}
\usepackage{physics}

\newcommand{\sigmabf}{\mbox{\boldmath $\sigma$}}

\newcommand{\rhobf}{\mbox{\boldmath $\rho$}}
\newcommand{\rhobfsm}{\small \mbox{\boldmath $\rho$}}

\begin{document}

\title{%
Search of cluster structure in nuclei via analysis of bremsstrahlung emission
}

\author{Sergei~P.~Maydanyuk}%
\email{maidan@kinr.kiev.ua}%
\affiliation{Institute of Modern Physics, Chinese Academy of Sciences, Lanzhou, 730000, China}
\affiliation{Institute for Nuclear Research, National Academy of Sciences of Ukraine, Kiev, 03680, Ukraine}

\author{Peng-Ming~Zhang}%
\email{zhpm@impcas.ac.cn} %
\affiliation{School of Physics and Astronomy, Sun Yat-Sen University, Zhuhai, China}
\affiliation{Institute of Modern Physics, Chinese Academy of Sciences, Lanzhou, 730000, China}

\author{Li-Ping~Zou}%
\email{zoulp@impcas.ac.cn} %
\affiliation{Institute of Modern Physics, Chinese Academy of Sciences, Lanzhou, 730000, China}


\date{\small\today}

\date{\small\today}

\begin{abstract}
We investigate emission of bremsstrahlung photons during scattering of $\alpha$-particles off nuclei.
For that, we construct bremsstrahlung model for $\alpha$-nucleus scattering,
where a new formalism for coherent and incoherent bremsstrahlung emissions in elastic scattering and mechanisms in inelastic scattering is added.
Basing of this approach, we analyze experimental bremsstrahlung cross-sections in the scattering of $\alpha$-particles off the \isotope[59]{Co}, \isotope[116]{Sn}, \isotope[\rm nat]{Ag} and \isotope[197]{Au} nuclei at 50 MeV of $\alpha$-particles beam measured at the Variable Energy Cyclotron Centre, Calcutta.
We observe oscillations in the calculated spectra for elastic scattering for each nucleus.
But, for \isotope[59]{Co}, \isotope[116]{Sn} and \isotope[\rm nat]{Ag} we obtain good agreement between calculated coherent spectrum with incoherent contribution for elastic scattering with experimental data in the full photon energy region.
For heavy nucleus \isotope[197]{Au} we find that
(1) Oscillating behavior of the calculated spectrum of coherent emission in elastic scattering is in disagreement with experimental data,
(2) Inclusion of incoherent emission improves description of the data, but summarized spectrum is in satisfactory agreement with the experimental data.
To understand unknown modification of wave function for scattering, we add new mechanisms of inelastic scattering to calculations and
extract information about unknown new amplitude of such mechanisms from experimental data analysis.
This amplitude has maxima at some energies, that characterizes existence of states of the most compact structures (clusters) in nucleus-target.
We explain origin of oscillations in the bremsstrahlung spectra for elastic scattering (at first time).
New information about coherent and incoherent contributions is extracted for studied reactions.
\end{abstract}




\maketitle

\section{Introduction
\label{sec.introduction}}

Bremsstrahlung photons emitted in nuclear reactions, has been used an independent tool for obtaining new information about internal mechanisms, interactions in such reactions.
In theory, such research requires to develop detailed models and complicated numerical calculations.
Experiments with measurements of photons are usually enough seldom, they requires to use additional special facilities (technique).
But, this way can allow to reach more accurate and rich information about studied nuclear objects~\cite{Maydanyuk_Zhang_Zou.2019.PRC.microscopy,Liu_Maydanyuk_Zhang_Liu.2019.PRC.hypernuclei}.

\vspace{2.0mm}
From literature one can find, that bremsstrahlung photons were studied the most often in fission of nuclei
(a main focus was given to nucleus \isotope[252]{Cm}, for example, see experimental study~\cite{Glassel.1989.NPA,Luke.1991.PRC,Ploeg.1992.PRL,Hofman.1993.PRC,Ploeg.1995.PRC,Pandit.2009.DAESNP,Pandit.2010.PLB,Kasagi.1989.JPSJ,Varlachev.2007.BRASP,Eremin.2010.IJMPE})
%
%
and scattering of protons off nuclei (for example, see measurements and analysis of data~\cite{Edington.1966.NP,Koehler.1967.PRL,Kwato_Njock.1988.PLB,Pinston.1989.PLB,Pinston.1990.PLB,Clayton.1992.PRC,
Pluiko.1987.PEPAN,Kamanin.1989.PEPAN,Clayton.1991.PhD,Chakrabarty.1999.PRC,Goethem.2002.PRL},
theoretical study in different models~\cite{Nakayama.1986.PRC,Nakayama.1989.PRC,Nakayama.1989.PRCv40,Knoll.1989.NPA,Herrmann.1991.PRC,Liou.1987.PRC,
Liou.1993.PRC,Liou.1995.PLB.v345,Liou.1995.PLB.v355,Liou.1996.PRC,Li.1998.PRC.v57,Li.1998.PRC.v58,Timmermans.2001.PRC,
Liou.2004.PRC,Li.2005.PRC,Timmermans.2006.PRC,Li.2011.PRC,Kurgalin.2001.IRAN}).
Comparing with last reaction, understanding of nucleus-nucleus interactions is more complicated problem.
The first task in this road is study of interactions between $\alpha$-particles and nuclei.
$\alpha$-Particle has already internal many-nucleon structure. On the other side, $\alpha$-particle is one of the most compact (i.e. strong) nuclear fragments, which can be used in scattering to investigate internal structure of nuclei.

\vspace{2.0mm}
Interactions between $\alpha$-particles and nuclei have been studied for a long time. Such interactions are included in three topics:
(1) $\alpha$-decay,
(2) inverse processes to decays (i.e., $\alpha$-capture),
(3) scattering of $\alpha$-particles off nuclei.
These interactions via $\alpha$ decays of nuclei have been investigated enough deeply and well
(see reviews, tables~\cite{Buck.1993.ADNDT,Akovali.1998.NDS,Duarte.2002.ADNDT,Audi.2003.NPA,Dasgupta-Schubert.2007.ADNDT,Silisteanu.2012.ADNDT,
Lovas.1998.PRep,Sobiczewski.2007.PPNP,Denisov.2009.ADNDT},
some papers~\cite{Denisov.2015.PRC,Stewart.1996.NPA,Xu.2006.PRC,Nazarewicz.2012.PRC,Delion.2013.PRC,Silisteanu.2015.RJP,Silisteanu.2017.PRC}
and database~\cite{www_library}).
Today, $\alpha$ decay is observed for more than 420 nuclei with $A > 105$ and $Z > 52$.
More than 1246 $\alpha$-emitters are tabulated, with half-lives in range of $10^{-9} < T_{1/2} < 10^{+38}$~sec \cite{Denisov.2009.ADNDT}.
Spin, parities, hexadecapole deformations, excited states, shell and cluster structures of decaying nuclei are studied.
%
However, energies of the $\alpha$-particles emitted from ground states of nuclei are inside low energy region with essentially restricted upper limit (it is usually not more 10 MeV).
This fact excludes many physical phenomena of $\alpha$-nucleus interactions from analysis.

\vspace{2.0mm}
In contrast to $\alpha$ decay, $\alpha$-captures cover more wide region of energies (7--30~MeV in measurements).
Fusion in this reaction attracts researchers for a long time, as open questions (related with understanding of nuclear forces, strong quantum phenomena, many-nucleon dynamics, etc.) exist
(see the current status of the experimental and theoretical investigations on topic of fusion in heavy-ion collisions in review~\cite{Back.2014.RMP},
also see Refs.~\cite{Birkelund.1979.PRep,Vaz.1981.PRep,Birkelund.1983.ARNPS,Beckerman.1985.PRep,Steadman.1986.ARNPS,Beckerman.1988.RPP,
Rowley.1991.PLB,Vandenbosch.1992.ARNPS,Reisdorf.1994.JPG,Dasgupta.1998.ARNPS,Balantekin.1998.RMP,
Liang.2005.IJMPE,Canto.2006.PRep,Keeley.2007.PPNP,Hagino.2012.PTP},
for the $\alpha$-capture see Refs.~\cite{Glas.1975.NPA,Glas.1974.PRC}).
In capture, there are indefinite places where models use different description (based on different understanding) but still give similar results in description of experimental cross-sections~\cite{Maydanyuk_Zhang_Zou.2017.PRC,Maydanyuk.2015.NPA}.
There are experimental capture cross-sections just for five nuclei \isotope[40]{Ca}, \isotope[44]{Ca} \cite{Eberhard.1979.PRL}, \isotope[59]{Co}~\cite{DAuria.1968.PR}, \isotope[208]{Pb}~\cite{Barnett.2000.PRC} and \isotope[209]{Bi}~\cite{Barnett.2000.PRC}.

\vspace{2.0mm}
In scattering of $\alpha$-particles off nuclei, energies of $\alpha$-particles in beams are inside really wide region,
such reactions are more suitable for experimental study than $\alpha$-capture.
So, scattering provides us possibility to investigate $\alpha$-nucleus interactions in the full volume, in contrast to $\alpha$-decay and $\alpha$-capture.
In scattering, many people put attention to different aspects of internal mechanisms, nucleon-nucleon relations, structure in $\alpha$-nucleus interactions (for example,
see Refs.~\cite{Karpeshin.2007.JPG,Gaul.1969.NPA}).
Understanding of clusters in nuclei is hot topic, it finds its natural place in investigations of scattering.

\vspace{2.0mm}
In such a situation, we put attention to bremsstrahlung photons which are emitted during scattering of $\alpha$-particles off nuclei.
We suppose that these photons give independent information about internal mechanisms during scattering.
Bremsstrahlung photons in energy range about 5--35~MeV were measured
in scattering of $\alpha + \isotope[197]{Au}, \isotope[159]{Tb}$ at $E_{\alpha}=40$~MeV and
$\alpha + \isotope[197]{Au}, \isotope[116]{Sn}, \isotope[\rm nat]{Ag}, \isotope[59]{Co}$ at $E_{\alpha}=50$~MeV
at Variable Energy Cyclotron Centre, Calcutta~\cite{Sharan.1993.PRC}.
Such experimental data were analyzed by statistical model for photon energies up to the giant dipole resonance region, and
were analyzed for hither photon energies in terms of the potential bremsstrahlung and incoherent nucleon-nucleon bremsstrahlung processes
[see Ref.~\cite{Sharan.1993.PRC}].
%
%
We are interesting in physics of these reactions via self-consistent analysis on the fully quantum basis.
In particular, we estimate larger role of quantum effects, related with tunneling phenomenon during scattering.
As a natural idea, one can suppose more important role of the resonant scattering at some above-barrier energies closer to barrier maximum and especially at some under-battier energies.
Note that for under-barrier energies quantum effects are important as their inclusion to model and calculations of cross-sections of nuclear reactions can changes results on some orders (see
Refs.~\cite{Maydanyuk.2015.NPA,Maydanyuk_Zhang_Zou.2017.PRC}, for details, explanations).
In particular, their role is larger even than role of nuclear deformations, some other nuclear characteristics.
This is especially important when we want to understand and estimate internal tiny mechanisms in nuclear reactions.

\vspace{2.0mm}
In this paper we generalize out bremsstrahlung theory~\cite{Liu_Maydanyuk_Zhang_Liu.2019.PRC.hypernuclei,Maydanyuk.2003.PTP,Maydanyuk.2006.EPJA,Maydanyuk.2008.EPJA,Maydanyuk.2008.MPLA,Maydanyuk.2009.JPS,%
Maydanyuk.2009.TONPPJ,Maydanyuk.2009.NPA,Maydanyuk.2010.PRC,Maydanyuk.2011.JPG,Maydanyuk.2011.JPCS,Maydanyuk.2012.PRC,Maydanyuk_Zhang.2015.PRC,%
Maydanyuk_Zhang_Zou.2016.PRC,Maydanyuk_Zhang_Zou.2018.PRC} and apply it for analysis of data~\cite{Sharan.1993.PRC}.
As energy region of the measured photons in the $\alpha$-nucleus scattering is essentially larger than in $\alpha$-decay
(there are experimental data for the bremsstrahlung in $\alpha$ decay for four nuclei:
\isotope[210]{Po}~\cite{D'Arrigo.1994.PHLTA,Kasagi.1997.JPHGB,Kasagi.1997.PRLTA,Boie.2007.PRL,Boie.2009.PhD},
\isotope[214]{Po}~\cite{Maydanyuk.2008.EPJA},
\isotope[226]{Ra} \cite{Maydanyuk.2008.MPLA} and \isotope[244]{Cm};
here energies of experimental bremsstrahlung probabilities are not larger than 500~keV),
we focus on extracting new information about coherent and incoherent emissions in elastic scattering, different electric and magnetic contributions, inelastic mechanisms in this reaction via analysis of bremsstrahlung emission.
At present, we do not see any alternative way to estimate (basin on experiments) ratio between coherent and incoherent mechanisms in scattering,
that gives bremsstrahlung analysis as unique tools~\cite{Maydanyuk_Zhang.2015.PRC,Maydanyuk_Zhang_Zou.2016.PRC}.
We explore (investigate) possibility of existence of more compact (cluster) structures of nuclei at some energies.

\section{Model
\label{sec.model}}

\subsection{Generalized Pauli equation for nucleons in the $\alpha$--nucleus system and operator of emission of photons
\label{sec.2.1}}


Let us consider $\alpha$-particle interacting with nucleus.
To describe evolution of nucleons of such a complicated system  in the laboratory frame
(we have $A+4$ nucleons of the system of nucleus and $\alpha$-particle), we shall use many-nucleon generalization of Pauli equation.
In this formulation, we follow to our previous idea in Ref.~\cite{Liu_Maydanyuk_Zhang_Liu.2019.PRC.hypernuclei} (see Eqs.(1)--(6) and explanations in that paper).
We define Hamiltonian $\hat{H}_{0}$ describing evolution of nucleons in the scattering of the $\alpha$-particle off nucleus-target (without photons) in form (4) in Ref.~\cite{Liu_Maydanyuk_Zhang_Liu.2019.PRC.hypernuclei},
and we define operator of emission $\hat{H}_{\gamma}$ of bremsstrahlung photon in such a reaction in form of (5)--(6) in Ref.~\cite{Liu_Maydanyuk_Zhang_Liu.2019.PRC.hypernuclei}
(we neglect terms at $\vb{A}_{j}^{2}$ and $A_{j,0}$, and use Coulomb gauge):
\begin{equation}
\begin{array}{lll}
  \hat{H}_{\gamma} =
  \displaystyle\sum_{i=1}^{4}
  \biggl\{
    - \displaystyle\frac{z_{i} e}{m_{i}c}\; \vb{A}_{i}\, \vu{p}_{i} -
    \mu_{i}^{\rm (an)}\, \sigmabf \cdot \vu{H}_{i}
  \biggr\} +
  \displaystyle\sum_{j=1}^{A}
  \biggl\{
    - \displaystyle\frac{z_{j} e}{m_{j}c}\; \vb{A}_{j}\, \vu{p}_{j} -
    \mu_{j}^{\rm (an)}\, \sigmabf \cdot \vu{H}_{j}
  \biggr\}, &
  \vu{H} = \vb{rot\: A} = \bigl[ \curl{\vb{A}} \bigr].
\end{array}
\label{eq.2.2.5}
\end{equation}
Here, $m_{i}$ and $z_{i}$ are mass and electric charge of nucleon with number $i$,
$\mu_{i}^{\rm (an)}$ are anomalous magnetic momenta for nucleons,
$\vu{p}_{i} = -i\hbar\, \vb{d}/\vb{dr}_{i} $ is momentum operator for nucleon with number $i$,
$V(\vb{r}_{1} \ldots \vb{r}_{A+4})$ is general form of the potential of interactions between nucleons,
$\sigmabf$ are Pauli matrixes,
$A_{i} = (\vb{A}_{i}, A_{i,0})$ is potential of electromagnetic field formed by moving nucleon with number $i$,
$A$ in summation is mass number of a nucleus-target.
We use $\mu_{\rm p}^{\rm (an)} = 2.79284734462\: \mu_{N}$ for proton, $\mu_{\rm n}^{\rm (an)} = -1.91304273\: \mu_{N}$ for neutron \cite{RewPartPhys_PDG.2018},
where $\mu_{N} = e\hbar / (2m_{\rm p}c)$ is nuclear magneton,
$m_{\rm p}$ is mass of proton.


We define the potential of electromagnetic field as (7) in Ref.~\cite{Liu_Maydanyuk_Zhang_Liu.2019.PRC.hypernuclei}:
\begin{equation}
\begin{array}{lcl}
  \vb{A} & = &
  \displaystyle\sum\limits_{\alpha=1,2}
    \sqrt{\displaystyle\frac{2\pi\hbar c^{2}}{w_{\rm ph}}}\; \vb{e}^{(\alpha),\,*}
    e^{-i\, \vb{k_{\rm ph}r}},
\end{array}
\label{eq.2.3.1}
\end{equation}
we obtain $\vu{H}$ in form (8) of~\cite{Liu_Maydanyuk_Zhang_Liu.2019.PRC.hypernuclei}.
%
%
Here, $\vb{e}^{(\alpha)}$ are unit vectors of polarization of the photon emitted [$\vb{e}^{(\alpha), *} = \vb{e}^{(\alpha)}$], $\vb{k}_{\rm ph}$ is wave vector of the photon and $w_{\rm ph} = k_{\rm ph} c = \bigl| \vb{k}_{\rm ph}\bigr|c$. Vectors $\vb{e}^{(\alpha)}$ are perpendicular to $\vb{k}_{\rm ph}$ in Coulomb calibration. We have two independent polarizations $\vb{e}^{(1)}$ and $\vb{e}^{(2)}$ for the photon with impulse $\vb{k}_{\rm ph}$ ($\alpha=1,2$). One can develop formalism simpler in the system of units where $\hbar = 1$ and $c = 1$, but we shall write constants $\hbar$ and $c$ explicitly.
Vectors $\vb{e}^{(\alpha)}$ satisfy to properties (9) ~\cite{Liu_Maydanyuk_Zhang_Liu.2019.PRC.hypernuclei}.
We substitute Eq.~(\ref{eq.2.3.1}) to formula~(\ref{eq.2.2.5}) for operator of emission and obtain:
\begin{equation}
\begin{array}{lcl}
  \hat{H}_{\gamma} & = &
  \sqrt{\displaystyle\frac{2\pi\hbar c^{2}}{w_{\rm ph}}}\;
  \displaystyle\sum_{i=1}^{4}
  \displaystyle\sum\limits_{\alpha=1,2}
    e^{-i\, \vb{k_{\rm ph}r}_{i}}\,
  \biggl\{
    i\, \mu_{N}\, \displaystyle\frac{2 z_{i} m_{\rm p}}{m_{\alpha i}}\: \vb{e}^{(\alpha)} \cdot \grad_{i} +
    \mu_{i}^{\rm (an)}\, \sigmabf \cdot \Bigl( i\, \bigl[ \vb{k_{\rm ph}} \times \vb{e}^{(\alpha)} \bigr] - \bigl[ \grad_{i} \times \vb{e}^{(\alpha)} \bigr] \Bigr)
  \biggr\}\; + \\

  & + &
  \sqrt{\displaystyle\frac{2\pi\hbar c^{2}}{w_{\rm ph}}}\;
  \displaystyle\sum_{j=1}^{A}
  \displaystyle\sum\limits_{\alpha=1,2}
    e^{-i\, \vb{k_{\rm ph}r}_{j}}\;
    \biggl\{
      i\, \mu_{N}\, \displaystyle\frac{2 z_{j} m_{\rm p}}{m_{Aj}}\: \vb{e}^{(\alpha)} \cdot \grad_{j} +
      \mu_{j}^{\rm (an)}\, \sigmabf \cdot \Bigl( i\, \bigl[ \vb{k_{\rm ph}} \times \vb{e}^{(\alpha)} \bigr] - \bigl[ \grad_{j} \times \vb{e}^{(\alpha)} \bigr] \Bigr)
    \biggr\}.
\end{array}
\label{eq.2.3.4}
\end{equation}

\subsection{Transition to coordinates of relative distances
\label{sec.2.4}}

We rewrite formalism above via coordinates of relative distances.
We define these coordinates and corresponding momenta,
along to formalism~\cite{Liu_Maydanyuk_Zhang_Liu.2019.PRC.hypernuclei} (see Eqs.~(11)--(14), and Appendix~A in that paper).
We define coordinate of centers of masses for the $\alpha$ particle  as $\vb{r}_{\alpha}$, for the nucleus-target as $\vb{R}_{A}$, and
for the complete system as $\vb{R}$:
%
%
\begin{equation}
\begin{array}{lll}
   \vb{r}_{\alpha} = \displaystyle\frac{1}{m_{\alpha}} \displaystyle\sum_{i=1}^{4} m_{i}\, \vb{r}_{\alpha i}, &
   \vb{R}_{A}      = \displaystyle\frac{1}{m_{A}} \displaystyle\sum_{j=1}^{A} m_{j}\, \vb{r}_{A j}, &
   \vb{R}          = \displaystyle\frac{m_{A}\vb{R}_{A} + m_{\alpha}\vb{r}_{\alpha}}{m_{A}+m_{\alpha}} =
     c_{A}\, \vb{R}_{A} + c_{\alpha}\, \vb{R}_{\alpha},
\end{array}
\label{eq.2.4.1.1}
\end{equation}
where $m_{\alpha}$ and $m_{A}$ are masses of the $\alpha$ particle and nucleus-target, and
we introduced new coefficients $c_{A} = \frac{m_{A}}{m_{A}+m_{\alpha}}$ and $c_{\alpha} = \frac{m_{\alpha}}{m_{A}+m_{\alpha}}$.
Introducing new relative coordinate $\vb{r}$,
new relative coordinates $\rhobf_{\alpha i}$ for nucleons of the $\alpha$-particle,
new relative coordinates $\rhobf_{A j}$ for nucleons (with possible hyperon) for the nucleus-target
as
\begin{equation}
\begin{array}{lll}
   \vb{r} = \vb{r}_{\alpha} - \vb{R}_{A}, &
   \rhobf_{\alpha i} = \vb{r}_{\alpha i} - \vb{r}_{\alpha}, &
   \rhobf_{A j} = \vb{r}_{j} - \vb{R}_{A},
\end{array}
\label{eq.2.4.1.2}
\end{equation}
we obtain new independent variables $\vb{R}$, $\vb{r}$,
$\rhobf_{\alpha j}$ ($i=1, 2, 3$),
$\rhobf_{Aj}$ ($j=1 \ldots A-1$), and corresponding momenta.
We find ($n=4$):
\begin{equation}
\begin{array}{lll}
  \vb{r}_{\alpha i} = \rhobf_{\alpha i} + \vb{R} + c_{A}\, \vb{r}, &
  \vb{r}_{Aj} = \rhobf_{A j} + \vb{R} - c_{\alpha}\, \vb{r}, \\
  \vb{r}_{\alpha n} = \vb{R} + c_{A} \vb{r} - \displaystyle\frac{1}{m_{n}} \displaystyle\sum_{k=1}^{n-1} m_{k}\, \rhobf_{\alpha k}, &
  \vb{r}_{AA} = \vb{R} - c_{\alpha} \vb{r} - \displaystyle\frac{1}{m_{AA}} \displaystyle\sum_{k=1}^{A-1} m_{k}\, \rhobf_{A k},
\end{array}
\label{eq.2.4.1.6}
\end{equation}
\begin{equation}
\begin{array}{lclll}
  \vspace{1mm}
  \vu{p}_{\alpha i} =
  \displaystyle\frac{m_{\alpha i}}{m_{A} + m_{\alpha}}\, \vu{P} +
    \displaystyle\frac{m_{\alpha i}}{m_{\alpha}}\,\vu{p} +
    \displaystyle\frac{m_{\alpha} - m_{\alpha i}}{m_{\alpha}}\, \vb{\tilde{p}}_{\alpha i} -
    \displaystyle\frac{m_{\alpha i}}{m_{\alpha}}\,
      \displaystyle\sum_{k=1, k \ne i}^{n-1} \vb{\tilde{p}}_{\alpha k} \quad
      {\rm at }\; i = 1 \ldots n-1, \\

  \vspace{1mm}
  \vu{p}_{\alpha n} =
  \displaystyle\frac{m_{\alpha n}}{m_{A} + m_{\alpha}}\, \vu{P} +
    \displaystyle\frac{m_{\alpha n}}{m_{\alpha}}\,\vu{p} -
    \displaystyle\frac{m_{\alpha n}}{m_{\alpha}}\,
      \displaystyle\sum_{k=1}^{n-1} \vb{\tilde{p}}_{\alpha k}, \\

  \vspace{1mm}
  \vu{p}_{Aj} =
  \displaystyle\frac{m_{Aj}}{m_{A} + m_{\alpha}}\, \vu{P} -
    \displaystyle\frac{m_{Aj}}{m_{A}}\,\vu{p} +
    \displaystyle\frac{m_{A} - m_{Aj}}{m_{A}}\, \vb{\tilde{p}}_{Aj} -
    \displaystyle\frac{m_{Aj}}{m_{A}}\,
      \displaystyle\sum_{k=1, k \ne j}^{A-1} \vb{\tilde{p}}_{Ak} \quad {\rm at }\; j = 1 \ldots A-1, \\

  \vspace{1mm}
  \vu{p}_{AA} =
  \displaystyle\frac{m_{AA}}{m_{A} + m_{\alpha}}\, \vu{P} -
    \displaystyle\frac{m_{AA}}{m_{A}}\,\vu{p} -
    \displaystyle\frac{m_{AA}}{m_{A}}\,
      \displaystyle\sum_{k=1}^{A-1} \vb{\tilde{p}}_{Ak},
\end{array}
\label{eq.2.4.3.3}
\end{equation}
where $\vb{\hat{P}}$, $\vb{\hat{p}}$, $\vb{\tilde{p}}_{\alpha i}$ and $\vb{\tilde{p}}_{Aj}$ are momenta corresponding to variables $\vb{R}$, $\vb{r}$, $\rhobf_{\alpha i}$, $\rhobf_{Aj}$.

\subsection{Operator of emission in relative coordinates
\label{sec.2.5}}

Now we will find operator of emission in relative coordinates.
We start from (\ref{eq.2.3.4}), rewriting this expression via relative momenta.
Substituting formulas (\ref{eq.2.4.3.3}) to these expressions, we find (see calculations in Appendix~\ref{sec.app.1}):
\begin{equation}
  \hat{H}_{\gamma} = \hat{H}_{P} + \hat{H}_{p} + \Delta \hat{H}_{\gamma E} + \Delta \hat{H}_{\gamma M} + \hat{H}_{k},
\label{eq.2.5.2}
\end{equation}
where
\begin{equation}
\begin{array}{lcl}
\vspace{-0.1mm}
  & & \hat{H}_{P} =
  -\, \sqrt{\displaystyle\frac{2\pi c^{2}}{\hbar w_{\rm ph}}}\;
  \mu_{N}\, \displaystyle\frac{2 m_{\rm p}}{m_{A} + m_{\alpha}}\;
  e^{-i\, \vb{k_{\rm ph}} \vb{R}}
  \displaystyle\sum\limits_{\alpha=1,2}
  \biggl\{
    e^{-i\, c_{A}\, \vb{k_{\rm ph}} \vb{r}}
      \displaystyle\sum_{i=1}^{4} z_{i}\, e^{-i\, \vb{k_{\rm ph}} \rhobf_{\alpha i}} +
    e^{i\, c_{\alpha}\, \vb{k_{\rm ph}} \vb{r} }
      \displaystyle\sum_{j=1}^{A} z_{j}\, e^{-i\, \vb{k_{\rm ph}} \rhobf_{Aj}}
  \biggr\}\, \vb{e}^{(\alpha)} \cdot \vu{P}\; - \\
  & - &
  \sqrt{\displaystyle\frac{2\pi c^{2}}{\hbar w_{\rm ph}}}\;
    \displaystyle\frac{i}{m_{A} + m_{\alpha}}\;
    e^{-i\, \vb{k_{\rm ph}} \vb{R}}\,
    \displaystyle\sum\limits_{\alpha=1,2}
    \biggl\{
      e^{-i\, c_{A}\, \vb{k_{\rm ph}} \vb{r}}\,
        \displaystyle\sum_{i=1}^{4} \mu_{i}^{\rm (an)}\, m_{\alpha i}\, e^{-i\, \vb{k_{\rm ph}} \rhobf_{\alpha i}}\, \sigmabf  +
      e^{i\, c_{\alpha}\, \vb{k_{\rm ph}} \vb{r}}\,
        \displaystyle\sum_{j=1}^{A} \mu_{j}^{\rm (an)}\, m_{Aj}\, e^{-i\, \vb{k_{\rm ph}} \rhobf_{Aj}}\, \sigmabf
  \biggr\}\, \cdot \bigl[ \vu{P} \times \vb{e}^{(\alpha)} \bigr],
\end{array}
\label{eq.2.5.3}
\end{equation}
\begin{equation}
\begin{array}{lll}
\vspace{-0.1mm}
  & & \hat{H}_{p} =
  -\, \sqrt{\displaystyle\frac{2\pi c^{2}}{\hbar w_{\rm ph}}}\;
  2\, \mu_{N}\,  m_{\rm p}\,
  e^{-i\, \vb{k_{\rm ph}} \vb{R}}
  \displaystyle\sum\limits_{\alpha=1,2}
  \biggl\{
    e^{-i\, c_{A} \vb{k_{\rm ph}} \vb{r}}\, \displaystyle\frac{1}{m_{\alpha}}\,
      \displaystyle\sum_{i=1}^{4} z_{i}\, e^{-i\, \vb{k_{\rm ph}} \rhobf_{\alpha i}} -
    e^{i\, c_{\alpha} \vb{k_{\rm ph}} \vb{r}}\,  \displaystyle\frac{1}{m_{A}}\,
      \displaystyle\sum_{j=1}^{A} z_{j}\, e^{-i\, \vb{k_{\rm ph}} \rhobf_{Aj}}
  \biggr\}\; \vb{e}^{(\alpha)} \cdot \vu{p}\; - \\
  & - &
  i\, \sqrt{\displaystyle\frac{2\pi c^{2}}{\hbar w_{\rm ph}}}\;
  e^{-i\, \vb{k_{\rm ph}} \vb{R}}
  \displaystyle\sum\limits_{\alpha=1,2}
  \biggl\{
    e^{-i\, c_{A} \vb{k_{\rm ph}} \vb{r}} \displaystyle\frac{1}{m_{\alpha}}\,
    \displaystyle\sum_{i=1}^{4}
      \mu_{i}^{\rm (an)}\, m_{\alpha i}\;
      e^{-i\, \vb{k_{\rm ph}} \rhobf_{\alpha i}}\, \sigmabf -
    e^{i\, c_{\alpha} \vb{k_{\rm ph}} \vb{r}} \displaystyle\frac{1}{m_{A}}
    \displaystyle\sum_{j=1}^{A}
      \mu_{j}^{\rm (an)}\, m_{Aj}\;
      e^{-i\, \vb{k_{\rm ph}} \rhobf_{Aj}}\, \sigmabf \biggr\}
  \cdot \bigl[ \vu{p} \times \vb{e}^{(\alpha)} \bigr].
\end{array}
\label{eq.2.5.4}
\end{equation}
\begin{equation}
\begin{array}{lcl}
  \hat{H}_{k} & = &
  i\, \hbar\,
  \sqrt{\displaystyle\frac{2\pi c^{2}}{\hbar w_{\rm ph}}}\:
  e^{-i\, \vb{k_{\rm ph}} \vb{R}}\,
  \displaystyle\sum\limits_{\alpha=1,2}
  \biggl\{
    e^{-i\, c_{A}\, \vb{k_{\rm ph}} \vb{r}}\,
    \displaystyle\sum_{i=1}^{4}
      \mu_{i}^{\rm (an)}\, e^{-i\, \vb{k_{\rm ph}} \rhobf_{\alpha i}}\, \sigmabf +
    e^{i\, c_{\alpha}\, \vb{k_{\rm ph}} \vb{r}}\,
    \displaystyle\sum_{j=1}^{A}
      \mu_{j}^{\rm (an)}\, e^{-i\, \vb{k_{\rm ph}} \rhobf_{Aj}}\, \sigmabf
  \biggr\}
  \cdot \bigl[ \vb{k_{\rm ph}} \times \vb{e}^{(\alpha)} \bigr].
\end{array}
\label{eq.2.5.5}
\end{equation}
A summation of expression (\ref{eq.2.5.4}) and $\hat{H}_{k}$ is many-nucleon generalization of operator of emission $\hat{W}$ in Eq.~(6) in Ref.~\cite{Maydanyuk.2012.PRC} with
included anomalous magnetic momenta for nucleons.
$\Delta \hat{H}_{\gamma E}$ and $\Delta \hat{H}_{\gamma M}$ are calculated in Appendix~\ref{sec.app.1} [see Eqs.~(\ref{eq.app.1.3.3})--(\ref{eq.app.1.3.4}) in this Section].
These expressions include only internal moments of nucleons, including and excluding spin:
\begin{equation}
\begin{array}{lcl}
\vspace{0.4mm}
  \Delta \hat{H}_{\gamma E} & = &
  -\, \sqrt{\displaystyle\frac{2\pi c^{2}}{\hbar w_{\rm ph}}}\:
    2\, \mu_{N}\, e^{-i\, \vb{k_{\rm ph}}\vb{R}}\,
    \displaystyle\sum\limits_{\alpha=1,2} \vb{e}^{(\alpha)}\; \times \\
\vspace{0.4mm}
  & \times &
  \biggl\{
  \biggl[
    e^{-i\, c_{A}\, \vb{k_{\rm ph}} \vb{r}}\,
    \displaystyle\sum_{i=1}^{n-1}
      \displaystyle\frac{z_{i}\,m_{\rm p}}{m_{\alpha i}}\, e^{-i\, \vb{k_{\rm ph}} \rhobf_{\alpha i}}\, \vb{\tilde{p}}_{\alpha i} +
    e^{i\, c_{\alpha}\, \vb{k_{\rm ph}} \vb{r}}\,
    \displaystyle\sum_{j=1}^{A-1}
      \displaystyle\frac{z_{j}\,m_{\rm p}}{m_{Aj}}\, e^{-i\, \vb{k_{\rm ph}} \rhobf_{Aj}}\, \vb{\tilde{p}}_{Aj}
  \biggr]\; - \\
  & - &
  \biggl[
    \displaystyle\frac{m_{\rm p}}{m_{\alpha}}\, e^{-i\, c_{A}\, \vb{k_{\rm ph}} \vb{r}}\,
      \displaystyle\sum_{i=1}^{n} z_{i}\, e^{-i\, \vb{k_{\rm ph}} \rhobf_{\alpha i}}\, \displaystyle\sum_{k=1}^{n-1} \vb{\tilde{p}}_{\alpha k} +
    \displaystyle\frac{m_{\rm p}}{m_{A}}\, e^{i\, c_{\alpha}\, \vb{k_{\rm ph}} \vb{r}}\,
      \displaystyle\sum_{j=1}^{A} z_{j}\, e^{-i\, \vb{k_{\rm ph}} \rhobf_{Aj}}\, \displaystyle\sum_{k=1}^{A-1} \vb{\tilde{p}}_{Ak}
  \biggr]
  \biggr\},
\end{array}
\label{eq.2.5.9}
\end{equation}
\begin{equation}
\begin{array}{lll}
\vspace{0.4mm}
  & \Delta \hat{H}_{\gamma M} =
  -\, i\, \sqrt{\displaystyle\frac{2\pi c^{2}}{\hbar w_{\rm ph}}}\;
    e^{-i\, \vb{k_{\rm ph}} \vb{R}}\,
  \displaystyle\sum\limits_{\alpha=1,2} \; \times \\

\vspace{0.4mm}
  \times &
  \biggl\{
  \biggl[
    e^{-i\, \vb{k_{\rm ph}} c_{A}\, \vb{r}}\,
    \displaystyle\sum_{i=1}^{n-1}
      \mu_{i}^{\rm (an)}\, e^{-i\, \vb{k_{\rm ph}} \rhobf_{\alpha i}}\, \sigmabf \cdot \bigl[ \vb{\tilde{p}}_{\alpha i} \times \vb{e}^{(\alpha)} \bigr] +

    e^{i\, \vb{k_{\rm ph}} c_{\alpha}\, \vb{r}}\,
    \displaystyle\sum_{j=1}^{A-1}
      \mu_{j}^{\rm (an)}\, e^{-i\, \vb{k_{\rm ph}} \rhobf_{Aj}}\, \sigmabf \cdot \bigl[ \vb{\tilde{p}}_{Aj} \times \vb{e}^{(\alpha)} \bigr]
  \biggr]\; - \\

  \times &
  \biggl[
    e^{-i\, \vb{k_{\rm ph}} c_{A}\, \vb{r}}\,
    \displaystyle\sum_{i=1}^{n}
      \mu_{i}^{\rm (an)}\, \displaystyle\frac{m_{\alpha i}}{m_{\alpha}}\,
      e^{-i\, \vb{k_{\rm ph}} \rhobf_{\alpha i}}\,
      \displaystyle\sum_{k=1}^{n-1} \sigmabf \cdot \bigl[ \vb{\tilde{p}}_{\alpha k} \times \vb{e}^{(\alpha)} \bigr] +

    e^{i\, \vb{k_{\rm ph}} c_{\alpha}\, \vb{r}}\,
    \displaystyle\sum_{j=1}^{A}
      \mu_{j}^{\rm (an)}\, \displaystyle\frac{m_{Aj}}{m_{A}}\,
      e^{-i\, \vb{k_{\rm ph}} \rhobf_{Aj}}\,
      \displaystyle\sum_{k=1}^{A-1} \sigmabf \cdot \bigl[ \vb{\tilde{p}}_{Ak} \times \vb{e}^{(\alpha)} \bigr]
  \biggr\}.
\end{array}
\label{eq.2.5.10}
\end{equation}
%

\subsection{Matrix element of emission in the $\alpha$-nucleus scattering
\label{sec.2.6}}

We define the wave function of the $\alpha$-nucleus system, following logic in Ref.~\cite{Liu_Maydanyuk_Zhang_Liu.2019.PRC.hypernuclei} (see Eqs.~(20) and (21) in that paper, and explanations):
%
%
\begin{equation}
  \Psi = \Phi (\vb{R}) \cdot \Phi_{\rm \alpha - nucl} (\vb{r}) \cdot \psi_{\rm nucl} (\beta_{A}) \cdot \psi_{\alpha} (\beta_{\alpha}) + \Delta \Psi.
\label{eq.2.6.1}
\end{equation}
%
%
%
%
Here, $\beta_{\alpha}$ is the set of numbers $1 \cdots 4$ of nucleons of the $\alpha$ particle,
$\beta_{A}$ is the set of numbers $1 \cdots A$ of nucleons of the nucleus,
$\Phi (\vb{R})$ is the function describing motion of center-of-mass of the full nuclear system,
$\Phi_{\rm \alpha - nucl} (\vb{r})$ is the function describing relative motion of the $\alpha$ particle concerning to nucleus (without description of internal relative motions of nucleons in the $\alpha$ particle and nucleus),
$\psi_{\alpha} (\beta_{\alpha})$ is the many-nucleon function dependent on nucleons of the $\alpha$ particle
defined in Eq.~(21) in Ref.~\cite{Liu_Maydanyuk_Zhang_Liu.2019.PRC.hypernuclei} (it determines space state on the basis of relative distances $\rhobf_{1} \cdots \rhobf_{4}$ of nucleons of the $\alpha$ particle concerning to its center-of-mass),
$\psi_{\rm nucl} (\beta_{A})$ is the many-nucleon function dependent on nucleons of the nucleus defined in Eq.~(21) in Ref.~\cite{Liu_Maydanyuk_Zhang_Liu.2019.PRC.hypernuclei},
$\Delta \Psi$ is correction from fully anti-symmetric formulation of wave function for all nucleons
(in this work we shall neglect by this correction).
One-nucleon functions $\psi_{\lambda_{s}}(s)$ in Eq.~(21) in Ref.~\cite{Liu_Maydanyuk_Zhang_Liu.2019.PRC.hypernuclei} represent the multiplication of space and spin-isospin functions as
$\psi_{\lambda_{s}} (s) = \varphi_{n_{s}} (\vb{r}_{s})\, \bigl|\, \sigma^{(s)} \tau^{(s)} \bigr\rangle$,
%
%
where
$\varphi_{n_{s}}$ is the space function of the nucleon with number $s$,
$n_{s}$ is the number of state of the space function of the nucleon with number $s$,
$\bigl|\, \sigma^{(s)} \tau^{(s)} \bigr\rangle$ is the spin-isospin function of the nucleon with number $s$.



We define matrix element of emission of the bremsstrahlung photons, using the wave functions $\Psi_{i}$ and $\Psi_{f}$ of
the full nuclear system in states before emission of photons ($i$-state) and after such emission ($f$-state),
as:
\begin{equation}
  F = \langle \Psi_{f} |\, \hat{H}_{\gamma} |\, \Psi_{i} \rangle.
\label{eq.2.7.1}
\end{equation}
In this matrix element we should integrate over all independent variables.
These variables are space variables $\vb{R}$, $\vb{r}$, $\rhobf_{\alpha n}$, $\rhobf_{Am}$.
Here, we should take into account space representation of all used momenta $\vu{p}$, $\vu{p}$, $\vb{\tilde{p}}_{\alpha n}$, $\vb{\tilde{p}}_{A m}$
(as
$\vu{p} = -i\hbar\, \vb{d/dR}$,
$\vu{p} = -i\hbar\, \vb{d/dr}$,
$\vb{\tilde{p}}_{\alpha n} = -i\hbar\, \vb{d/d} \rhobf_{\alpha n}$,
$\vb{\tilde{p}}_{A m} = -i\hbar\, \vb{d/d} \rhobf_{Am}$).

\vspace{2mm}
Substituting formulas (\ref{eq.2.5.2})--(\ref{eq.2.5.5}) and (\ref{eq.2.5.9})--(\ref{eq.2.5.10}) for operator of emission to (\ref{eq.2.7.1}),
we obtain:
\begin{equation}
  \langle \Psi_{f} |\, \hat{H}_{\gamma} |\, \Psi_{i} \rangle \;\; = \;\;
  \sqrt{\displaystyle\frac{2\pi\, c^{2}}{\hbar w_{\rm ph}}}\,
  \Bigl\{ M_{P} + M_{p} + M_{k} + M_{\Delta E} + M_{\Delta M} \Bigr\},
\label{eq.2.7.2}
\end{equation}
where
\begin{equation}
\begin{array}{lcl}
\vspace{1mm}
  M_{P} & = &
  \sqrt{\displaystyle\frac{\hbar w_{\rm ph}} {2\pi c^{2}}}\;
  \mel{\Psi_{f}\,} {\hat{H}_{P}} {\,\Psi_{i}}\; = \\

  & = &
  -\, \displaystyle\frac{1}{m_{A} + m_{\alpha}}\,
  \displaystyle\sum\limits_{\alpha=1,2}
  \biggl\langle \Psi_{f}\, \biggl|\,
    2\, \mu_{N}\, m_{\rm p}\;
    e^{-i\, \vb{k_{\rm ph}} \vb{R}}
    \biggl\{
      e^{-i\, c_{A}\, \vb{k_{\rm ph}} \vb{r}}
        \displaystyle\sum_{i=1}^{4} z_{i}\, e^{-i\, \vb{k_{\rm ph}} \rhobf_{\alpha i}} +
      e^{i\, c_{\alpha}\, \vb{k_{\rm ph}} \vb{r} }
        \displaystyle\sum_{j=1}^{A} z_{j}\, e^{-i\, \vb{k_{\rm ph}} \rhobf_{Aj}}
    \biggr\}\, \vb{e}^{(\alpha)} \cdot \vu{P} + \\
  & + &
    i\:
    e^{-i\, \vb{k_{\rm ph}} \vb{R}}\,
    \biggl\{
      e^{-i\, c_{A}\, \vb{k_{\rm ph}} \vb{r}}\,
        \displaystyle\sum_{i=1}^{4} \mu_{i}^{\rm (an)}\, m_{\alpha i}\, e^{-i\, \vb{k_{\rm ph}} \rhobf_{\alpha i}}\, \sigmabf  +
      e^{i\, c_{\alpha}\, \vb{k_{\rm ph}} \vb{r}}\,
        \displaystyle\sum_{j=1}^{A} \mu_{j}^{\rm (an)}\, m_{Aj}\, e^{-i\, \vb{k_{\rm ph}} \rhobf_{Aj}}\, \sigmabf
  \biggr\}\, \cdot \bigl[ \vu{P} \cp \vb{e}^{(\alpha)} \bigr]\,
  \biggr|\, \Psi_{i} \biggr\rangle,
\end{array}
\label{eq.2.7.3.a}
\end{equation}
\begin{equation}
\begin{array}{lcl}
\vspace{1mm}
  M_{p} & = &
  \sqrt{\displaystyle\frac{\hbar w_{\rm ph}} {2\pi c^{2}}}\;
  \mel{\Psi_{f}\,} {\hat{H}_{p}} {\,\Psi_{i}}\; = \\

  & = &
  - \displaystyle\sum\limits_{\alpha=1,2}
    \biggl\langle
      \Psi_{f}\,
    \biggl|\,
  2\, \mu_{N}\,  m_{\rm p}\,
  e^{-i\, \vb{k_{\rm ph}} \vb{R}}
  \biggl\{
    e^{-i\, c_{A} \vb{k_{\rm ph}} \vb{r}}\, \displaystyle\frac{1}{m_{\alpha}}\,
      \displaystyle\sum_{i=1}^{4} z_{i}\, e^{-i\, \vb{k_{\rm ph}} \rhobf_{\alpha i}} -
    e^{i\, c_{\alpha} \vb{k_{\rm ph}} \vb{r}}\,  \displaystyle\frac{1}{m_{A}}\,
      \displaystyle\sum_{j=1}^{A} z_{j}\, e^{-i\, \vb{k_{\rm ph}} \rhobf_{Aj}}
  \biggr\}\; \vb{e}^{(\alpha)} \cdot \vu{p}\; + \\

  & + &
  i\,
  e^{-i\, \vb{k_{\rm ph}} \vb{R}}
  \biggl\{
    e^{-i\, c_{A} \vb{k_{\rm ph}} \vb{r}} \displaystyle\frac{1}{m_{\alpha}}\,
    \displaystyle\sum_{i=1}^{4}
      \mu_{i}^{\rm (an)}\, m_{\alpha i}\;
      e^{-i\, \vb{k_{\rm ph}} \rhobf_{\alpha i}}\, \sigmabf -
    e^{i\, c_{\alpha} \vb{k_{\rm ph}} \vb{r}} \displaystyle\frac{1}{m_{A}}
    \displaystyle\sum_{j=1}^{A}
      \mu_{j}^{\rm (an)}\, m_{Aj}\;
      e^{-i\, \vb{k_{\rm ph}} \rhobf_{Aj}}\, \sigmabf \biggr\}
    \cdot \bigl[ \vu{p} \times \vb{e}^{(\alpha)} \bigr]
  \biggr|\, \Psi_{i}\, \biggr\rangle ,
\end{array}
\label{eq.2.7.3.b}
\end{equation}
\begin{equation}
\begin{array}{lcl}
\vspace{1mm}
  M_{k} & = &
  \sqrt{\displaystyle\frac{\hbar w_{\rm ph}} {2\pi c^{2}}}\;
  \mel{\Psi_{f}\,} {\hat{H}_{k}} {\,\Psi_{i}}\; =  \\
& = &
  i\, \hbar\,
  \displaystyle\sum\limits_{\alpha=1,2}
  \biggl\langle \Psi_{f}\, \biggl|\,
    e^{-i\, \vb{k_{\rm ph}} \vb{R}}
  \biggl\{
    e^{-i\, c_{A}\, \vb{k_{\rm ph}} \vb{r}}\,
    \displaystyle\sum_{i=1}^{4}
      \mu_{i}^{\rm (an)}\, e^{-i\, \vb{k_{\rm ph}} \rhobf_{\alpha i}}\, \sigmabf +
    e^{i\, c_{\alpha}\, \vb{k_{\rm ph}} \vb{r}}\,
    \displaystyle\sum_{j=1}^{A}
      \mu_{j}^{\rm (an)}\, e^{-i\, \vb{k_{\rm ph}} \rhobf_{Aj}}\, \sigmabf
  \biggr\}
  \cdot \bigl[ \vb{k_{\rm ph}} \cp \vb{e}^{(\alpha)} \bigr]
  \biggr|\, \Psi_{i} \biggr\rangle,
\end{array}
\label{eq.2.7.3.c}
\end{equation}
\begin{equation}
\begin{array}{lcl}
\vspace{0.1mm}
  M_{\Delta E} & = &
  \sqrt{\displaystyle\frac{\hbar w_{\rm ph}} {2\pi c^{2}}}\: \mel{\Psi_{f}\,} {\hat{H}_{\gamma E}} {\,\Psi_{i}} =

  -\, \displaystyle\sum\limits_{\alpha=1,2} \vb{e}^{(\alpha)}
    \biggl\langle \Psi_{f}\, \biggl|\, 2\, \mu_{N}\, e^{-i\, \vb{k_{\rm ph}}\vb{R}}\; \times \\

\vspace{0.1mm}
  & \times &
  \biggl\{
  \biggl[
    e^{-i\, c_{A}\, \vb{k_{\rm ph}} \vb{r}}\,
    \displaystyle\sum_{i=1}^{3}
      \displaystyle\frac{z_{i} m_{\rm p}}{m_{\alpha i}}\,
      e^{-i\, \vb{k_{\rm ph}} \rhobf_{\alpha i}}\,
      \vb{\tilde{p}}_{\alpha i} +
    e^{i\, c_{\alpha}\, \vb{k_{\rm ph}} \vb{r}}\,
    \displaystyle\sum_{j=1}^{A-1}
      \displaystyle\frac{z_{j} m_{\rm p}}{m_{Aj}}\,
      e^{-i\, \vb{k_{\rm ph}} \rhobf_{Aj}}\,
      \vb{\tilde{p}}_{Aj}
  \biggr]\; - \\

  & - &
  \biggl[
    \displaystyle\frac{m_{\rm p}}{m_{\alpha}}\,
    e^{-i\, c_{A}\, \vb{k_{\rm ph}} \vb{r}}\,
    \displaystyle\sum_{i=1}^{4}
      z_{i}\, e^{-i\, \vb{k_{\rm ph}} \rhobf_{\alpha i}}\,
    \Bigl( \displaystyle\sum_{k=1}^{n-1} \vb{\tilde{p}}_{\alpha k} \Bigr) +

    \displaystyle\frac{m_{\rm p}}{m_{A}}\,
    e^{i\, c_{\alpha}\, \vb{k_{\rm ph}} \vb{r}}\,
    \displaystyle\sum_{j=1}^{A}
      z_{j}\, e^{-i\, \vb{k_{\rm ph}} \rhobf_{Aj}}
    \Bigl( \displaystyle\sum_{k=1}^{A-1} \vb{\tilde{p}}_{Ak} \Bigr)\,
  \biggr]
  \biggr\}
  \biggr|\, \Psi_{i} \biggr\rangle,
\end{array}
\label{eq.2.7.3.d}
\end{equation}
\begin{equation}
\begin{array}{lll}
\vspace{0.1mm}
  & M_{\Delta M} =
  \sqrt{\displaystyle\frac{\hbar w_{\rm ph}} {2\pi c^{2}}}\: \mel{\Psi_{f}\,} {\hat{H}_{\gamma M}} {\,\Psi_{i}}\; = 
  -\, i\, \displaystyle\sum\limits_{\alpha=1,2}
    \biggl\langle \Psi_{f}\, \biggl|\, e^{-i\, \vb{k_{\rm ph}} \vb{R}}\; \times \\

\vspace{0.3mm}
  \times &
  \biggl\{
  \biggl[
    e^{-i\, \vb{k_{\rm ph}} c_{A}\, \vb{r}}\,
    \displaystyle\sum_{i=1}^{n-1}
      \mu_{i}^{\rm (an)}\, e^{-i\, \vb{k_{\rm ph}} \rhobf_{\alpha i}}\, \sigmabf \cdot \bigl[ \vb{\tilde{p}}_{\alpha i} \times \vb{e}^{(\alpha)} \bigr] +
    e^{i\, \vb{k_{\rm ph}} c_{\alpha}\, \vb{r}}\,
    \displaystyle\sum_{j=1}^{A-1}
      \mu_{j}^{\rm (an)}\, e^{-i\, \vb{k_{\rm ph}} \rhobf_{Aj}}\, \sigmabf \cdot \bigl[ \vb{\tilde{p}}_{Aj} \times \vb{e}^{(\alpha)} \bigr]
  \biggr]\; - \\

  - &
  \Bigl[
    e^{-i\, \vb{k_{\rm ph}} c_{A}\, \vb{r}}\,
    \displaystyle\sum_{i=1}^{n}
      \mu_{i}^{\rm (an)}\, \displaystyle\frac{m_{\alpha i}}{m_{\alpha}}\,
      e^{-i\, \vb{k_{\rm ph}} \rhobf_{\alpha i}}\,
      \displaystyle\sum_{k=1}^{n-1} \sigmabf \cdot \bigl[ \vb{\tilde{p}}_{\alpha k} \times \vb{e}^{(\alpha)} \bigr] +

    e^{i\, \vb{k_{\rm ph}} c_{\alpha}\, \vb{r}}\,
    \displaystyle\sum_{j=1}^{A}
      \mu_{j}^{\rm (an)}\, \displaystyle\frac{m_{Aj}}{m_{A}}\,
      e^{-i\, \vb{k_{\rm ph}} \rhobf_{Aj}}\,
      \displaystyle\sum_{k=1}^{A-1} \sigmabf \cdot \bigl[ \vb{\tilde{p}}_{Ak} \times \vb{e}^{(\alpha)} \bigr]
  \Bigr]
  \biggr\}
  \biggr|\, \Psi_{i} \biggr\rangle.
\end{array}
\label{eq.2.7.3.e}
\end{equation}

\subsection{Integration over momentum $K_{f}$ in scattering problem and form factors
\label{sec.2.12}}

We will calculate cross-sections of emission of photons, not dependent on vector $\vb{K}_{f}$ (i.e., momentum of the full nuclear system after emission of photon in the laboratory frame). Therefore, we have to average all matrix elements over all degrees of freedom related with $\mathbf{K}_{f}$, i.e. we integrate these matrix elements over $\mathbf{K}_{f}$.
After calculations (see Appendixes~\ref{sec.2.8}, \ref{sec.2.9}, \ref{sec.2.12}, for details),
we obtain:
\begin{equation}
\begin{array}{lll}
\vspace{-0.2mm}
  M_{p} & = &
  i \hbar\, (2\pi)^{3} \displaystyle\frac{1}{\mu}\;
  \displaystyle\sum\limits_{\alpha=1,2}
  \displaystyle\int\limits_{}^{}
    \Phi_{\rm \alpha - nucl, f}^{*} (\vb{r})\; e^{-i\, \vb{k}_{\rm ph} \vb{r}}\; \times \\
  & \times &
  \biggl\{
  2\, \mu_{N}\,  m_{\rm p} \cdot Z_{\rm eff} (\vb{k}_{\rm ph}, \vb{r}) \cdot \vb{e}^{(\alpha)}\, \vb{\displaystyle\frac{d}{dr}} +

  i\,
  \vb{M}_{\rm eff} (\vb{k}_{\rm ph}, \vb{r}) \cdot \Bigl[ \vb{\displaystyle\frac{d}{dr}} \times \vb{e}^{(\alpha)} \Bigr]
  \biggr\} \cdot
  \Phi_{\rm \alpha - nucl, i} (\vb{r})\; \vb{dr}.
\end{array}
\label{eq.2.12.3}
\end{equation}
\begin{equation}
\begin{array}{lcl}
\vspace{-0.1mm}
  M_{P} & = &
  \displaystyle\frac{\hbar\, (2\pi)^{3}}{m_{A} + m_{\alpha}}\,
  \displaystyle\sum\limits_{\alpha=1,2}
  \displaystyle\int\limits_{}^{}
    \Phi_{\rm \alpha - nucl, f}^{*} (\vb{r})\:

  \biggl\{
    2\, \mu_{N}\, m_{\rm p}\;
    \Bigl[
      e^{-i\, c_{A}\, \vb{k_{\rm ph}} \vb{r}} F_{\alpha,\, {\rm el}} +
      e^{i\, c_{\alpha}\, \vb{k_{\rm ph}} \vb{r}} F_{A,\, {\rm el}}
    \Bigr]\, \vb{e}^{(\alpha)} \cdot \vb{K}_{i} + \\

  & + &
  i\,
  \Bigl[
    e^{-i\, c_{A}\, \vb{k_{\rm ph}} \vb{r}}\, \vb{F}_{\alpha,\, {\rm mag}} +
    e^{i\, c_{\alpha}\, \vb{k_{\rm ph}} \vb{r}}\, \vb{F}_{A,\, {\rm mag}}
  \Bigr] \cdot
    \bigl[ \vb{K}_{i} \cp \vb{e}^{(\alpha)} \bigr]
  \biggr\} \cdot
  \Phi_{\rm \alpha - nucl, i} (\vb{r})\; \vb{dr}.
\end{array}
\label{eq.2.12.4}
\end{equation}
\begin{equation}
\begin{array}{lcl}
  M_{k} & = &
  i\, \hbar\, (2\pi)^{3}\,
  \displaystyle\sum\limits_{\alpha=1,2}
    \bigl[ \vb{k_{\rm ph}} \cp \vb{e}^{(\alpha)} \bigr]
  \displaystyle\int\limits_{}^{}
    \Phi_{\rm \alpha - nucl, f}^{*} (\vb{r})\;
  \biggl\{
    e^{-i\, c_{A}\, \vb{k_{\rm ph}} \vb{r}}\, \vb{D}_{\alpha,\, {\rm k}} +
    e^{i\, c_{\alpha}\, \vb{k_{\rm ph}} \vb{r}}\, \vb{D}_{A,\, {\rm k}}
  \biggr\} \cdot
  \Phi_{\rm \alpha - nucl, i} (\vb{r})\; \vb{dr},
\end{array}
\label{eq.2.12.5}
\end{equation}
\begin{equation}
\begin{array}{lll}
\vspace{-0.1mm}
  M_{\Delta E} & = &
  -\, (2\pi)^{3}\; 2\, \mu_{N}
  \displaystyle\sum\limits_{\alpha=1,2} \vb{e}^{(\alpha)}
  \displaystyle\int\limits_{}^{}
    \Phi_{\rm \alpha - nucl, f}^{*} (\vb{r})\;
  \biggl\{
    \Bigl[
      e^{-i\, c_{A}\, \vb{k_{\rm ph}} \vb{r}}\, \vb{D}_{\alpha 1,\, {\rm el}} +
      e^{i\, c_{\alpha}\, \vb{k_{\rm ph}} \vb{r}}\, \vb{D}_{A 1,\, {\rm el}}
    \Bigr]\; - \\

  & - &
  \Bigl[
    \displaystyle\frac{m_{\rm p}}{m_{\alpha}}\, e^{-i\, c_{A}\, \vb{k_{\rm ph}} \vb{r}}\, \vb{D}_{\alpha 2,\, {\rm el}} +
    \displaystyle\frac{m_{\rm p}}{m_{A}}\, e^{i\, c_{\alpha}\, \vb{k_{\rm ph}} \vb{r}}\, \vb{D}_{A 2,\, {\rm el}}
  \Bigr]
  \biggr\} \cdot
  \Phi_{\rm \alpha - nucl, i} (\vb{r})\; \vb{dr},
\end{array}
\label{eq.2.12.6}
\end{equation}
\begin{equation}
\begin{array}{lll}
\vspace{-0.1mm}
  M_{\Delta M} & = &
  -\, i\, (2\pi)^{3}\,
  \displaystyle\sum\limits_{\alpha=1,2}
  \displaystyle\int\limits_{}^{}
    \Phi_{\rm \alpha - nucl, f}^{*} (\vb{r})\;
  \biggl\{
    \Bigl[
      e^{-i\, c_{A}\, \vb{k_{\rm ph}} \vb{r}}\; D_{\alpha 1,\, {\rm mag}} (\vb{e}^{(\alpha)}) +
      e^{i\, c_{\alpha}\, \vb{k_{\rm ph}} \vb{r}}\; D_{A 1,\, {\rm mag}} (\vb{e}^{(\alpha)})
    \Bigr]\; - \\
  & - &
  \Bigl[
    e^{-i\, c_{A}\, \vb{k_{\rm ph}} \vb{r}}\; D_{\alpha 2,\, {\rm mag}} (\vb{e}^{(\alpha)}) +
    e^{i\, c_{\alpha}\, \vb{k_{\rm ph}} \vb{r}}\; D_{A 2,\, {\rm mag}} (\vb{e}^{(\alpha)})
  \Bigr]
  \biggr\} \cdot
  \Phi_{\rm \alpha - nucl, i} (\vb{r})\; \vb{dr},
\end{array}
\label{eq.2.12.7}
\end{equation}
Here, electric and magnetic form factors of the $\alpha$-particle and nucleus, effective electric charge, effective magnetic momentum are defined in Appendix~\ref{sec.2.9}.
Also we obtain (after integration)
\begin{equation}
  \mathbf{K}_{i} = \mathbf{K}_{f} + \mathbf{k}.
\label{eq.2.12.8}
\end{equation}

\subsection{Leading matrix element of emission on the basis of $M_{p}$
\label{sec.2.13}}

Let us consider different approximations, simplifying calculations on computer, but reducing accuracy in determination of the spectrum not much.
In Eq.~(\ref{eq.2.12.3}) we obtained matrix element, which forms the largest contribution to the full bremsstrahlung spectrum:
%
\begin{equation}
  M_{p} = M_{p}^{(E)} + M_{p}^{(M)},
\label{eq.2.13.1.2}
\end{equation}
where
%
\begin{equation}
\begin{array}{lll}
\vspace{-0.2mm}
  M_{p}^{(E)} & = &
  i \hbar\, (2\pi)^{3} \displaystyle\frac{2\, \mu_{N}\,  m_{\rm p}}{\mu}\;
  \displaystyle\sum\limits_{\alpha=1,2}
    \vb{e}^{(\alpha)}
    \biggl\langle\: \Phi_{\rm \alpha - nucl, f} (\vb{r})\; \biggl|\,
    Z_{\rm eff} (\vb{k}_{\rm ph}, \vb{r}) \,
    e^{-i\, \vb{k}_{\rm ph} \vb{r}}\;
    \vb{\displaystyle\frac{d}{dr}}
    \biggr|\: \Phi_{\rm \alpha - nucl, i} (\vb{r})\: \biggr\rangle, \\

  M_{p}^{(M)} & = &
  -\, \hbar\, (2\pi)^{3} \displaystyle\frac{1}{\mu}\;
  \displaystyle\sum\limits_{\alpha=1,2}
  \biggl\langle\: \Phi_{\rm \alpha - nucl, f} (\vb{r})\; \biggl|\,
  \vb{M}_{\rm eff} (\vb{k}_{\rm ph}, \vb{r}) \cdot
  e^{-i\, \vb{k}_{\rm ph} \vb{r}} \cdot
  \Bigl[ \vb{\displaystyle\frac{d}{dr}} \times \vb{e}^{(\alpha)} \Bigr]
  \biggr|\: \Phi_{\rm \alpha - nucl, i} (\vb{r})\: \biggr\rangle.
\end{array}
\label{eq.2.13.1.3}
\end{equation}

\subsubsection{Dipole approximation for effective electric charge
\label{sec.2.13.2}}

The effective charge of nuclear system (\ref{eq.2.10.3})
%
\[
\begin{array}{lll}
  Z_{\rm eff} (\vb{k}_{\rm ph}, \vb{r}) =
  e^{i\, \vb{k_{\rm ph}} \vb{r}}\,
  \Bigl[
    e^{-i\, c_{A} \vb{k_{\rm ph}} \vb{r}}\, \displaystyle\frac{m_{A}}{m_{\alpha} + m_{A}}\, F_{\alpha,\, {\rm el}} -
    e^{i\, c_{\alpha} \vb{k_{\rm ph}} \vb{r}}\,  \displaystyle\frac{m_{\alpha}}{m_{\alpha} + m_{A}}\, F_{A,\, {\rm el}}
  \Bigr]
\end{array}
\]
in first approximation $\exp(i\mathbf{kr}) \to 1$ (i.e., at $\mathbf{kr} \to 0$, called as dipole one) obtain form
%
\begin{equation}
  Z_{\rm eff}^{\rm (dip)} (\vb{k}_{\rm ph}) =
  \displaystyle\frac{m_{A}\, z_{\alpha}(\vb{k}_{\rm ph}) - m_{\alpha}\, Z_{\rm A}(\vb{k}_{\rm ph})}{m_{\alpha} + m_{A}}.
\label{eq.2.13.2.1}
\end{equation}
In such approximation, the effective charge is independent on relative distance between the $\alpha$-particle and nucleus.

For the first calculations one can omit relative displacements of nucleons inside nucleus.
In such approximation one can use $e^{-i \mathbf{k} \rhobfsm_{Aj}} \to 1$ for each nucleon in calculations, and form factor of nucleus is just its electric charge,
where dependence on photon emitted is lost (the same is for the $\alpha$-particle):
%
%
\begin{equation}
  Z_{\rm A} (\vb{k}_{\rm ph}) \to
  \Bigl\langle \psi_{\rm nucl, f} (\rhobf_{A1} \ldots \rhobf_{AA-1})\: \Bigl|\;
    \displaystyle\sum\limits_{j=1}^{A}
      z_{Aj}\:
  \Bigr|\: \psi_{\rm nucl, i} (\rhobf_{A1} \ldots \rhobf_{AA-1})\, \Bigr\rangle =
  \displaystyle\sum\limits_{j=1}^{A}\, z_{Aj} = z_{\rm A},
\label{eq.2.13.2.2}
\end{equation}
as functions $\psi_{\rm nucl, s}$ are normalized. We write
%
\begin{equation}
  Z_{\rm eff}^{\rm (dip, 0)} = \displaystyle\frac{m_{A}\, z_{\alpha} - m_{\alpha}\, z_{\rm A}}{m_{\alpha} + m_{A}},
\label{eq.2.13.2.3}
\end{equation}
\begin{equation}
\begin{array}{lll}
\vspace{-0.2mm}
  M_{p}^{(E,\, {\rm dip})} & = &
  i \hbar\, (2\pi)^{3} \displaystyle\frac{2\, \mu_{N}\,  m_{\rm p}}{\mu}\;
  Z_{\rm eff}^{\rm (dip, 0)}\;
  \displaystyle\sum\limits_{\alpha=1,2}
    \vb{e}^{(\alpha)}
    \biggl\langle\: \Phi_{\rm \alpha - nucl, f} (\vb{r})\; \biggl|\,
    e^{-i\, \vb{k}_{\rm ph} \vb{r}}\;
    \vb{\displaystyle\frac{d}{dr}}
    \biggr|\: \Phi_{\rm \alpha - nucl, i} (\vb{r})\: \biggr\rangle.
\end{array}
\label{eq.2.13.2.4}
\end{equation}
Here, we use upper index ``dip'' denoting the used dipole approximation.

\subsubsection{Emission of photons formed due to displacements of nucleons of nucleus
\label{sec.2.13.3}}

We shall find correction to the matrix element $M_{p}^{(E,\, {\rm dip})}$ в (\ref{eq.2.13.2.4}), which takes into account displacements of nucleons of nucleus.
We write matrix element as
%
%
\begin{equation}
\begin{array}{lcl}
  M_{p}^{(E)} = M_{p}^{(E,\, {\rm dip})} + M_{p,\, {\rm corr}}^{(E,\, {\rm dip})}.
\end{array}
\label{eq.2.13.3.1}
\end{equation}
We rewrite the full matrix element (\ref{eq.2.13.1.3}) as
%
\begin{equation}
\begin{array}{lll}
  M_{p}^{(E)} & = &
  i \hbar\, (2\pi)^{3} \displaystyle\frac{2\, \mu_{N}\,  m_{\rm p}}{\mu}\;
  \displaystyle\sum\limits_{\alpha=1,2}
    \vb{e}^{(\alpha)}
    \biggl\langle\: \Phi_{\rm \alpha - nucl, f} (\vb{r})\; \biggl|\,
      e^{i\, \vb{k_{\rm ph}} \vb{r}}\,
      \Bigl[
        e^{-i\, c_{A} \vb{k_{\rm ph}} \vb{r}}\, \displaystyle\frac{m_{A}\, F_{\alpha,\, {\rm el}}}{m_{\alpha} + m_{A}} - \\
  & - &
        e^{i\, c_{\alpha} \vb{k_{\rm ph}} \vb{r}}\,  \displaystyle\frac{m_{\alpha}\, F_{A,\, {\rm el}}}{m_{\alpha} + m_{A}}
      \Bigr]
    e^{-i\, \vb{k}_{\rm ph} \vb{r}}\;
    \vb{\displaystyle\frac{d}{dr}}
    \biggr|\: \Phi_{\rm \alpha - nucl, i} (\vb{r})\: \biggr\rangle.
\end{array}
\label{eq.2.13.3.2}
\end{equation}
From here we find correction in Eq.~(\ref{eq.2.13.3.1})
%
\begin{equation}
  M_{p,\, {\rm corr}}^{(E,\, {\rm dip})} = M_{p}^{(E)} - M_{p}^{(E,\, {\rm dip})},
\label{eq.2.13.3.3}
\end{equation}
where
\begin{equation}
\begin{array}{lll}
  M_{p,\, {\rm corr}}^{(E,\, {\rm dip})} =

  i \hbar\, (2\pi)^{3} \displaystyle\frac{2\, \mu_{N}\,  m_{\rm p}}{\mu}\;
  \displaystyle\sum\limits_{\alpha=1,2}
    \vb{e}^{(\alpha)}
    \biggl\langle\: \Phi_{\rm \alpha - nucl, f} (\vb{r})\; \biggl|\,
      Z_{\rm eff}^{2} (\vb{k}, \vb{r})\;

    e^{-i\, \vb{k}_{\rm ph} \vb{r}}\; \vb{\displaystyle\frac{d}{dr}}\;
      \biggr|\: \Phi_{\rm \alpha - nucl, i} (\vb{r})\: \biggr\rangle,
\end{array}
\label{eq.2.13.3.4}
\end{equation}
\begin{equation}
\begin{array}{lll}
  Z_{\rm eff}^{(2)} (\vb{k}, \vb{r}) & = &
    \displaystyle\frac{m_{A}}{m_{\alpha} + m_{A}}\;
      \Bigl( e^{i\, \vb{k_{\rm ph}} \vb{r}}\, e^{-i\, c_{A} \vb{k_{\rm ph}} \vb{r}}\, F_{\alpha,\, {\rm el}}(\vb{k}_{\rm ph}) - z_{\alpha} \Bigr) -

    \displaystyle\frac{m_{\alpha}}{m_{\alpha} + m_{A}}\;
      \Bigl( e^{i\, \vb{k_{\rm ph}} \vb{r}}\, e^{i\, c_{\alpha} \vb{k_{\rm ph}} \vb{r}}\, F_{A,\, {\rm el}}(\vb{k}_{\rm ph}) - z_{A} \Bigr).
\end{array}
\label{eq.2.13.3.5}
\end{equation}
In the dipole approximation of effective charge ($ \vb{k}_{\rm ph} \cdot \vb{r} \to 0$) we obtain:
%
%
%
\begin{equation}
\begin{array}{lcl}
  Z_{\rm eff}^{\rm (dip,\, 2)} (\vb{k}_{\rm ph}) & = &
  -\: \displaystyle\frac{m_{\rm p}}{M+m_{\rm p}}\:
  \Bigl\langle\:
    \psi_{\rm nucl, f} (\rhobf_{A1} \ldots \rhobf_{AA})\:
  \Bigl|\;
    \displaystyle\sum\limits_{j=1}^{A} z_{Aj}\, \bigl( e^{-i \mathbf{k} \rhobfsm_{Aj}} - 1 \bigr)
  \Bigr|\:
    \psi_{\rm nucl, i} (\rhobf_{A1} \ldots \rhobf_{AA})\: \Bigr\rangle
\end{array}
\label{eq.2.13.3.7}
\end{equation}
and formula (\ref{eq.2.13.3.4}) for $M_{p,\, {\rm corr}}^{(E,\, {\rm dip})}$ is simplified as
%
\begin{equation}
\begin{array}{lll}
  M_{p,\, {\rm corr}}^{(E,\, {\rm dip},\, 0)} =
  i \hbar\, (2\pi)^{3} \displaystyle\frac{2\, \mu_{N}\,  m_{\rm p}}{\mu}\;
  Z_{\rm eff}^{(2,\, {\rm dip})} (\vb{k}_{\rm ph})\;
  \displaystyle\sum\limits_{\alpha=1,2}
    \vb{e}^{(\alpha)}
    \biggl\langle\: \Phi_{\rm \alpha - nucl, f} (\vb{r})\; \biggl|\;
      e^{-i\, \vb{k}_{\rm ph} \vb{r}}\; \vb{\displaystyle\frac{d}{dr}}\;
    \biggr|\: \Phi_{\rm \alpha - nucl, i} (\vb{r})\: \biggr\rangle.
\end{array}
\label{eq.2.13.3.8}
\end{equation}
%
One can see from Eq.~(\ref{eq.2.13.3.7}) that correction $Z_{\rm eff}^{(2,\, {\rm dip})} (\vb{k})$ in such an approximation changes the electric charge of nucleus.
As exponential factor in the matrix element is less unity, correction of charge is less than this charge (that confirms it as correction to charge of nucleus).
One can see that this exponential factor suppresses electric charge of each proton, due to not-central distribution of the full charge inside nucleus.
In general, the last matrix element can be considered as a factor of normalization.
We will study nucleons of nucleus to be in bound states, therefore the matrix element should be calculated without divergencies.


\subsection{Emission of electric type, formed due to relative motions of nucleons in nucleus (on the basis of $M_{\Delta E}$)
\label{sec.2.14}}

Let us consider the matrix element $M_{\Delta E}$ in Eq.~(\ref{eq.2.12.6}).
As functions $\vb{D}_{\alpha 1,\, {\rm el}}$, $\vb{D}_{\alpha 2,\, {\rm el}}$, $\vb{D}_{A 1,\, {\rm el}}$, $\vb{D}_{A 2,\, {\rm el}}$ do not depend on variable $\vb{r}$, we rewrite this formula as multiplication of two independent integrals as
\begin{equation}
\begin{array}{lll}
\vspace{1.0mm}
  M_{\Delta E} & = &
  -\, (2\pi)^{3}\; 2\, \mu_{N}
  \displaystyle\sum\limits_{\alpha=1,2} \vb{e}^{(\alpha)}\;
  \Bigl\{
  \Bigl\langle
    \Phi_{\rm \alpha - nucl, f} (\vb{r})\;
    \Bigl|\,
      e^{-i\, c_{A}\, \vb{k_{\rm ph}} \vb{r}}
    \Bigr|\,
    \Phi_{\rm \alpha - nucl, i} (\vb{r})\:
  \Bigr\rangle\,
  \Bigl( \vb{D}_{\alpha 1,\, {\rm el}} - \displaystyle\frac{m_{\rm p}}{m_{\alpha}}\, \vb{D}_{\alpha 2,\, {\rm el}} \Bigr)\; + \\

  & + &
  \Bigl\langle
    \Phi_{\rm \alpha - nucl, f} (\vb{r})\;
    \Bigl|\,
      e^{i\, c_{\alpha}\, \vb{k_{\rm ph}} \vb{r}}\,
    \Bigr|\,
    \Phi_{\rm \alpha - nucl, i} (\vb{r})\:
  \Bigr\rangle\,
  \Bigl( \vb{D}_{A 1,\, {\rm el}} - \displaystyle\frac{m_{\rm p}}{m_{A}}\, \vb{D}_{A 2,\, {\rm el}} \Bigr)
  \Bigr\}.
\end{array}
\label{eq.2.14.2}
\end{equation}


One can simplify the obtained formulas, applying expansion over partial waves on such a basis:
\begin{equation}
\begin{array}{lcl}
  e^{i\, c_{\alpha}\, \mathbf{kr}} =
  e^{i\, c_{\alpha}\, kr\, \cos \beta} =
  \displaystyle\sum\limits_{l=0}^{+\infty}
    i^{l}\, (2l+1)\, P_{l} (\cos \beta)\:
    j_{l} \Bigl(c_{\alpha} kr \Bigr),
\end{array}
\label{eq.2.14.3}
\end{equation}
%
we rewrite (\ref{eq.2.14.2}) as
\begin{equation}
\begin{array}{lll}
\vspace{0.9mm}
  M_{\Delta E} & = &
  -\, (2\pi)^{3}\; 2\, \mu_{N}
  \displaystyle\sum\limits_{\alpha=1,2} \vb{e}^{(\alpha)}\;
  \displaystyle\sum\limits_{l=0}^{+\infty}
    i^{l}\, (2l+1)\, P_{l} (\cos \beta)\:
  \Bigl\{
    M_{\Delta E}^{(l)} (- c_{A} k_{\rm ph})\;
    \Bigl( \vb{D}_{\alpha 1,\, {\rm el}} - \displaystyle\frac{m_{\rm p}}{m_{\alpha}}\, \vb{D}_{\alpha 2,\, {\rm el}} \Bigr)\; + \\

  & + &
    M_{\Delta E}^{(l)} (c_{\alpha} k_{\rm ph})\;
    \Bigl( \vb{D}_{A 1,\, {\rm el}} - \displaystyle\frac{m_{\rm p}}{m_{A}}\, \vb{D}_{A 2,\, {\rm el}} \Bigr)
  \Bigr\},
\end{array}
\label{eq.2.14.5}
\end{equation}
where we introduce partial components as
\begin{equation}
\begin{array}{lll}
  M_{\Delta E}^{(l)} (k_{\rm ph}) & = &
  \Bigl\langle \Phi_{\rm \alpha - nucl, f} (\vb{r})\; \Bigl|\, j_{l} (k_{\rm ph}r) \Bigr|\, \Phi_{\rm \alpha - nucl, i} (\vb{r})\: \Bigr\rangle.
\end{array}
\label{eq.2.14.6}
\end{equation}
%
Calculation of are given in Appendix~\ref{sec.app.3}:
%
\begin{equation}
\begin{array}{llll}
  \vb{D}_{\alpha 1,\, {\rm el}} = \displaystyle\frac{3\, \hbar}{8}\; \vb{k}_{\rm ph}\; Z_{\rm \alpha} (\vb{k}_{\rm ph}), &
  \vb{D}_{A 1,\, {\rm el}} = \displaystyle\frac{\hbar}{2}\; \displaystyle\frac{A-1}{A}\; \vb{k}_{\rm ph}\; Z_{\rm A} (\vb{k}_{\rm ph}), &
  \vb{D}_{\alpha 2,\, {\rm el}} \sim \vb{k}_{\rm ph}, &
  \vb{D}_{A 2,\, {\rm el}} \sim \vb{k}_{\rm ph}.
\end{array}
\label{eq.2.14.7}
\end{equation}
%
For such solutions and taking into account $\vb{e}^{(\alpha)} \cdot \vb{k}_{\rm ph} = 0$ (as ortogonality of vectors $\vb{e}^{(\alpha)}$ and $\vb{k}_{\rm ph}$), we obtain:
\begin{equation}
\begin{array}{llllllll}
  \vb{e}^{(\alpha)} \cdot \vb{D}_{\alpha 1,\, {\rm el}} = 0, &
  \vb{e}^{(\alpha)} \cdot \vb{D}_{A 1,\, {\rm el}} = 0, &
  \vb{e}^{(\alpha)} \cdot \vb{D}_{\alpha 2,\, {\rm el}} = 0, &
  \vb{e}^{(\alpha)} \cdot \vb{D}_{A 2,\, {\rm el}} = 0, &
  M_{\Delta E} = 0.
\end{array}
\label{eq.2.14.8}
\end{equation}

\subsection{Emission of magnetic type, formed due to relative motions of nucleons in nucleus (on the basis of $M_{\Delta M}$)
\label{sec.2.15}}

Let us consider the matrix element $M_{\Delta M}$ in Eq.~(\ref{eq.2.12.7}).
As functions $D_{\alpha 1,\, {\rm mag}}$, $D_{\alpha 2,\, {\rm mag}}$, $D_{A 1,\, {\rm mag}}$, $D_{A 2,\, {\rm mag}}$ do not depend on variable $\vb{r}$, we rewrite $M_{\Delta M}$ as multiplication of two independent integrals as
\begin{equation}
\begin{array}{lll}
\vspace{1.0mm}
  M_{\Delta M} & = &
  -\, i\, (2\pi)^{3}\:
  \displaystyle\sum\limits_{\alpha=1,2}\;
  \Bigl\{
  \Bigl\langle
    \Phi_{\rm \alpha - nucl, f} (\vb{r})\;
    \Bigl|\,
      e^{-i\, c_{A}\, \vb{k_{\rm ph}} \vb{r}}
    \Bigr|\,
    \Phi_{\rm \alpha - nucl, i} (\vb{r})\:
  \Bigr\rangle\,
  \Bigl( D_{\alpha 1,\, {\rm mag}} (\vb{e}^{(\alpha)}) - D_{\alpha 2,\, {\rm mag}} (\vb{e}^{(\alpha)}) \Bigr)\; + \\

  & + &
  \Bigl\langle
    \Phi_{\rm \alpha - nucl, f} (\vb{r})\;
    \Bigl|\,
      e^{i\, c_{\alpha}\, \vb{k_{\rm ph}} \vb{r}}\,
    \Bigr|\,
    \Phi_{\rm \alpha - nucl, i} (\vb{r})\:
  \Bigr\rangle\,
  \Bigl( D_{A 1,\, {\rm mag}} (\vb{e}^{(\alpha)}) - D_{A 2,\, {\rm mag}} (\vb{e}^{(\alpha)}) \Bigr)
  \Bigr\}.
\end{array}
\label{eq.2.15.2}
\end{equation}


After summation over spin states, for even number of spin states we obtain:
\begin{equation}
\begin{array}{lll}
  D_{\alpha 2,\, {\rm mag}} (\vb{e}^{(\alpha)}) = 0, &
  D_{A 2,\, {\rm mag}} (\vb{e}^{(\alpha)}) = 0
\end{array}
\label{eq.2.15.3}
\end{equation}
and the matrix element (\ref{eq.2.15.2}) is simplified as
\begin{equation}
\begin{array}{lll}
\vspace{1.0mm}
  M_{\Delta M} & = &
  -\, i\, (2\pi)^{3}\:
  \displaystyle\sum\limits_{\alpha=1,2}\;
  \Bigl\{
  \Bigl\langle
    \Phi_{\rm \alpha - nucl, f} (\vb{r})\;
    \Bigl|\,
      e^{-i\, c_{A}\, \vb{k_{\rm ph}} \vb{r}}
    \Bigr|\,
    \Phi_{\rm \alpha - nucl, i} (\vb{r})\:
  \Bigr\rangle \cdot
  D_{\alpha 1,\, {\rm mag}} (\vb{e}^{(\alpha)})\; + \\

  & + &
  \Bigl\langle
    \Phi_{\rm \alpha - nucl, f} (\vb{r})\;
    \Bigl|\,
      e^{i\, c_{\alpha}\, \vb{k_{\rm ph}} \vb{r}}\,
    \Bigr|\,
    \Phi_{\rm \alpha - nucl, i} (\vb{r})\:
  \Bigr\rangle \cdot
  D_{A 1,\, {\rm mag}} (\vb{e}^{(\alpha)})
  \Bigr\}.
\end{array}
\label{eq.2.15.4}
\end{equation}

We use representation for one nucleon full wave function as
\begin{equation}
  \psi_{\lambda_{s}} (s) = \varphi_{n_{s}} (\vb{r}_{s}) \cdot \bigl|\, \sigma^{(s)} \bigr\rangle \cdot \bigl|\, \tau^{(s)} \bigr\rangle.
\label{eq.2.15.5}
\end{equation}
Using it, we calculate the one-nucleon matrix element in $D_{\alpha 1,\, {\rm mag}} (\vb{e}^{(\alpha)})$ form (\ref{eq.2.15.4}) for nucleons with numbers $s$ and $s'$ (for example, for the $\alpha$-particle;
we will omit index indicating on the $\alpha$-particle):
\begin{equation}
\begin{array}{lll}
\vspace{1mm}
  & &
    \Bigl\langle \psi_{\alpha, \lambda_{s}, f} (s)\, \Bigl|\,
      e^{-i\, \vb{k_{\rm ph}} \rhobf_{\alpha i}}\; \sigmabf \cdot
      \bigl[ \vb{\tilde{p}}_{\alpha i} \times \vb{e}^{(\alpha)} \bigr]
    \Bigr|\,  \psi_{\alpha, \lambda_{s'}, i} (s') \Bigr\rangle = \\

  & = &
    \Bigl\langle
      \varphi_{n_{s}, f} (\vb{r}_{s}) \cdot \sigma^{(s)}\, \Bigl|\,
        e^{-i\, \vb{k_{\rm ph}} \rhobf_{i}}\; \sigmabf \cdot
      \bigl[ \vb{\tilde{p}}_{i} \times \vb{e}^{(\alpha)} \bigr]
    \Bigr|\, \varphi_{n_{s'}, i} (\vb{r}_{s'}) \cdot \sigma^{(s')} \Bigr\rangle \cdot \delta_{\tau_{s=f}, \tau_{s'=i}}.
\end{array}
\label{eq.2.15.6}
\end{equation}
Here, we suppose that after emission of photon neutron is not transferred on proton (and there is no inverse process),
and we have (due to normalization of isospin functions):
\begin{equation}
\begin{array}{lll}
  \bigl\langle\, \tau_{f}^{(s)} \bigr|\, \tau_{i}^{(s')} \bigr\rangle = \delta_{\tau_{s=f}, \tau_{s'=i}}.
\end{array}
\label{eq.2.15.7}
\end{equation}
Here, last $\delta$-function indicates just that it needs to use neutrons in final state $f$ for neutrons in initial state $i$, and protons in final state $f$ for protons in initial state $i$.
We calculate the matrix element (\ref{eq.2.15.6}) further:
\begin{equation}
\begin{array}{lll}
\vspace{1mm}
  &
    \Bigl\langle
      \varphi_{n_{s}, f} (\vb{r}_{s}) \cdot \sigma^{(s)}\, \Bigl|\,
        e^{-i\, \vb{k_{\rm ph}} \rhobf_{i}}\; \sigmabf \cdot
      \bigl[ \vb{\tilde{p}}_{i} \times \vb{e}^{(\alpha)} \bigr]
    \Bigr|\, \varphi_{n_{s'}, i} (\vb{r}_{s'}) \cdot \sigma^{(s')} \Bigr\rangle \cdot \delta_{\tau_{s=f}, \tau_{s'=i}}\; = \\

  = &
    \bigl\langle \sigma^{(s_{i})} \bigl|\: \sigmabf\, \bigr|\, \sigma^{(s_{f})} \bigr\rangle \cdot
    \Bigl[
    \Bigl\langle
      \varphi_{n_{s}, f} (\vb{r}_{s})\, \Bigl|\, e^{-i\, \vb{k_{\rm ph}} \rhobf_{i}}\: \vb{\tilde{p}}_{i}
    \Bigr|\, \varphi_{n_{s'}, i} (\vb{r}_{s'}) \Bigr\rangle
    \times
      \vb{e}^{(\alpha)}
    \Bigr] \cdot \delta_{\tau_{s=f}, \tau_{s'=i}}.
\end{array}
\label{eq.2.15.8}
\end{equation}

In this paper, we will use action of spin operator on spinor wave function as
\begin{equation}
\begin{array}{lll}
  \sigmabf\, \bigr|\, \sigma_{i}^{(s)} \bigr\rangle = \pm \displaystyle\frac{1}{2}\, \bar{\sigmabf} \cdot \bigr|\, \sigma_{i}^{(s)} \bigr\rangle
  & \to &
  \bigl\langle \sigma_{f}^{(s)} \bigl|\: \sigmabf\, \bigr|\, \sigma_{i}^{(s)} \bigr\rangle\; =
  \pm \displaystyle\frac{1}{2}\, \bar{\sigmabf} \cdot \bigl\langle \sigma_{f}^{(s)} \bigr|\, \sigma_{i}^{(s)} \bigr\rangle.
\end{array}
\label{eq.2.15.9}
\end{equation}
Here, we introduced vector $\bar{\sigmabf}$ of eigenvalues of spin operator $\sigmabf$ acting on spin eigenfunction $\bigr|\, \sigma_{i}^{(s)} \bigr\rangle$.
We suppose that after emission of photon spin of nucleon is not changed,
and we have (due to normalization of spin functions):
\begin{equation}
\begin{array}{lll}
  \bigl\langle\, \sigmabf_{f}^{(s)} \bigr|\, \sigmabf_{i}^{(s)} \bigr\rangle = 1.
\end{array}
\label{eq.2.15.10}
\end{equation}
So, we obtain:
\begin{equation}
\begin{array}{lll}
  \bigl\langle \sigma_{f}^{(s)} \bigl|\: \sigmabf\, \bigr|\, \sigma_{i}^{(s)} \bigr\rangle\; =
  \pm \displaystyle\frac{1}{2}\, \bar{\sigmabf}
\end{array}
\label{eq.2.15.11}
\end{equation}
and
\begin{equation}
\begin{array}{lll}
\vspace{1mm}
  &
    \bigl\langle \sigma^{(s)} \bigl|\: \sigmabf\, \bigr|\, \sigma^{(s)} \bigr\rangle \cdot
    \Bigl[
    \Bigl\langle
      \varphi_{n_{s}, f} (\vb{r}_{s})\, \Bigl|\,
      e^{-i\, \vb{k_{\rm ph}} \rhobf_{\alpha i}}\: \vb{\tilde{p}}_{\alpha i}
    \Bigr|\, \varphi_{n_{s}, i} (\vb{r}_{s}) \Bigr\rangle
    \times
      \vb{e}^{(\alpha)}
    \Bigr]\; = \\

 = &
  \pm \displaystyle\frac{1}{2}\, \bar{\sigmabf} \cdot
    \Bigl[
    \Bigl\langle
      \varphi_{n_{s}, f} (\vb{r}_{s})\, \Bigl|\,
      e^{-i\, \vb{k_{\rm ph}} \rhobf_{\alpha i}}\: \vb{\tilde{p}}_{\alpha i}
    \Bigr|\, \varphi_{n_{s}, i} (\vb{r}_{s}) \Bigr\rangle
    \times
      \vb{e}^{(\alpha)}
    \Bigr].
\end{array}
\label{eq.2.15.12}
\end{equation}

We calculate summation in the matrix element $D_{\alpha 1,\, {\rm mag}} (\vb{e}^{(\alpha)})$ for $\alpha$ particle:
\begin{equation}
\begin{array}{lll}
\vspace{1mm}
  D_{\alpha 1,\, {\rm mag}} (\vb{e}^{(\alpha)}) & = &
  \displaystyle\sum\limits_{i=1}^{3}
    \mu_{i}^{\rm (an)}\,
    \Bigl\langle \psi_{\alpha, f} (\beta_{\alpha})\, \Bigl|\,
      e^{-i\, \vb{k_{\rm ph}} \rhobf_{\alpha i}}\; \sigmabf \cdot
      \bigl[ \vb{\tilde{p}}_{\alpha i} \times \vb{e}^{(\alpha)} \bigr]
    \Bigr|\,  \psi_{\alpha, i} (\beta_{\alpha}) \Bigr\rangle\; = \\

\vspace{1mm}
  & = &
  -\, \mu_{i=3}^{\rm (an)}\,
  \displaystyle\frac{1}{2}\, \bar{\sigmabf} \cdot
    \Bigl[
    \Bigl\langle
      \varphi_{n_{s}, f} (\vb{r}_{s})\, \Bigl|\,
      e^{-i\, \vb{k_{\rm ph}} \rhobf_{\alpha i}}\: \vb{\tilde{p}}_{\alpha i}
    \Bigr|\, \varphi_{n_{s}, i} (\vb{r}_{s}) \Bigr\rangle
    \times
      \vb{e}^{(\alpha)}
    \Bigr].
\end{array}
\label{eq.2.15.13}
\end{equation}
Here, we assume that the last nucleon for the $\alpha$-particle has spin $-\,1/2$ (according to scheme of states of nucleons of nucleus).
We have
%
\begin{equation}
\begin{array}{lll}
\vspace{1mm}
  \vb{D}_{\alpha 1,\, {\rm el}} = &
    \displaystyle\sum\limits_{i=1}^{3}
      \displaystyle\frac{z_{i} m_{\rm p}}{m_{\alpha i}}\,
      \Bigl\langle \psi_{\alpha, f} (\beta_{\alpha})\, \Bigl|\,
        e^{-i \vb{k}_{\rm ph} \rhobf_{\alpha i}} \vb{\tilde{p}}_{\alpha i}
      \Bigr|\,  \psi_{\alpha, i} (\beta_{\alpha}) \Bigr\rangle \simeq

      -i\,\hbar\, \vb{k}_{\rm ph}\, Z_{\alpha} (k_{\rm ph}), \\
\end{array}
\label{eq.2.15.14}
\end{equation}
Taking it into account, we rewrite Eq.~(\ref{eq.2.15.13}) as
\begin{equation}
\begin{array}{lll}
  \displaystyle\sum\limits_{\alpha=1,2} D_{\alpha 1,\, {\rm mag}} (\vb{e}^{(\alpha)}) & = &
  i\,\hbar\, |\vb{k}_{\rm ph}|\, Z_{\alpha} (k_{\rm ph}) \cdot f_{1}, \\
\end{array}
\label{eq.2.15.15}
\end{equation}
and we obtain solution for nucleus (with even number of protons and neutrons).
Here, we introduce a new unknown function $f_{1}$ (which is independent on characteristics of the emitted photon).
Using such a formulation, we find final solution for the matrix element of emission (\ref{eq.2.15.4}):
\begin{equation}
\begin{array}{lll}
\vspace{1.0mm}
  M_{\Delta M} & = &
  \hbar\, (2\pi)^{3} \cdot k_{\rm ph} \cdot
  \Bigl\{
    f_{1} \cdot Z_{\alpha} (k_{\rm ph}) \cdot
  \Bigl\langle
    \Phi_{\rm \alpha - nucl, f} (\vb{r})\;
    \Bigl|\,
      e^{-i\, c_{A}\, \vb{k_{\rm ph}} \vb{r}}
    \Bigr|\,
    \Phi_{\rm \alpha - nucl, i} (\vb{r})\:
  \Bigr\rangle\; + \\

  & + &
  f_{2} \cdot Z_{A} (k_{\rm ph}) \cdot
  \Bigl\langle
    \Phi_{\rm \alpha - nucl, f} (\vb{r})\;
    \Bigl|\,
      e^{i\, c_{\alpha}\, \vb{k_{\rm ph}} \vb{r}}\,
    \Bigr|\,
    \Phi_{\rm \alpha - nucl, i} (\vb{r})\:
  \Bigr\rangle
  \Bigr\}.
\end{array}
\label{eq.2.15.16}
\end{equation}

One can apply expansion over partial waves (\ref{eq.2.14.3}).
We obtain:
\begin{equation}
\begin{array}{lll}
  M_{\Delta M} & = &
  \hbar\, (2\pi)^{3} \cdot k_{\rm ph} \cdot
  \displaystyle\sum\limits_{l=0}^{+\infty} i^{l}\, (2l+1)\, P_{l} (\cos \beta)\:
  \Bigl\{
    f_{1} \cdot Z_{\alpha} (k_{\rm ph}) \cdot M_{\Delta E}^{(l)} (- c_{A} k_{\rm ph}) +
    f_{2} \cdot Z_{A} (k_{\rm ph}) \cdot  M_{\Delta E}^{(l)} (c_{\alpha} k_{\rm ph})
  \Bigr\}.
\end{array}
\label{eq.2.15.17}
\end{equation}

\section{Calculations, analysis, discussions
\label{sec.results}}

\subsection{Coherent bremsstrahlung in elastic scattering: multipole and dipole approaches, different normalisations of wave function of scattering
\label{sec.results.1}}

\begin{figure}[htbp]
\centerline{\includegraphics[width=88mm]{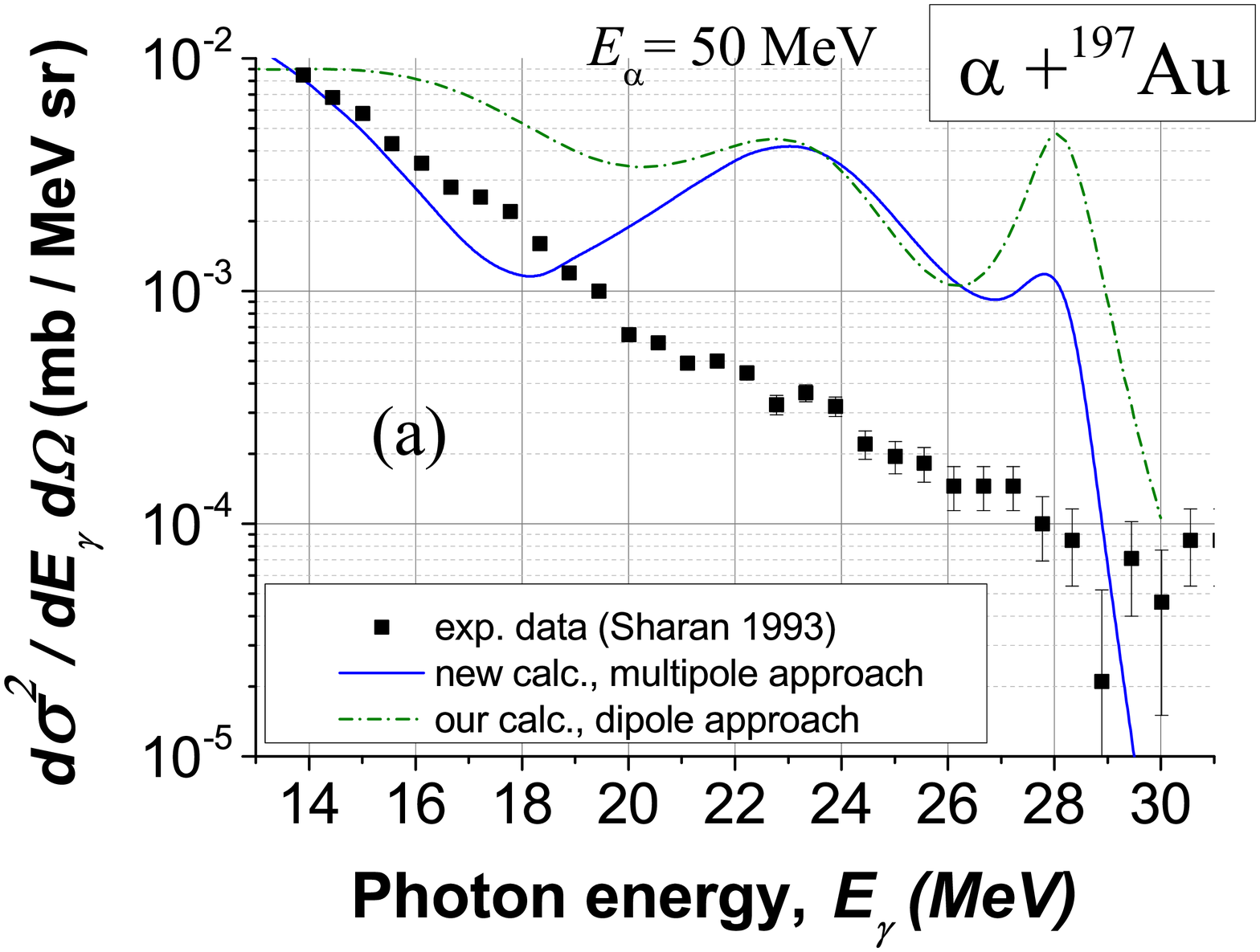}
\hspace{-1mm}\includegraphics[width=88mm]{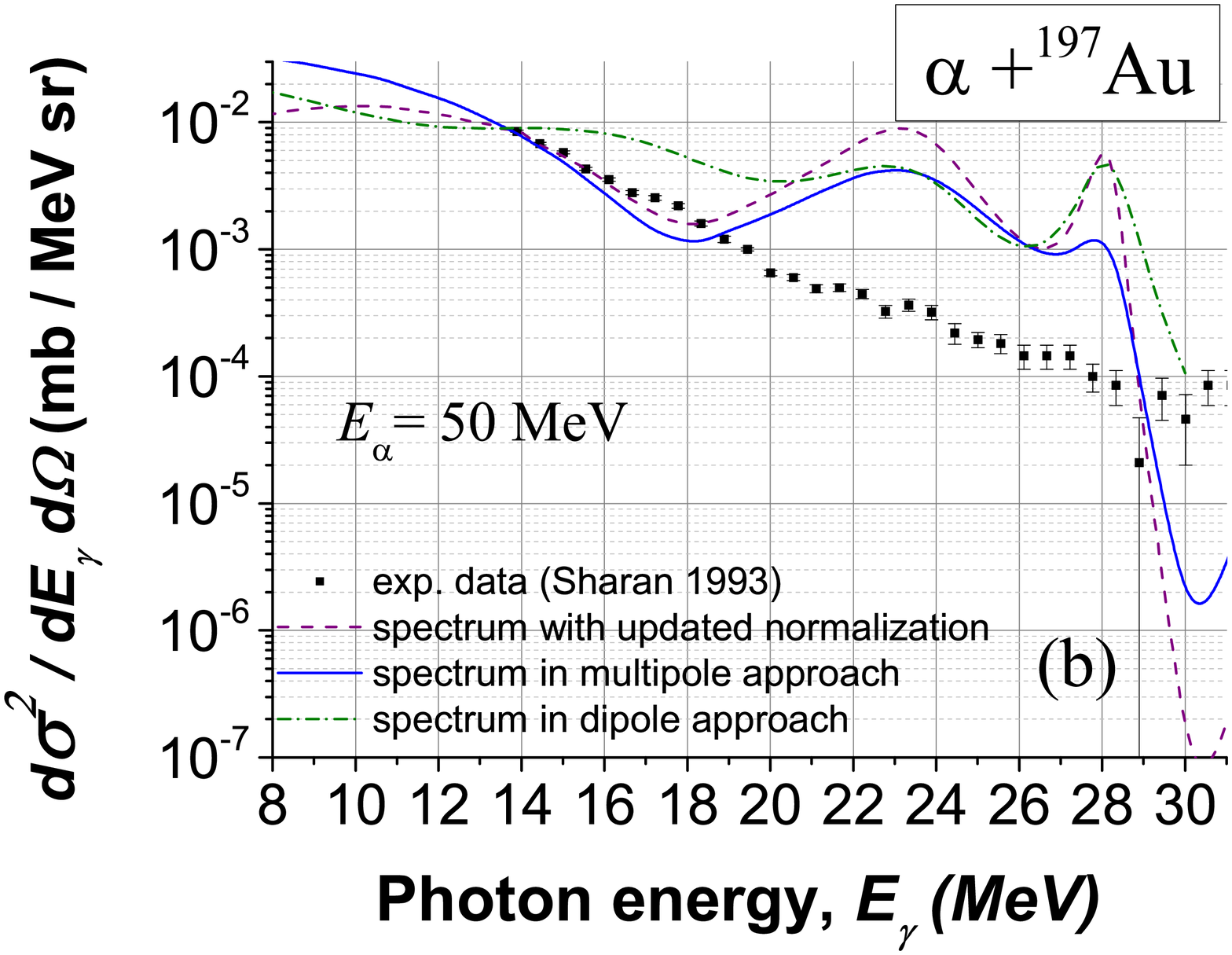}}
\vspace{-3mm}
\caption{\small (Color online)
The coherent bremsstrahlung cross-sections of photons of electric type emitted during the elastic scattering of $\alpha$-particles off the \isotope[197]{Au} nuclei in comparison with experimental data
at energy of beam of the $\alpha$-particles of 50~MeV.
[The bremsstrahlung probabilities are based on the matrix element $M_{p}^{(E,\, {\rm dip})}$ defined in Eqs.~(\ref{eq.2.12.3}), (\ref{eq.2.13.1.2}), (\ref{eq.2.13.1.3}), (\ref{eq.2.13.3.1}),
without correction $M_{p,\, {\rm corr}}^{(E,\, {\rm dip})}$,
effective charge is used in dipole approximation in Eq.~(\ref{eq.2.13.2.3}).
In numeric calculations of integral in Eq.~(\ref{eq.2.12.3}) over variable $\vb{r}$ we vary upper limit $r_{\rm max}$ of this integrals,
number of intervals of integration, and obtain convergent spectra presented in these figures and further ones
(current accuracy is at $r_{\rm max}=20000$~fm, number of intervals = 1000000).]
Here,
black rectangles 
are experimental data~(Sharan 1993: Ref.~\cite{Sharan.1993.PRC}),
blue solid line is the calculated spectrum in the multipole approach,
green dash-dotted line is the calculated spectrum in the dipole approach,
purple dashed line is calculation with new normalization (see text for explanation).
%
Panel (a):
Calculations on multipole VS dipole approaches.
One can see that calculated spectra have maximums at similar energies of photons (minimums are at different photon energies).
This indicates on that maximums are not dependent on way of expansion ow wave function of photons (included to matrix element of emission).
Panel (b):
Calculations at different normalizations of wave function of scattering in the final state.
\label{fig.1}}
\end{figure}

We start from analysis of the coherent bremsstrahlung of photons emitted during elastic scattering.
As studied reaction, we choose the scattering of $\alpha$-particles off the \isotope[197]{Au} nuclei at energy of beam of the $\alpha$-particles of 50~MeV.
For estimations, we use the multipolar approach (in expansion of wave function of photons;
for example, see Eqs.~(46)--(49) in Ref.~\cite{Liu_Maydanyuk_Zhang_Liu.2019.PRC.hypernuclei}) and
define cross-section of emission of bremsstrahlung photons according to Eq.~(58) in Ref.~\cite{Liu_Maydanyuk_Zhang_Liu.2019.PRC.hypernuclei}.
We normalize each calculated spectrum on one data-point of existed experimental data for the corresponding reaction.
Such calculations in comparison with experimental data are presented in Fig.~\ref{fig.1}~(a).

From that figure one can see that agreement between calculations on the basis of multipole approach (see blue solid line in that figure) and experimental data is not satisfactory.
This situation looks to be enough strange, as we supposed that the multipolar model could give the most accurate description of bremsstrahlung,
as we estimated previously on the basis of other nuclear reactions
(see Refs.~\cite{Maydanyuk.2003.PTP,Maydanyuk.2006.EPJA,Maydanyuk.2008.EPJA,Maydanyuk.2008.MPLA,Maydanyuk.2009.NPA,
Maydanyuk.2010.PRC,Maydanyuk.2011.JPG,Maydanyuk.2012.PRC,Maydanyuk_Zhang.2015.PRC,Maydanyuk_Zhang_Zou.2016.PRC}, also see Ref.~\cite{Tkalya.1999.PHRVA}).
It turns out that difference is essential.
However, further detailed analysis has shown the following.
%
\begin{enumerate}
\item
In general, calculations of the coherent bremsstrahlung for the elastic $\alpha$-nucleus scattering at the different expansions of wave function of photons does not describe experimental data well.

\item
In the multipole approach we see clear existence of oscillations with maxima in the spectra.
In order to understand why such maxima are appeared, we repeat calculations, using dipole approach (see also \cite{Papenbrock.1998.PRLTA}) instead of multipole one.
Such calculations are shown in Fig.~\ref{fig.1} by the green dashed line.
Once again, we do not reach good description of the experimental data, but maxima are at similar energies of the emitted photons.

\item
In calculations above, we normalized wave function for scattering in the state after emission of photons related to $\delta$-function, as for scattering in quantum mechanics [see Eqs.~(21.9) and (21.11) in Ref.~\cite{Landau.v3.1989}, p.~89--90].
But, we can change this normalization, where we require that flux of the incident $\alpha$-particles on the barrier should be fixed (i.e. the same) for different energies of these $\alpha$-particles.
In particular, such a normalization is used in order to obtain better accuracy in calculations of cross-sections of capture of $\alpha$-particle by nucleus-target in scattering (that is based on estimations of penetrability of the barrier in dependence on energy).
For higher accuracy, it is not enough to include just tunneling region to calculations, and resulting cross-sections are changed essentially in dependence on additional parameters
(which characterize internal quantum processes inside nuclear region, see Refs.~\cite{Maydanyuk_Zhang_Zou.2017.PRC,Maydanyuk.2015.NPA}].
So, there are indefinities in description of internal quantum processes during scattering (even without fusion, for the same full wave function, with the same boundary conditions).
But, for calculations of the matrix elements of emission it needs to include outgoing flux also to calculations, and a problem of such indefinities is resolved.
It turns out, that the calculated spectrum basing on such different type of normalization [see the purple solid line in Fig.~\ref{fig.1}~(b)] once again has oscillations at similar energies.

\item
One can suppose that maximums in the spectra could reflect possible effects from resonant scattering.
But, resonant and potential components of full scattering are included to wave functions in our calculations.
So, we have taken into account interference between such two effects also. But, it is not enough for good description of experimental data.
We conclude that of mechanisms of another nature should be added for better description of experimental data.
\end{enumerate}
A general tendencies of the all spectra calculated above do not coincide with tendency of the experimental points.
This situation requires to include more details from other mechanisms into the model and calculations.

\subsection{Influence of effective charge $Z_{\rm eff}$ on the coherent bremsstrahlung
\label{sec.results.2}}

\begin{figure}[htbp]
\centerline{\includegraphics[width=88mm]{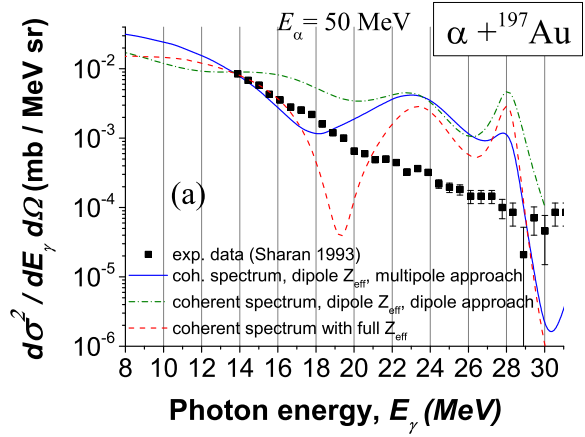}
}
\vspace{-3mm}
\caption{\small (Color online)
Role of the effective charge $Z_{\rm eff}$ in calculations of the coherent bremsstrahlung cross-sections of photons of electric type emitted during the elastic scattering of $\alpha$-particles off the \isotope[197]{Au} nuclei in comparison with experimental data
at energy of beam of the $\alpha$-particles of 50~MeV
[the bremsstrahlung probabilities are based on the matrix element $M_{p}^{(E)}$ defined in Eqs.~(\ref{eq.2.13.1.3})].
Here,
black rectangles are experimental data~(Sharan 1993: Ref.~\cite{Sharan.1993.PRC}),
red dashed line is the calculated spectrum for full effective charge $Z_{\rm eff}$ defined in Eq.~(\ref{eq.2.10.3}) in multipole approach,
blue solid line is the calculated spectrum for effective charge $Z_{\rm eff}^{\rm (dip, 0)}$ at dipole approximation defined in Eq.~(\ref{eq.2.13.2.3}) in the multipole approach,
green dash-dotted line is the calculated spectrum for effective charge $Z_{\rm eff}^{\rm (dip, 0)}$ at dipole approximation defined in Eq.~(\ref{eq.2.13.2.3}) in the dipole approach.
One can see that calculated spectra have maximums at similar energies of photons (minimums are at different photon energies).
This indicates on that different representations of the effective charge do not shift maximums in the spectra much.
\label{fig.2}}
\end{figure}

Before next analysis, at first we shall analyze role of the effective electric charge in calculations of the bremsstrahlung spectrum.
In particular, we will be interesting in how the spectrum is changed if to change dipole approximation of the effective charge on its full definition.
We define full effective charge $Z_{\rm eff} (\vb{k}_{\rm ph}, \vb{r})$ in Eq.~(\ref{eq.2.10.3}) and effective charge $Z_{\rm eff}^{\rm (dip, 0)}$
in dipole approximation in Eq.~(\ref{eq.2.13.2.3}):
\[ 
\begin{array}{lll}
  Z_{\rm eff} (\vb{k}_{\rm ph}, \vb{r}) =
  e^{i\, \vb{k_{\rm ph}} \vb{r}}\,
  \Bigl[
    e^{-i\, c_{A} \vb{k_{\rm ph}} \vb{r}}\, \displaystyle\frac{m_{A}}{m_{\alpha} + m_{A}}\, F_{\alpha,\, {\rm el}} -
    e^{i\, c_{\alpha} \vb{k_{\rm ph}} \vb{r}}\,  \displaystyle\frac{m_{\alpha}}{m_{\alpha} + m_{A}}\, F_{A,\, {\rm el}}
  \Bigr],
\end{array}
\] 
\[ 
  Z_{\rm eff}^{\rm (dip, 0)} = \displaystyle\frac{m_{A}\, z_{\alpha} - m_{\alpha}\, z_{\rm A}}{m_{\alpha} + m_{A}},
\] 
We calculate the coherent bremsstrahlung emission for the elastic scattering on the basis of the matrix element (\ref{eq.2.13.1.3}):
\[ 
\begin{array}{lll}
  M_{p}^{(E)} & = &
  i \hbar\, (2\pi)^{3} \displaystyle\frac{2\, \mu_{N}\,  m_{\rm p}}{\mu}\;
  \displaystyle\sum\limits_{\alpha=1,2}
    \vb{e}^{(\alpha)}
    \biggl\langle\: \Phi_{\rm \alpha - nucl, f} (\vb{r})\; \biggl|\,
    Z_{\rm eff} (\vb{k}_{\rm ph}, \vb{r}) \,
    e^{-i\, \vb{k}_{\rm ph} \vb{r}}\;
    \vb{\displaystyle\frac{d}{dr}}
    \biggr|\: \Phi_{\rm \alpha - nucl, i} (\vb{r})\: \biggr\rangle,
\end{array}
\] 
and we neglect by $M_{p}^{(M)} = 0$ in such an analysis.
Such calculations are presented in Fig.~\ref{fig.2}.
From this figure one can see that the full definition of the effective charge changes the spectrum visibly.
But we obtain maximums in the spectra at similar energies for both approaches (minimums are at different energies).
This indicates on that probabilities of existence of $\alpha$-nuclear system at energies of such maximums are the largest
(and these energies are not dependent much on the effective electric charge).
Also the full definition of effective charge suppresses minimums in the spectrum essentially.

\subsection{Mechanisms producing incoherent bremsstrahlung in elastic scattering
\label{sec.results.3}}

Results of analysis of bremsstrahlung in proton nucleus scattering~\cite{Maydanyuk.2012.PRC,Maydanyuk_Zhang.2015.PRC} clearly show, that inclusion of incoherent bremsstrahlung emission can essentially change the full bremsstrahlung spectrum, even its general tendency.
Moreover, analysis of experimental data \cite{Goethem.2002.PRL} measured by the TAPS collaboration with good accuracy has shown, that mechanisms producing the incoherent bremsstrahlung emission can be not small (see Fig.~8~(a) in Ref.~\cite{Maydanyuk_Zhang.2015.PRC}, for details).

On such a motivation, as a next step, we include mechanisms producing incoherent bremsstrahlung emission of photons during the elastic scattering to our model and calculations.
In particular, for the elastic scattering we do not include imaginary part of $\alpha$-nucleus potential from optical model formalism in calculations of wave functions.
We calculate the spectrum including the matrix element $M_{\Delta M}$ in Eq.~(\ref{eq.2.15.17}):
\[
\begin{array}{lll}
  M_{\Delta M} & = &
  \hbar\, (2\pi)^{3} \cdot k_{\rm ph} \cdot
  \displaystyle\sum\limits_{l=0}^{+\infty} i^{l}\, (2l+1)\, P_{l} (\cos \beta)\:
  \Bigl\{
    f_{1} \cdot Z_{\alpha} (k_{\rm ph}) \cdot M_{\Delta E}^{(l)} (- c_{A} k_{\rm ph}) +
    f_{2} \cdot Z_{A} (k_{\rm ph}) \cdot  M_{\Delta E}^{(l)} (c_{\alpha} k_{\rm ph})
  \Bigr\}.
\end{array}
\]
In calculations, we use unknown $f_{1}$ and $f_{2}$ as a free parameters (we restrict ourselves by case $f_{1}=f_{2}$).
Such parameters characterize of incoherent emission on the basis of coherent emission in elastic scattering, and
we extract these parameters from analysis of experimental data.

\vspace{3mm}
In order to extract unknown information about incoherent emission in elastic scattering, emission of photons due to inelastic mechanisms from experimental data, 
we introduce the following new receipt:
\begin{enumerate}
\item
To estimate coherent bremsstrahlung contribution in elastic scattering in description of experimental data.

\item
To add incoherent contribution in elastic scattering to calculations, to find free parameter $f_{1}$, at which agreement between calculated full spectrum and experimental data will be the best.

\item
To add inelastic mechanisms to calculations, to extract information about them from analysis of experimental data.
\end{enumerate}

\begin{figure}[htbp]
\centerline{\includegraphics[width=88mm]{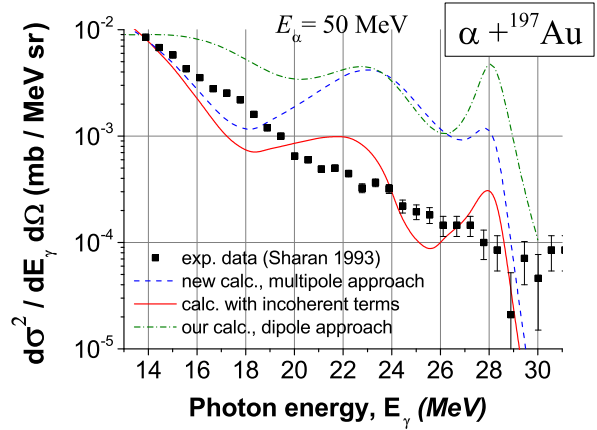}
}
\vspace{-3mm}
\caption{\small (Color online)
The bremsstrahlung cross-sections of coherent and incoherent photons emitted during elastic scattering of $\alpha$-particles off the \isotope[197]{Au} nuclei in comparison with experimental data
at energy of beam of the $\alpha$-particles of 50~MeV
[we use $f_{1} = f_{2}$ in the matrix element (\ref{eq.2.15.17})].
Here,
black rectangles 
are experimental data~(Sharan 1993: Ref.~\cite{Sharan.1993.PRC}),
red solid line is the calculated spectrum with incoherent contribution in the multipole approach,
blue dashed line is the calculated spectrum without incoherent contribution in the multipole approach,
green dash-dotted line is the calculated spectrum without incoherent contribution in the dipole approach.
One can see that inclusion of mechanisms producing of incoherent emission of photons to model allows to improve essentially a general tendency of calculated spectrum (see red solid line in figure) in description of experimental data.
But, oscillations in new calculated spectrum exist. Their maximums are at energies similar to energies of maxima of previous calculations without incoherent emission
(see blue solid line and green dash-dotted line in figure).
\label{fig.3}}
\end{figure}

\vspace{3mm}
We apply such a receipt for analysis of reaction with \isotope[197]{Au}.
Results of such calculations for the elastic scattering of $\alpha$-particles off the \isotope[197]{Au} nuclei are presented in Fig.~\ref{fig.3}.

\vspace{2.0mm}
Conclusion from analysis of such results is the following.
Inclusion of mechanisms of incoherent emission to calculations improves essentially a general tendency in description of the experimental data.
However, it cannot reduce oscillations in calculated spectra which are absent in experimental data.
So, we need in further inclusion of other mechanisms in order to reach better description of experimental data.

\subsection{Inclusion of inelastic mechanisms
\label{sec.results.4}}

\begin{figure}[htbp]
\centerline{\includegraphics[width=88mm]{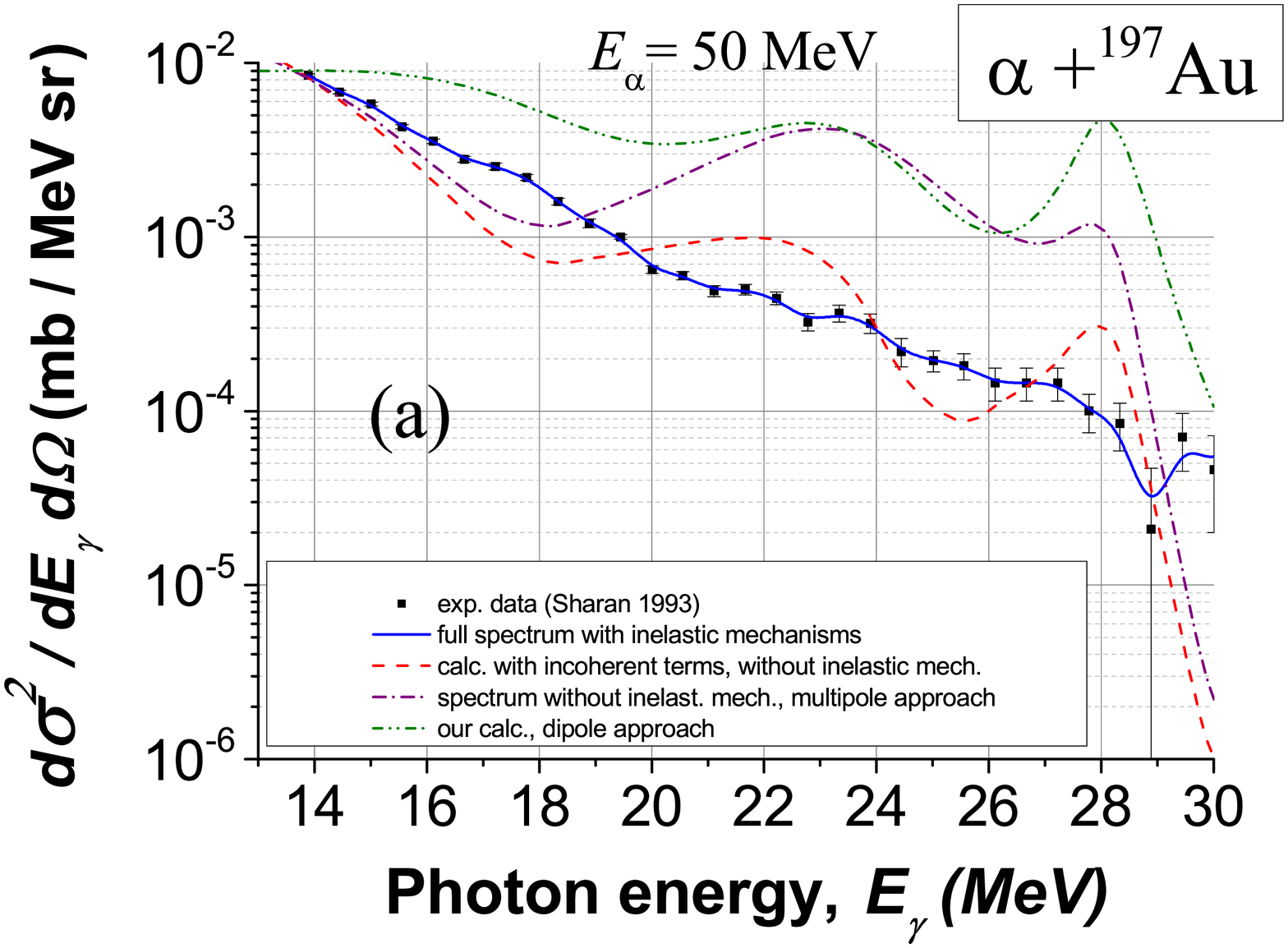}
\hspace{-1mm}\includegraphics[width=88mm]{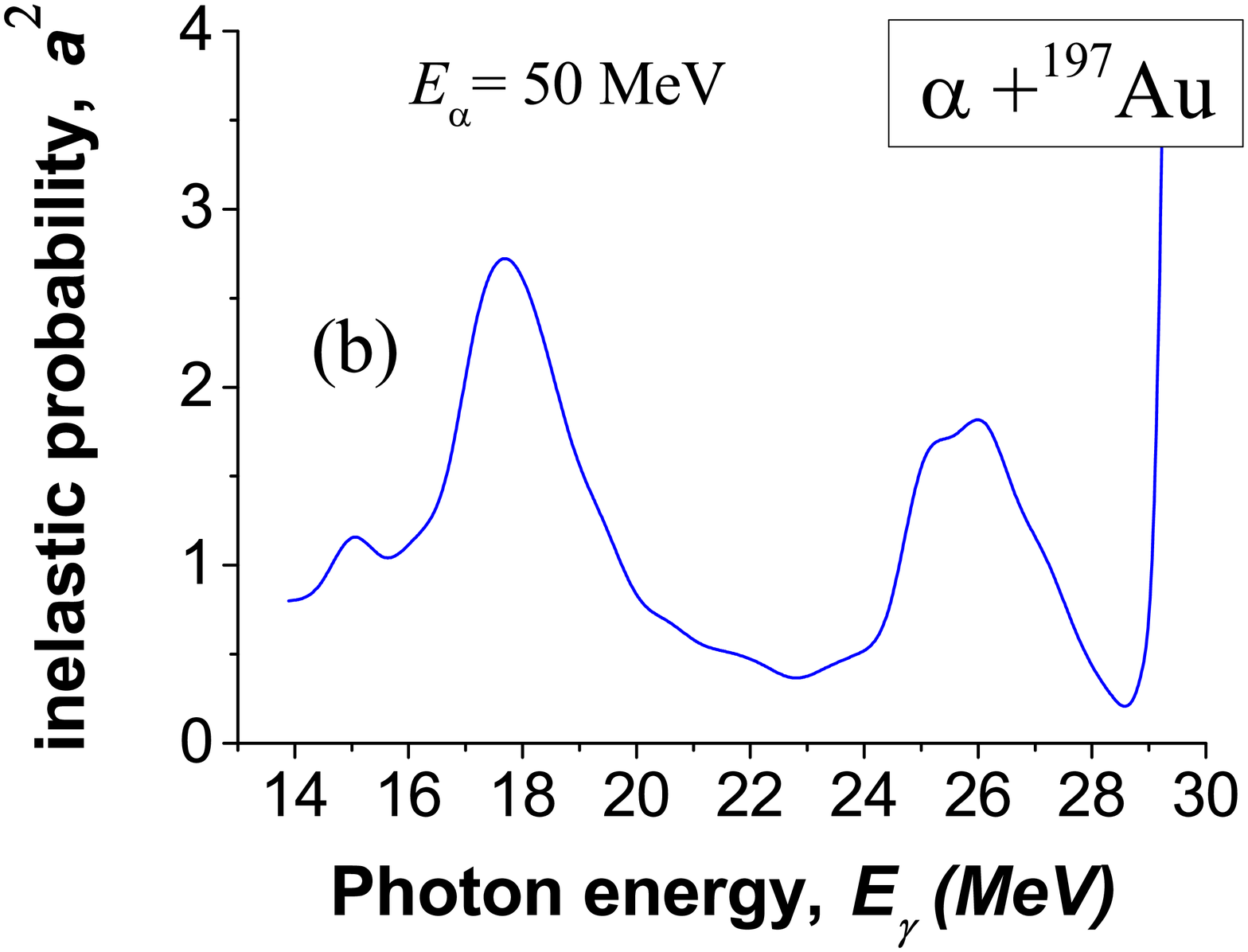}}
\vspace{-3mm}
\caption{\small (Color online)
Panel (a):
The calculated bremsstrahlung cross-sections of photons emitted during scattering of $\alpha$-particles off the \isotope[197]{Au} nuclei with inclusion of inelastic mechanisms in comparison with experimental data
at energy of beam of the $\alpha$-particles of 50~MeV.
Here, black rectangles are experimental data~(Sharan 1993: Ref.~\cite{Sharan.1993.PRC}),
blue solid line is calculated spectrum for inelastic scattering with inclusion of amplitude defined in Eq.~(\ref{eq.result.4.1}) and presented in figure~(b),
red dashed line is the calculated spectrum with incoherent contribution in the multipole approach,
brown dash-dotted line is the calculated spectrum in the multipole approach without incoherent contribution and without inelastic mechanisms,
green dash-dotted line is the calculated spectrum in the dipole approach without incoherent contribution and without inelastic mechanisms.
%
Panel (b):
Amplitude of inelastic mechanisms defined in Eq.~(\ref{eq.result.4.1}) and extracted from analysis of the experimental data.
\label{fig.4}}
\end{figure}

In order to study the inelastic mechanisms, we redefine (introduce) $\alpha$-nucleus wave function as
\begin{equation}
\begin{array}{lcl}
  \varphi_{\rm inelastic} (E, \vb{r}) & = & a_{\rm inelastic} (E) \cdot \varphi_{\rm elastic} (E, \vb{r}),
\end{array}
\label{eq.result.4.1}
\end{equation}
where
$\varphi_{\rm inelastic} (E, \vb{r})$ is a new $\alpha$-nucleus wave function for inelastic scattering,
$a_{\rm inelastic} (E)$ is unknown amplitude, which takes into account inelastic mechanisms.
This amplitude should not include properties of resonant and potential elastic scatterings.
We do not change $\varphi_{\rm elastic} (E, \vb{r})$ in calculations (we use its normalization obtained above for the elastic scattering).
Calculations are presented in Fig.~\ref{fig.4}.
Amplitude $a_{\rm inelastic} (E)$ presented in Fig.~\ref{fig.4}~(b) provides us information about inelastic mechanisms in scattering.
Such amplitude is obtained via analysis of the bremsstrahlung photon emission.

\subsection{Analysis of wave function of relative motion of the $\alpha$-nucleus system
\label{sec.results.5}}

\begin{figure}[htbp]
\centerline{\includegraphics[width=88mm]{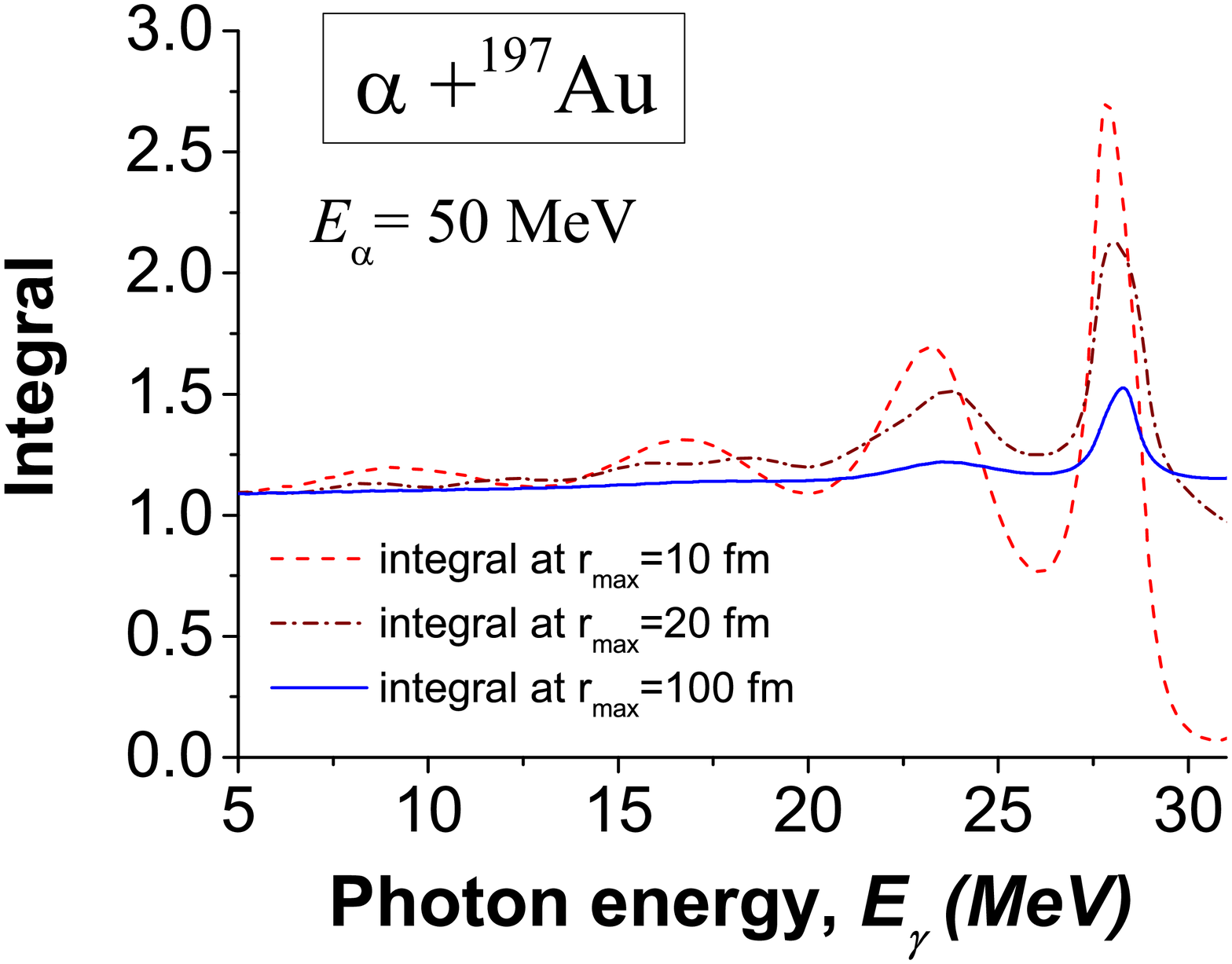}
}
\vspace{-3mm}
\caption{\small (Color online)
The integral defined in Eq.~(\ref{eq.results.5.1}) for the elastic scattering of $\alpha$-particles off the \isotope[197]{Au} nuclei at energy of beam of the $\alpha$-particles of 50~MeV
[
we have fixed normalization of wave function at asymptotic distances as $|\chi(r_{\rm asymp})| = 2$ with high accuracy in the full studied energy region].
Here,
red dashed line is integral calculated at $r_{\rm max} = 10$~fm (multiplied on factor of 9.7),
brown dash-dotted line is integral calculated at $r_{\rm max} = 20$~fm (multiplied on factor of 4.5),
blue solid line is integral calculated at $r_{\rm max} = 100$~fm.
One can see clear peaks at some energies for the same normalization of wave function. These peaks give own contribution in the bremsstrahlung spectrum, forming oscillations of this spectrum.
By such a way, we explain origin of oscillations in the bremsstrahlung spectrum even for the elastic scattering.
\label{fig.5}}
\end{figure}

In order to obtain deeper insight to origin of oscillations in the spectra in Figs.~\ref{fig.1}--\ref{fig.3},
we analyse $\alpha$-nucleus wave function in the final state (after emission of photon) via the following integral:
\begin{equation}
\begin{array}{lll}
  I (E) = \displaystyle\int\limits_{r_{\rm min}}^{r_{\rm max}} |\chi_{l}(r, E)|^{2}\; dr, &
  \Phi_{\rm \alpha - nucl} (\vb{r}) = \displaystyle\frac{\chi_{l}(r)}{r}\; Y_{lm} (\theta, \phi).
\end{array}
\label{eq.results.5.1}
\end{equation}
Here,
$\Phi_{\rm \alpha - nucl} (\vb{r})$ is the wave function of relative motion of the $\alpha$-particle concerning to nucleus-target, defined in Eq.~(\ref{eq.2.6.1}),
$\chi_{l}(r, E)$ is the radial wave function after emission of photon used in different calculations of matrix elements of emission,
$\vb{r}$ is the relative distance between centers of masses of the $\alpha$-particle and nucleus ($r = |\vb{r}|$).
Calculations are presented in Fig.~\ref{fig.5} in dependence on external boundary $r_{\rm max}$ and energy $E$.
According to quantum mechanics, a physical sense of such an integral can be understood as probability of formation of compound nuclear system (from the $\alpha$-particle and nucleus-target) with external boundary at $r_{\rm max}$ during elastic scattering.
From this figure one can see that integral has oscillating behavior in dependence on energy of the incident $\alpha$-particles, this oscillations have maximums at some energies.

One can understand origin of such oscillations from the following logics.
The compound nuclear system is formed due to penetration of the $\alpha$-particle through barrier including all further oscillations inside internal nuclear region.
While coefficient of penetrability through the barrier has monotonous dependence on energy, but oscillations and condition of the wave function to be finite at zero provide oscillating behavior in dependence on energy of the incident $\alpha$-particles
[see Ref.~\cite{Maydanyuk_Zhang_Zou.2017.PRC}, the simplest clear understanding can be obtained from Fig.~1 in that paper].
In result, we obtain oscillations with maximums of wave function for scattering (which includes both tunneling and oscillating phenomena).
Maximums of integral (\ref{eq.results.5.1}) correspond to the largest probabilities of formation of compound nuclear system, i.e. they indicate on resonant scattering at such energies.
In particular, from Fig.~\ref{fig.5} one can see that the closer the external boundary $r_{\rm max}$ to the size of nucleus (i.e., about 12--15~fm),
the larger (more important) resonating effects in scattering.
But, external region suppresses such resonating effects.
The matrix element of emission (and the corresponding bremsstrahlung spectrum) includes such a wave function in the final state with oscillating behavior, so it has such an oscillating behavior also.
So, we have explained origin of oscillations in the bremsstrahlung spectra on the basis of nuclear scattering theory.

\begin{figure}[htbp]
\centerline{\includegraphics[width=88mm]{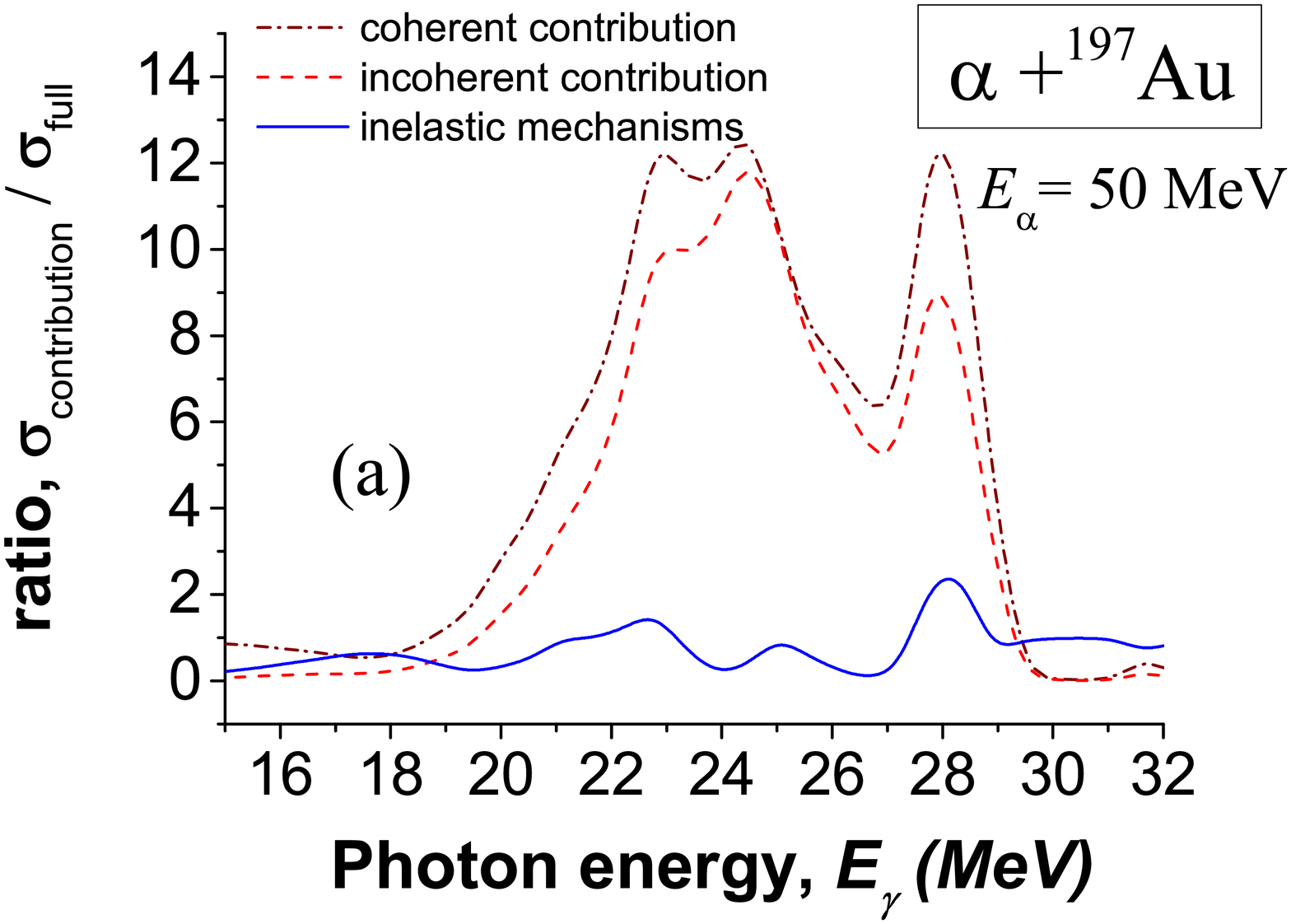}
\hspace{-1mm}\includegraphics[width=88mm]{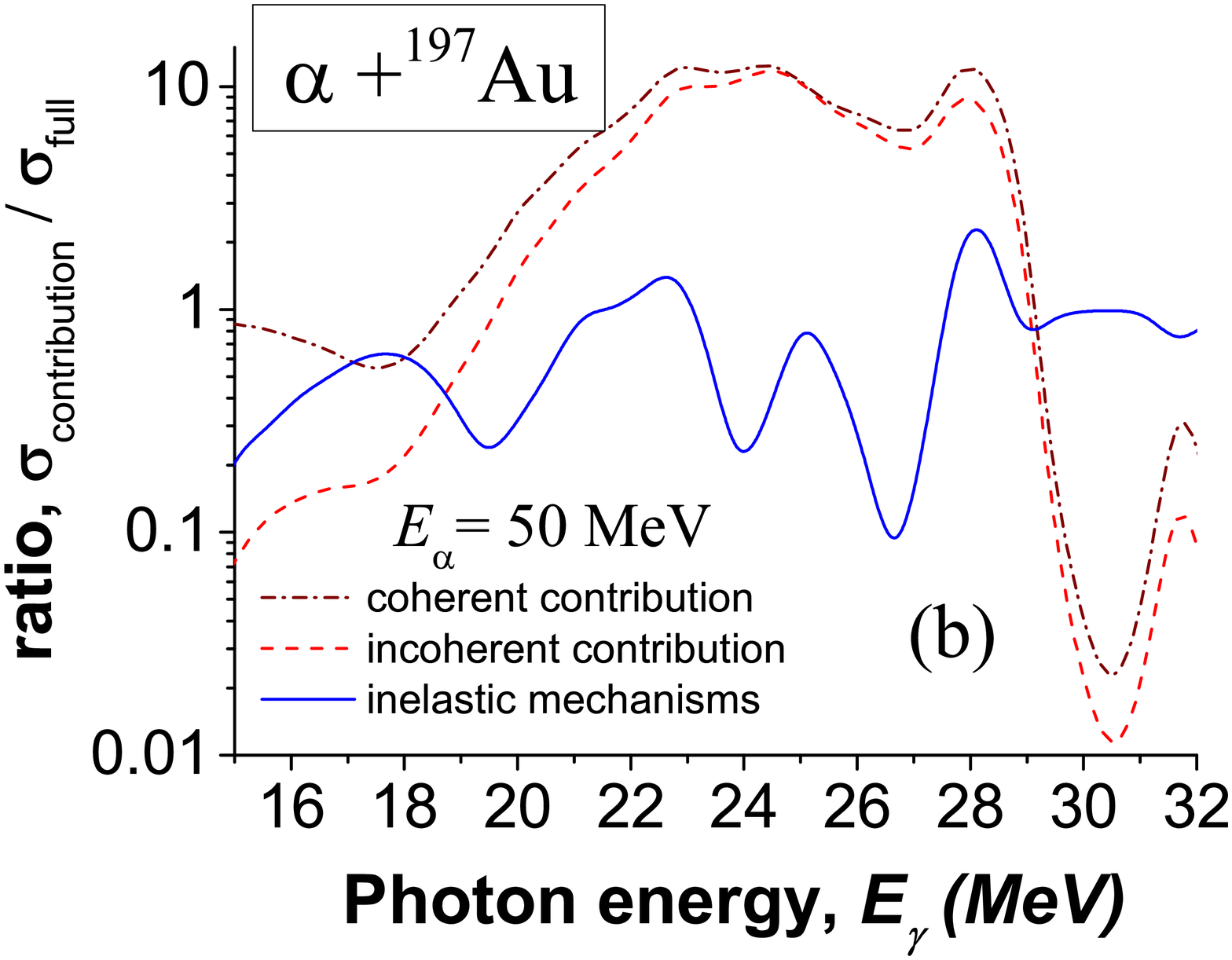}}
\vspace{-3mm}
\caption{\small (Color online)
Ratios between the coherent contribution, incoherent contribution and contribution caused by inelastic mechanisms
concerning to the full bremsstrahlung spectrum for full scattering of $\alpha$-particles off the \isotope[197]{Au} nuclei
at energy of beam of the $\alpha$-particles of 50~MeV.
Here,
brown dash-dotted line is extracted contribution of coherent emission during elastic scattering (without incoherent contribution and without inelastic mechanisms),
red dashed line is extracted contribution of incoherent emission during elastic scattering  (without coherent contribution and without inelastic mechanisms),
blue solid line is extracted contribution for inelastic mechanisms during scattering,
As this is a new type of information, we present it in both linear scale (a) and logarithmic one (b).
In particular, from figure (b) one can better see that internal structure of nucleus (related with inelastic mechanisms) is more visible at higher photon energies, starting from 29~MeV.
Inside photon energy region from 18 to 29 MeV, elastic scattering is more important.
Practically, inside the full photon energy region, coherent emission is larger (i.e., more important) than incoherent one.
\label{fig.6}}
\end{figure}

In Fig.~\ref{fig.6} we add ratios for the coherent contribution in elastic scattering, incoherent contribution in elastic scattering and contribution of emitted photons due to inelastic mechanisms in scattering concerning to the full bremsstrahlung spectrum for this reaction.


\subsection{Bremsstrahlung emission for \isotope[59]{Co}, \isotope[116]{Sn} and \isotope[107]{Ag}
\label{sec.results.6}}

\begin{figure}[htbp]
\centerline{\includegraphics[width=88mm]{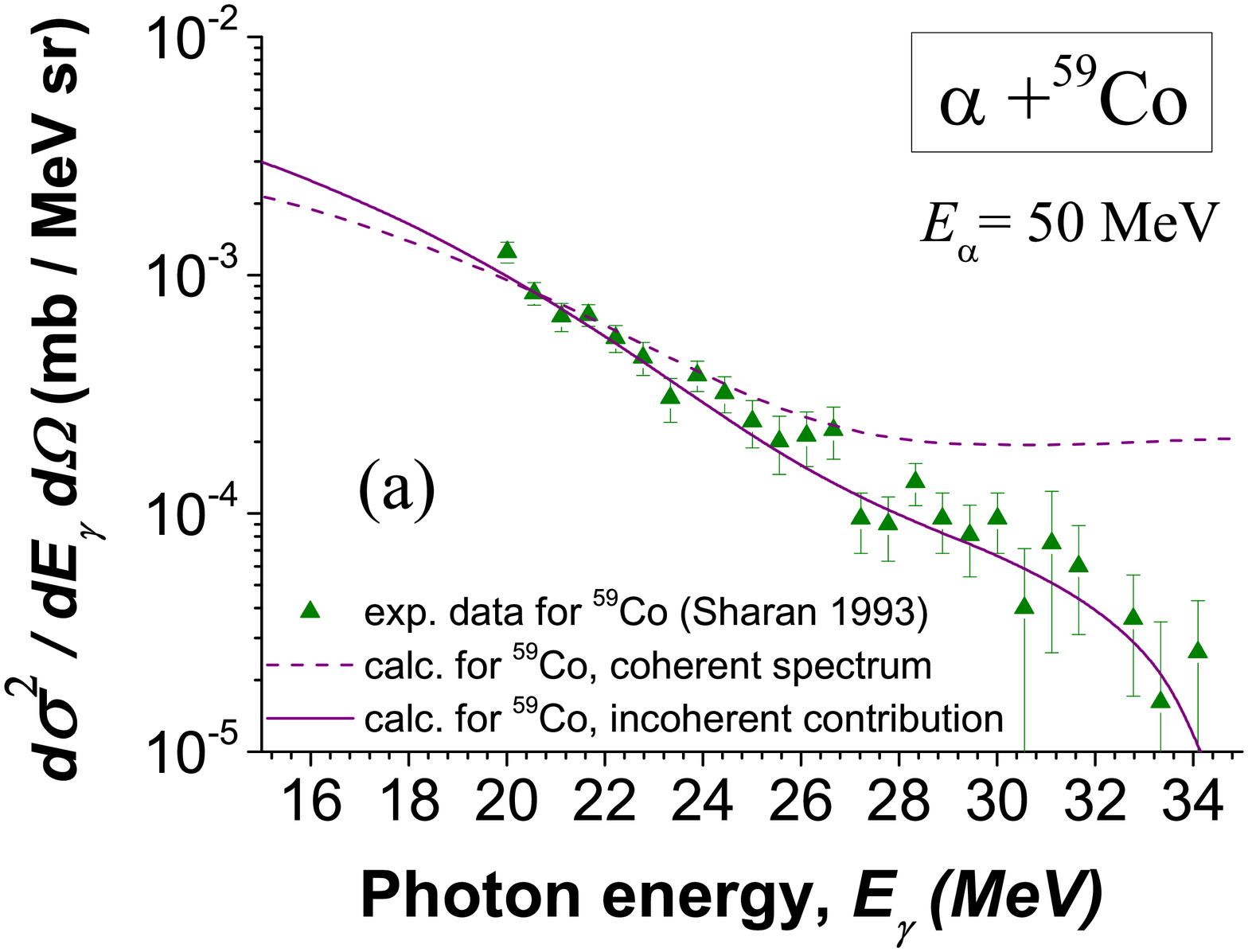}
\hspace{-1mm}\includegraphics[width=88mm]{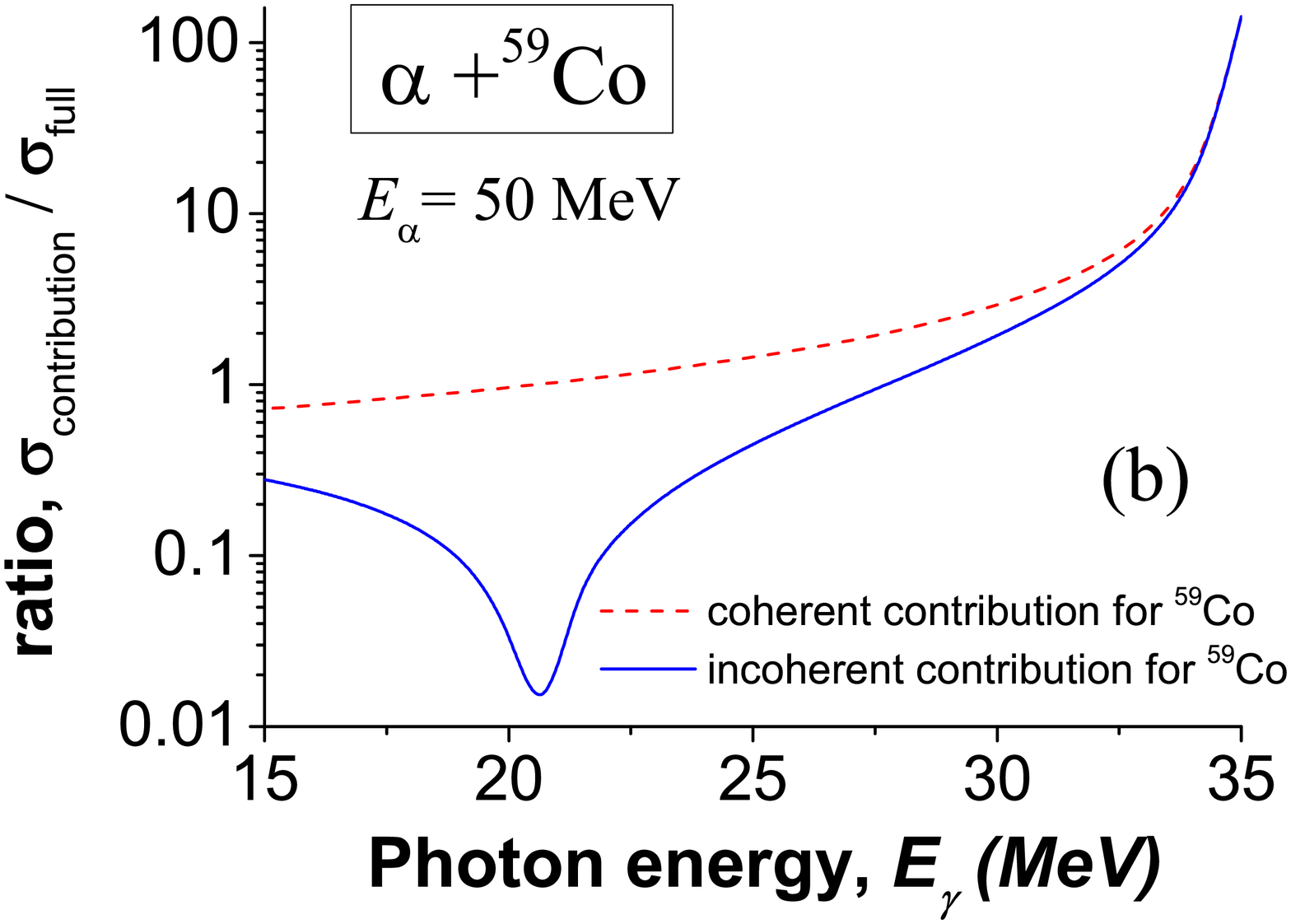}}
\vspace{-3mm}
\caption{\small (Color online)
Panel (a):
The calculated bremsstrahlung cross-sections of photons emitted during elastic scattering of $\alpha$-particles off the \isotope[59]{Co} nuclei with inclusion of incoherent emission and without it in comparison with experimental data
at energy of beam of the $\alpha$-particles of 50~MeV.
Here,
green triangles are experimental data~(Sharan 1993: Ref.~\cite{Sharan.1993.PRC}),
purple solid line is the calculated spectrum with inclusion of coherent and incoherent bremsstrahlung contributions,
purple dash-dotted line is the calculated spectrum for coherent bremsstrahlung without incoherent emission.
Panel (b):
Ratios between coherent or incoherent contribution and full bremsstrahlung spectrum for this reaction.
One can see that at higher energies each contribution is larger than full spectrum, that is explained by destructive interference between coherent and incoherent emissions.
In general, the coherent emission is larger than the incoherent emission inside full energy region.
\label{fig.7}}
\end{figure}

At finishing, we add calculations of the bremsstrahlung spectra in the elastic scattering of $\alpha$-particles off the \isotope[59]{Co}, \isotope[116]{Sn} and \isotope[107]{Ag} nuclei
at 50~MeV of energy of $\alpha$-particles beam.
Results of such calculations for \isotope[59]{Co} are presented in Fig.~\ref{fig.7}.
From this figure one can see that even without inelastic mechanisms, inclusion of incoherent emission to the model and calculations allows to describe experimental data really well in full energy region!
A presence of minor disagreement between theory and experimental data at for two data points (at $E_{\gamma}=26.6$~MeV and $E_{\gamma}=28.3$~MeV) indicates on minor role of inelastic mechanisms at such energies (which also shown in tiny oscillations in full spectrum).
They can be further studied and estimated as this was obtained for reaction $\alpha + \isotope[197]{Au}$ above (we omit this analysis in this paper).

\begin{figure}[htbp]
\centerline{\includegraphics[width=88mm]{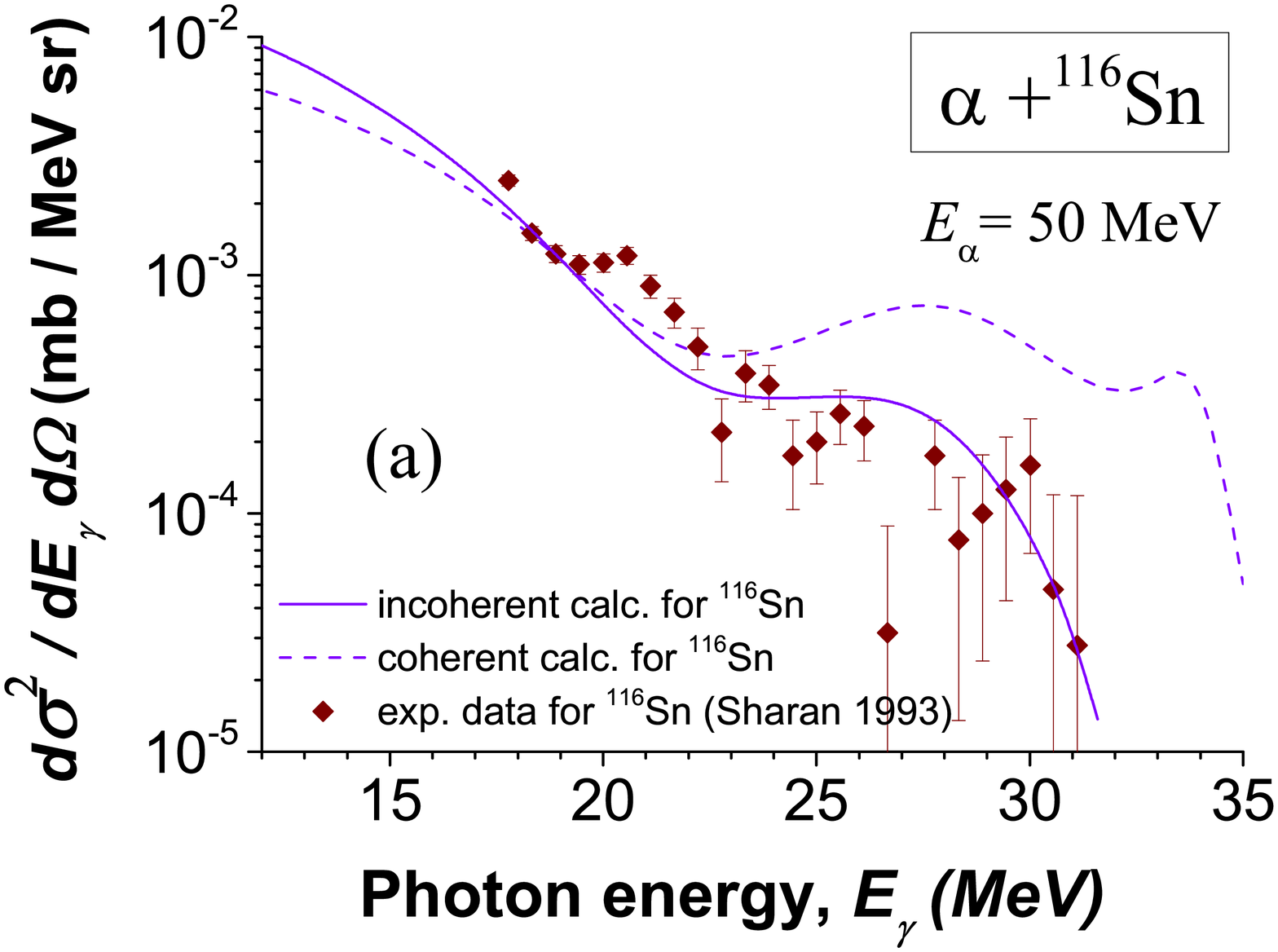}
\hspace{-1mm}\includegraphics[width=88mm]{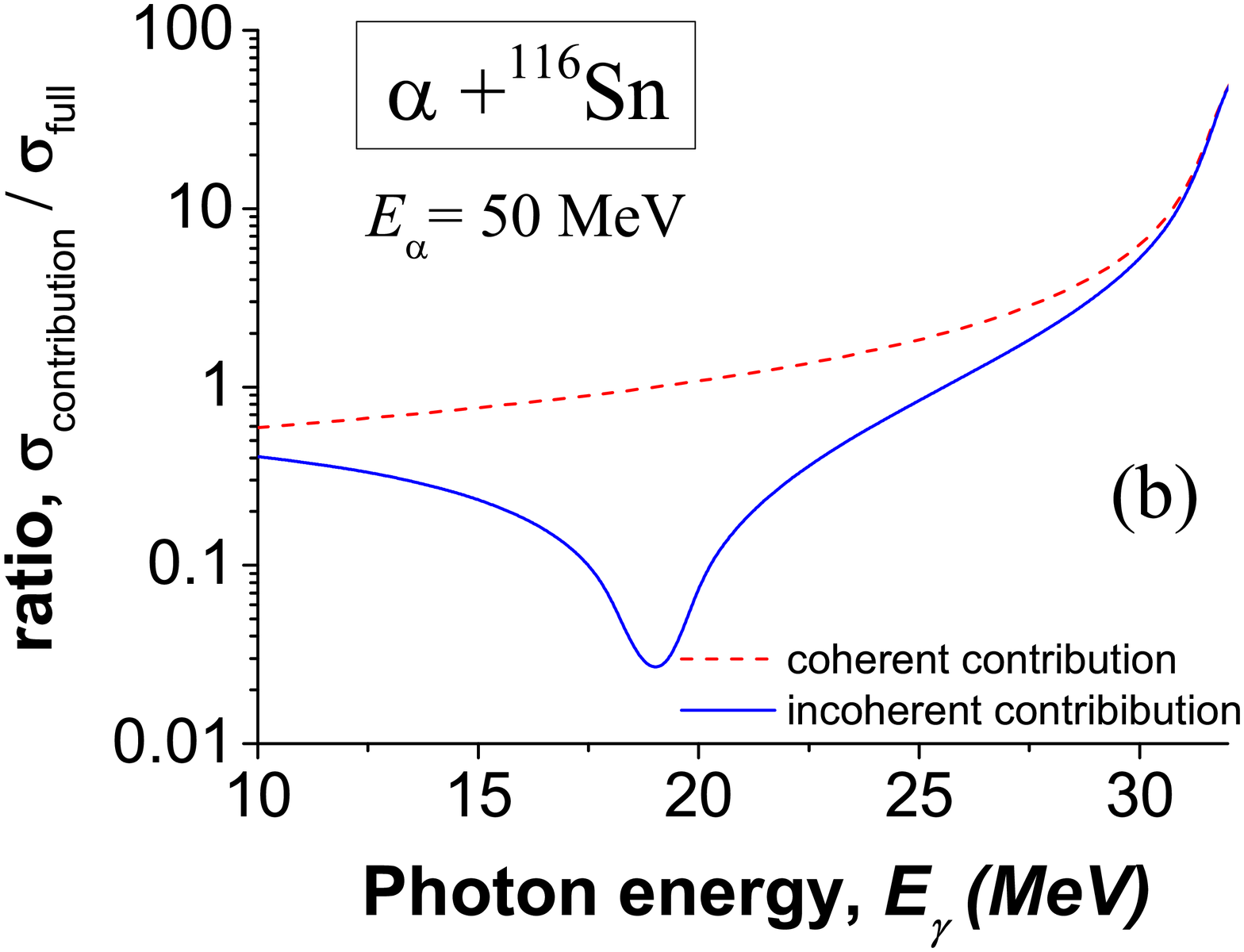}}
\vspace{-3mm}
\caption{\small (Color online)
Panel (a):
The calculated bremsstrahlung cross-sections of photons emitted during elastic scattering of $\alpha$-particles off the \isotope[116]{Sn} nuclei with inclusion of incoherent emission and without it in comparison with experimental data
at energy of beam of the $\alpha$-particles of 50~MeV.
Here,
brown rhombuses are experimental data~(Sharan 1993: Ref.~\cite{Sharan.1993.PRC}),
violet solid line is the calculated spectrum with inclusion of coherent and incoherent bremsstrahlung contributions,
violet dash-dotted line is the calculated spectrum for coherent bremsstrahlung without incoherent emission.
Panel (b):
Ratios between coherent or incoherent contribution and full bremsstrahlung spectrum for this reaction.
One can see that at higher energies each contribution is larger than full spectrum, that is explained by destructive interference between coherent and incoherent emissions.
The coherent emission is larger than the incoherent emission inside full energy region.
\label{fig.8}}
\end{figure}

\begin{figure}[htbp]
\centerline{\includegraphics[width=88mm]{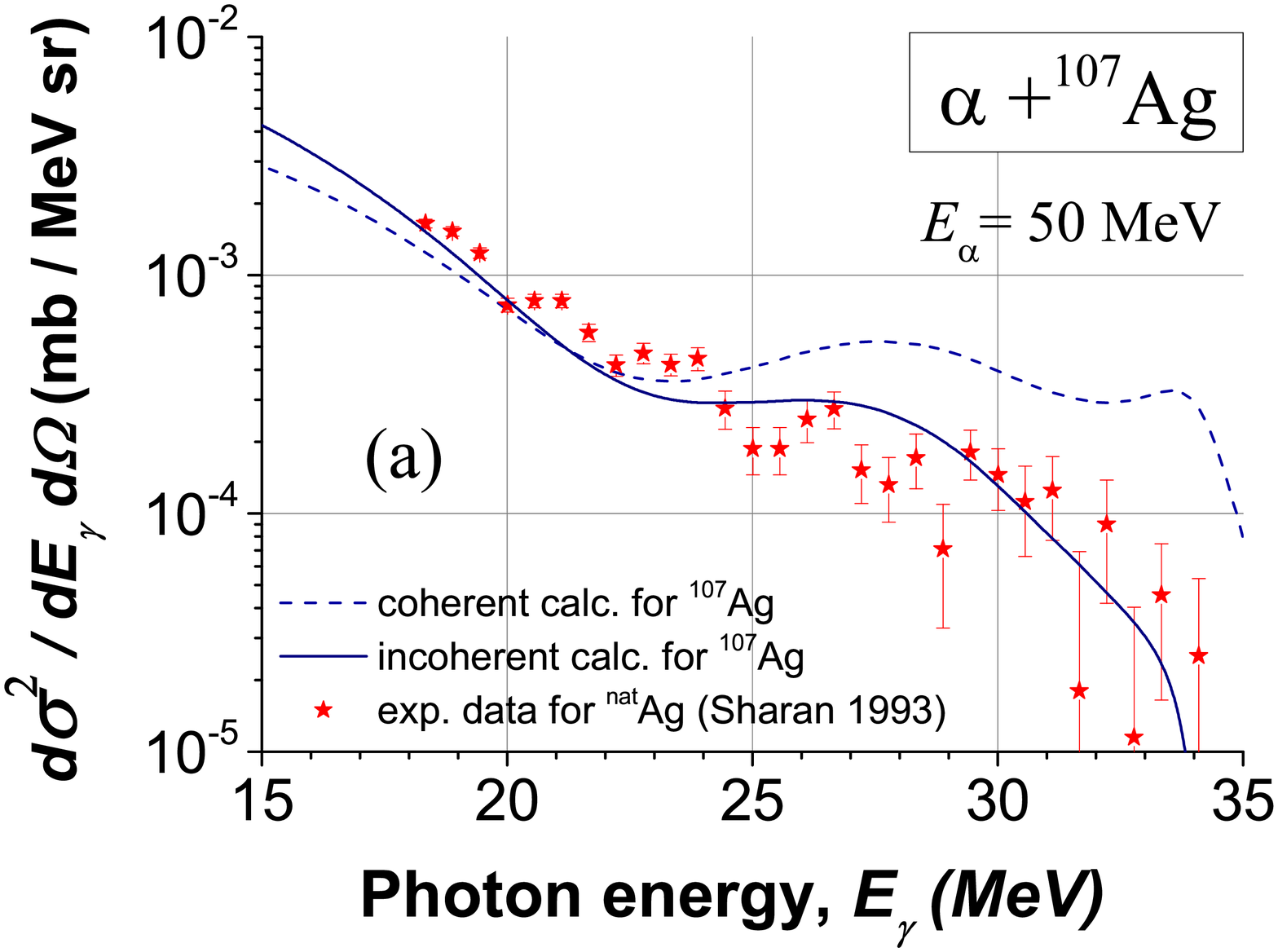}
\hspace{-1mm}\includegraphics[width=88mm]{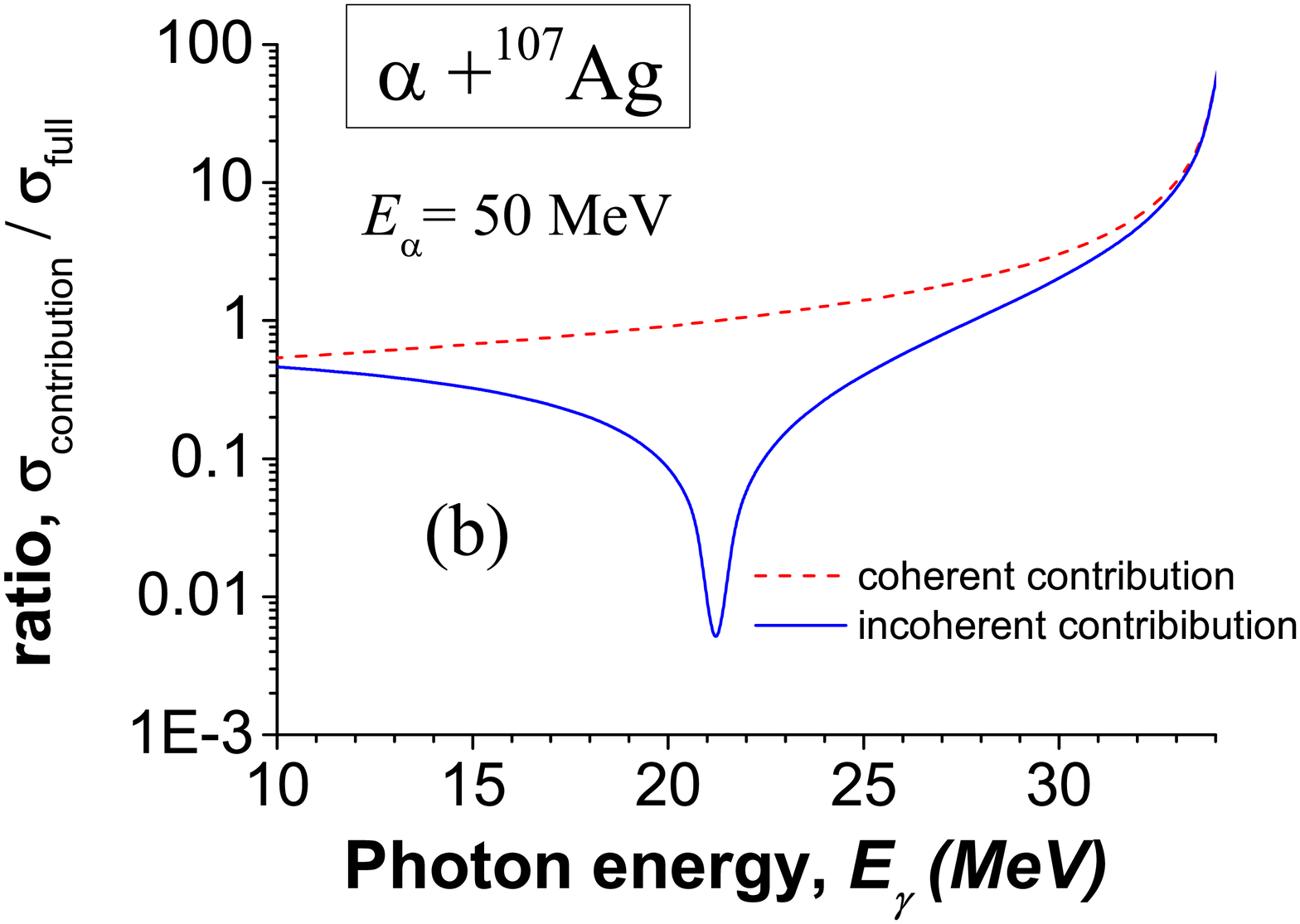}}
\vspace{-3mm}
\caption{\small (Color online)
Panel (a):
The calculated bremsstrahlung cross-sections of photons emitted during elastic scattering of $\alpha$-particles off the \isotope[107]{Ag} nuclei with inclusion of incoherent emission and without it in comparison with experimental data
at energy of beam of the $\alpha$-particles of 50~MeV.
Here,
red stars are experimental data~(Sharan 1993: Ref.~\cite{Sharan.1993.PRC}),
dark blue solid line is the calculated spectrum with inclusion of coherent and incoherent bremsstrahlung contributions,
dark blue dash-dotted line is the calculated spectrum for coherent bremsstrahlung without incoherent emission.
Panel (b):
Ratios between coherent or incoherent contribution and full bremsstrahlung spectrum for this reaction.
\label{fig.9}}
\end{figure}

In Fig.~\ref{fig.8} we show our calculations for \isotope[116]{Sn}.
Here, we observe larger oscillations in the calculated spectra, than for nucleus \isotope[59]{Co}, and smaller than for \isotope[197]{Au}.
But, general agreement between experimental data and our calculations is nod bad.
Once again, inclusion of incoherent emission to calculations improves essentially such an agreement.
Also we estimated the bremsstrahlung spectra for scattering of $\alpha$-particles off the \isotope[107]{Ag} nuclei at energy of beam of the $\alpha$-particles of 50~MeV
[for calculations we choose nucleus \isotope[107]{Ag},
while \isotope[nat]{Ag} was used in experiments, see Ref.~\cite{Sharan.1993.PRC}].
For \isotope[107]{Ag} we observe similar pictures (see Fig.~\ref{fig.9}) as results given in Fig.~\ref{fig.8} for \isotope[116]{Sn}.
Also we provide the corresponding ratios between coherent emission or incoherent emission and full spectrum for each reaction
[see Figs.~\ref{fig.7}~(b), \ref{fig.8}~(b) and \ref{fig.9}~(b)].
From here one can see the largest role for coherent emission,
but incoherent emission is not small.
Such an analysis demonstrates advance of our approach to extract accurate information about incoherent emission for the $\alpha$-nucleus scattering for the full energy region, basing on existed experimental data.


\section{Conclusions
\label{sec.conclusions}}

In this paper we investigate bremsstrahlung emission of photons during scattering of $\alpha$-particles off nuclei.
We have generalized our bremsstrahlung theory [see Refs.~\cite{Maydanyuk.2003.PTP,Maydanyuk.2009.NPA,Maydanyuk.2009.TONPPJ,Maydanyuk.2009.JPS,Maydanyuk.2010.PRC,Maydanyuk.2011.JPCS,Maydanyuk.2011.JPG,
Maydanyuk.2006.EPJA,Maydanyuk.2008.EPJA,Maydanyuk.2008.MPLA,Maydanyuk.2012.PRC,Maydanyuk_Zhang.2015.PRC,Maydanyuk_Zhang_Zou.2016.PRC,Maydanyuk_Zhang_Zou.2018.PRC,
Maydanyuk_Zhang_Zou.2019.PRC.microscopy,Liu_Maydanyuk_Zhang_Liu.2019.PRC.hypernuclei},
and references therein] for that reaction, where
we add new formalism for coherent and incoherent emission of photons for elastic scattering and mechanisms of inelastic scattering.
For the analysis of the bremsstrahlung spectra we have chosen the scattering of $\alpha$-particles off the \isotope[59]{Co}, \isotope[116]{Sn}, \isotope[\rm nat]{Ag} and \isotope[197]{Au} nuclei at energy of beam of the $\alpha$-particles of 50~MeV, for which the bremsstrahlung photons were measured at the Variable Energy Cyclotron Centre, Calcutta~\cite{Sharan.1993.PRC}.

\begin{itemize}
\item
We start analysis from heavy nucleus \isotope[197]{Au}.
As we estimated, the leading contribution to the full spectrum is given by coherent emission for elastic scattering for this nucleus.
It turns out, that calculations of such emission is in disagreement with the experimental data.
In particular, we observe clear presence of oscillations with large amplitudes (and different periods) in the spectrum (which are absent in the experimental data, see Fig.~\ref{fig.1}).
We conclude that it is not enough to describe bremsstrahlung in scattering only on the basis of coherent emission for elastic processes.

\item
In order to understand origin of oscillations in the spectra and test such results, we applied different methods of expansion of wave function of photons (we used multipole and dipole approaches),
and different types of normalization of $\alpha$-nucleus wave function for elastic scattering.
But, all calculated spectra have oscillations with maxima at similar energies of photons (see Fig.~\ref{fig.1}).
We conclude that such results are not much dependent on the way of description of coherent emission of photons, but they are more related with used description of $\alpha$-nucleus wave function for elastic scattering (without photons).

\item
Note that resonant and potential components of full elastic scattering are included to wave functions in our calculations.
So, we have taken into account interference between such two effects. But, it is not enough for good description of experimental data.
We conclude that other mechanisms of inelastic (nature) origin should be added for better description of experimental data.

\item
Analyzing different approximations of effective electric charge $Z_{\rm eff}$, we find that this effective charge can change the spectrum visibly (see Fig.~\ref{fig.2}).
This is indication on that different scenarios of dynamical evolution of the $\alpha$-particle concerning to nucleus (which can be constructed via different evolutions of distribution of full electric charge inside the nuclear system in dependence of relative distance between $\alpha$-particle and nucleus) can be studied via analysis of the bremsstrahlung spectra.
But, we observe that variations of effective electric charge (related with evolution of $\alpha$-particle concerning to nucleus) do not change essentially energies of maximums of the spectra (minimums are at different energies).
Maximums of the spectra indicate on the largest probabilities of existence of $\alpha$-nuclear system, i.e. some the most stable states with energies corresponding to such maximums.
To follow to such a logic, such stable states do not dependent essentially on the effective electric charge.


\item
Inclusion of incoherent emission for elastic scattering to calculations allows to improve essentially a general tendency of calculated spectrum in description of experimental data.
But, the full spectrum has oscillations with large amplitudes, that is in disagreement with experimental data (see Fig.~\ref{fig.3}).
We conclude that inclusion of incoherent emission is not enough to describe experimental data.
Also we find that inclusion of the incoherent emission to calculations does not shift energies of maximums of oscillations in spectrum much.
This reinforces our supposition that oscillations in the spectra are more related with used form of $\alpha$-nucleus wave function for elastic scattering.

\item
After analysis of results above, we conclude that good agreement with experimental data can be obtained if to suppose more important role of inelastic mechanisms during scattering at some energies.
We introduce a new unknown amplitude of inelastic mechanisms, which should be taken into account in wave function of the $\alpha$-nucleus scattering.
Before constructing (searching) of analytical formula for this unknown amplitude, we would like to obtain the first understanding about it.
From analysis of experimental data, we extract information about such an amplitude [see Fig.~\ref{fig.4}~(b)].
We observe oscillating behavior of modulus of this amplitude with maxima at energies related with maxima of the bremsstrahlung for elastic scattering.
Maxima of modulus of this amplitude should indicate maximal probabilities of formed compound nucleus system (at energy after emission of photon).
We suppose that such compact structures is indication of formation of cluster structures in nucleus-target.
%
%
We give predictions for these energies for the studied scattering of $\alpha + \isotope[197]{Au}$.

\item
We explain origin of oscillations in the bremsstrahlung spectra via resonating effects in elastic scattering (see Fig.~\ref{fig.5} and explanations in text).
We provide detailed information about contributions for coherent emission and incoherent emission during elastic scattering, also emission of photons due to inelastic mechanisms during inelastic scattering for the nucleus \isotope[197]{Au} (see Fig.~\ref{fig.6}).


\item
As next step, we analyze bremsstrahlung emission for scattering on more light nucleus \isotope[59]{Co} at 50~MeV of the $\alpha$-particle beam energy.
We observe that our approach describes experimental data with really well agrement on the full photon energy region
even without inclusion of inelastic mechanisms to calculations (see Fig.~\ref{fig.7},
these our calculations improve previous theoretical description of experimental data in Ref.~\cite{Sharan.1993.PRC}).
We conclude that role of incoherent emission in full bremsstrahlung is very important and not small for good description of these experimental data.
We provide the corresponding information about coherent and incoherent contributions to the full bremsstrahlung sopectrum.
This confirms the largest role of coherent emission. But at higher energies there is destructive interference between coherent and incoherent contributions.

\item
Finally, we analyze bremsstrahlung emission for elastic scattering on middle nucleus \isotope[116]{Sn} and \isotope[107]{Ag} at 50~MeV of the $\alpha$-particle beam energy.
For these reactions, we observe larger oscillations in the bremsstrahlung spectrum than for \isotope[59]{Co}, but smaller oscillations than for \isotope[197]{Au}.
In general, we obtain enough good agreement between calculations and experimental data
(see Figs.~\ref{fig.8} and ~\ref{fig.9}).

\item
For possible future experimental checking and study of possible oscillations in the bremsstrahlung spectra of photons emitted during the scattering of the $\alpha$-particle off nuclei, 
we recommend to use heavier nuclei and lower energies of the $\alpha$-particle beams as possible.
Possible measurements of photons at higher energies will provide more detailed and rich information about internal nuclear processes.
Experimental information about energies of photons at maximums in the spectra will be also very useful for further theoretical investigations.

\end{itemize}

In order to test our calculations and formalism (giving oscillations in the bremsstrahlung spectra for elastic $\alpha$-nucleus scattering,
and especially our results for nucleus \isotope[197]{Au}),
we have reconstructed our previous results for bremsstrahlung in $\alpha$ decay
(see Fig.~1~(b) in Ref.~\cite{Liu_Maydanyuk_Zhang_Liu.2019.PRC.hypernuclei}).
By such a way, we have constructed an unified bremsstrahlung theory form $\alpha$-nucleus scattering and $\alpha$ decay.
The developed approach allows to extract from experimental data accurate information about coherent, incoherent emissions, emission of photons caused by inelastic mechanisms during scattering (see Fig.~\ref{fig.6}).
This approach can be used for detailed investigations of such compact structures in nuclei during scattering and estimating corresponding energies.

\section*{Acknowledgements
\label{sec.acknowledgements}}

Authors are highly appreciated to
Profs. A.~G.~Magner, S.~N.~Fedotkin, F.~A.~Ivanyuk, J.~Balog,
Drs. A.~V.~Pimikov, N.~S.~Korchagin for interesting, fruitful and useful discussions.
S.~P.~Maydanyuk thanks the Institute of Modern Physics of Chinese Academy of Sciences for warm hospitality and support.
This work was supported by
the Major State Basic Research Development Program in China (No. 2015CB856903),
the National Natural Science Foundation of China (Grant No. 11575254, No.~11447105, No.~11175215, 
and No.~11805242), 
the National Key Research and Development Program of China (No. 2016YFE0130800), 
the Chinese Academy of Sciences fellowships for researchers from developing countries (No. 2014FFJA0003).

\appendix
\section{Operator of emission in relative coordinates
\label{sec.app.1}}


In this Appendix we find operator of emission in relative coordinates.
We start from Eq.~(\ref{eq.2.3.4}) rewritten via momenta:
\begin{equation}
\begin{array}{lcl}
  \hat{H}_{\gamma} & = &
  -\, \sqrt{\displaystyle\frac{2\pi c^{2}}{\hbar w_{\rm ph}}}\;
  \displaystyle\sum_{i=1}^{4}
  \displaystyle\sum\limits_{\alpha=1,2}
    e^{-i\, \vb{k_{\rm ph}r}_{i}}\,
  \biggl\{
    \mu_{N}\, \displaystyle\frac{2 z_{i} m_{\rm p}}{m_{\alpha i}}\: \vb{e}^{(\alpha)} \cdot \vu{p}_{\alpha i} +
    i\, \mu_{i}^{\rm (an)}\, \sigmabf \cdot \Bigl( - \hbar \bigl[ \vb{k_{\rm ph}} \times \vb{e}^{(\alpha)} \bigr] + \bigl[ \vu{p}_{\alpha i} \times \vb{e}^{(\alpha)} \bigr] \Bigr)
  \biggr\}\; - \\

  & - &
  \sqrt{\displaystyle\frac{2\pi c^{2}}{\hbar w_{\rm ph}}}\;
  \displaystyle\sum_{j=1}^{A}
  \displaystyle\sum\limits_{\alpha=1,2}
    e^{-i\, \vb{k_{\rm ph}r}_{j}}\;
    \biggl\{
      \mu_{N}\, \displaystyle\frac{2 z_{j} m_{\rm p}}{m_{Aj}}\: \vb{e}^{(\alpha)} \cdot \vu{p}_{Aj} +
      i\, \mu_{j}^{\rm (an)}\, \sigmabf \cdot \Bigl( - \hbar \bigl[ \vb{k_{\rm ph}} \times \vb{e}^{(\alpha)} \bigr] + \bigl[ \vu{p}_{Aj} \times \vb{e}^{(\alpha)} \bigr] \Bigr)
    \biggr\}.
\end{array}
\label{eq.app.1.1}
\end{equation}
Substituting here formulas (\ref{eq.2.4.3.3}) for $\vu{p}_{\alpha i}$, $\vu{p}_{\alpha n}$, $\vu{p}_{Aj}$ and $\vu{p}_{AA}$, we find:
\begin{equation}
\begin{array}{lcl}
\vspace{0.2mm}
  & & \hat{H}_{\gamma} =
  -\, \sqrt{\displaystyle\frac{2\pi c^{2}}{\hbar w_{\rm ph}}}\;
  \displaystyle\sum\limits_{\alpha=1,2}
  \displaystyle\sum_{i=1}^{4}
    e^{-i\, \vb{k_{\rm ph}r}_{i}}\,
  \biggl\{
    \mu_{N}\, \displaystyle\frac{2 z_{i} m_{\rm p}}{m_{A} + m_{\alpha}}\, \vb{e}^{(\alpha)} \cdot \vu{P} +
    i\, \mu_{i}^{\rm (an)}\, \displaystyle\frac{m_{\alpha i}}{m_{A} + m_{\alpha}}\, \sigmabf \cdot \bigl[ \vu{P} \times \vb{e}^{(\alpha)} \bigr]\; + \\
\vspace{2.0mm}
  & + &
    \mu_{N}\, \displaystyle\frac{2 z_{i} m_{\rm p}}{m_{\alpha}}\, \vb{e}^{(\alpha)} \cdot \vu{p} +
    i\, \mu_{i}^{\rm (an)}\, \displaystyle\frac{m_{\alpha i}}{m_{\alpha}}\, \sigmabf \cdot \bigl[ \vu{p} \times \vb{e}^{(\alpha)} \bigr]\; + \\

\vspace{-0.2mm}
  & + &
    \Bigl( \mu_{N}\, \displaystyle\frac{2 z_{i} m_{\rm p}}{m_{\alpha i}}\, \displaystyle\frac{m_{\alpha} - m_{\alpha i}}{m_{\alpha}}\, \Bigr)_{i \ne n}\,
      \vb{e}^{(\alpha)} \cdot \vb{\tilde{p}}_{\alpha i} -
    \mu_{N}\, \displaystyle\frac{2 z_{i} m_{\rm p}}{m_{\alpha}}\, \vb{e}^{(\alpha)} \cdot \displaystyle\sum_{k=1, k \ne i}^{n-1} \vb{\tilde{p}}_{\alpha k}\; -
    i\, \hbar\, \mu_{i}^{\rm (an)}\, \sigmabf \cdot \bigl[ \vb{k_{\rm ph}} \times \vb{e}^{(\alpha)} \bigr]\; + \\
\vspace{3.5mm}
  & + &
    i\, \mu_{i}^{\rm (an)}\,
      \Bigl( \displaystyle\frac{m_{\alpha} - m_{\alpha i}}{m_{\alpha}}\, \sigmabf \cdot \bigl[ \vb{\tilde{p}}_{\alpha i} \times \vb{e}^{(\alpha)} \bigr] \Bigr)_{i \ne n} -
    i\, \mu_{i}^{\rm (an)}\, \displaystyle\frac{m_{\alpha i}}{m_{\alpha}}\, \sigmabf \cdot
      \Bigl[ \displaystyle\sum_{k=1, k \ne i}^{n-1} \vb{\tilde{p}}_{\alpha k} \times \vb{e}^{(\alpha)} \Bigr]
  \biggr\}\; - \\

\vspace{0.2mm}
  & - &
  \sqrt{\displaystyle\frac{2\pi c^{2}}{\hbar w_{\rm ph}}}\;
  \displaystyle\sum\limits_{\alpha=1,2}
  \displaystyle\sum_{j=1}^{A}
    e^{-i\, \vb{k_{\rm ph}r}_{j}}\,
  \biggl\{
    \mu_{N}\, \displaystyle\frac{2 z_{j} m_{\rm p}}{m_{A} + m_{\alpha}}\, \vb{e}^{(\alpha)} \cdot \vu{P} +
    i\, \mu_{j}^{\rm (an)}\, \displaystyle\frac{m_{Aj}}{m_{A} + m_{\alpha}}\, \sigmabf \cdot \bigl[ \vu{P} \times \vb{e}^{(\alpha)} \bigr]\; - \\
\vspace{2.0mm}
  & - &
    \mu_{N}\, \displaystyle\frac{2 z_{j} m_{\rm p}}{m_{A}}\; \vb{e}^{(\alpha)} \cdot \vu{p} -
    i\, \mu_{j}^{\rm (an)}\, \displaystyle\frac{m_{Aj}}{m_{A}}\, \sigmabf \cdot \bigl[ \vu{p} \times \vb{e}^{(\alpha)} \bigr]\; + \\

\vspace{-0.2mm}
  & + &
    \Bigl( \mu_{N}\, \displaystyle\frac{2 z_{j} m_{\rm p}}{m_{Aj}}\, \displaystyle\frac{m_{A} - m_{Aj}}{m_{A}}\, \Bigr)_{j \ne A}\,
      \vb{e}^{(\alpha)} \cdot \vb{\tilde{p}}_{Aj} -
    \mu_{N}\, \displaystyle\frac{2 z_{j} m_{\rm p}}{m_{A}}\, \vb{e}^{(\alpha)} \cdot \displaystyle\sum_{k=1, k \ne j}^{A-1} \vb{\tilde{p}}_{Ak}\; -
    i\, \hbar\, \mu_{j}^{\rm (an)}\, \sigmabf \cdot \bigl[ \vb{k_{\rm ph}} \times \vb{e}^{(\alpha)} \bigr]\; + \\
  & + &
    i\, \mu_{j}^{\rm (an)}\,
      \Bigl( \displaystyle\frac{m_{A} - m_{Aj}}{m_{A}}\, \sigmabf \cdot \bigl[ \vb{\tilde{p}}_{Aj} \times \vb{e}^{(\alpha)} \bigr] \Bigr)_{j \ne A} -
    i\, \mu_{j}^{\rm (an)}\, \displaystyle\frac{m_{Aj}}{m_{A}}\, \sigmabf \cdot \Bigl[ \displaystyle\sum_{k=1, k \ne j}^{A-1} \vb{\tilde{p}}_{Ak} \times \vb{e}^{(\alpha)} \Bigr]
  \biggr\}.
\end{array}
\label{eq.app.1.4}
\end{equation}
This formula can be rewritten as
\begin{equation}
  \hat{H}_{\gamma} = \hat{H}_{P} + \hat{H}_{p} + \Delta\,\hat{H}_{\gamma} + \hat{H}_{k},
\label{eq.app.1.5}
\end{equation}
where
\begin{equation}
\begin{array}{lcl}
\vspace{0.2mm}
  \Delta \hat{H}_{\gamma} & = &
  -\, \sqrt{\displaystyle\frac{2\pi c^{2}}{\hbar w_{\rm ph}}}\;
  \displaystyle\sum\limits_{\alpha=1,2}
  \displaystyle\sum_{i=1}^{4}
    e^{-i\, \vb{k_{\rm ph}r}_{i}}\,
  \biggl\{
    \Bigl( \mu_{N}\, \displaystyle\frac{2 z_{i} m_{\rm p}}{m_{\alpha i}}\, \displaystyle\frac{m_{\alpha} - m_{\alpha i}}{m_{\alpha}}\, \Bigr)_{i \ne n}\,
      \vb{e}^{(\alpha)} \cdot \vb{\tilde{p}}_{\alpha i} -
    \mu_{N}\, \displaystyle\frac{2 z_{i} m_{\rm p}}{m_{\alpha}}\, \vb{e}^{(\alpha)} \cdot \displaystyle\sum_{k=1, k \ne i}^{n-1} \vb{\tilde{p}}_{\alpha k}\; + \\
\vspace{2.5mm}
  & + &
    i\, \mu_{i}^{\rm (an)}\,
      \Bigl( \displaystyle\frac{m_{\alpha} - m_{\alpha i}}{m_{\alpha}}\, \sigmabf \cdot \bigl[ \vb{\tilde{p}}_{\alpha i} \times \vb{e}^{(\alpha)} \bigr] \Bigr)_{i \ne n} -
    i\, \mu_{i}^{\rm (an)}\, \displaystyle\frac{m_{\alpha i}}{m_{\alpha}}\, \sigmabf \cdot
      \Bigl[ \displaystyle\sum_{k=1, k \ne i}^{n-1} \vb{\tilde{p}}_{\alpha k} \times \vb{e}^{(\alpha)} \Bigr]
  \biggr\}\; - \\

\vspace{0.2mm}
  & - &
  \sqrt{\displaystyle\frac{2\pi c^{2}}{\hbar w_{\rm ph}}}\;
  \displaystyle\sum\limits_{\alpha=1,2}
  \displaystyle\sum_{j=1}^{A}
    e^{-i\, \vb{k_{\rm ph}r}_{j}}\,
  \biggl\{
    \Bigl( \mu_{N}\, \displaystyle\frac{2 z_{j} m_{\rm p}}{m_{Aj}}\, \displaystyle\frac{m_{A} - m_{Aj}}{m_{A}}\, \Bigr)_{j \ne A}\,
      \vb{e}^{(\alpha)} \cdot \vb{\tilde{p}}_{Aj} -
    \mu_{N}\, \displaystyle\frac{2 z_{j} m_{\rm p}}{m_{A}}\, \vb{e}^{(\alpha)} \cdot \displaystyle\sum_{k=1, k \ne j}^{A-1} \vb{\tilde{p}}_{Ak}\; + \\
  & + &
    i\, \mu_{j}^{\rm (an)}\,
      \Bigl( \displaystyle\frac{m_{A} - m_{Aj}}{m_{A}}\, \sigmabf \cdot \bigl[ \vb{\tilde{p}}_{Aj} \times \vb{e}^{(\alpha)} \bigr] \Bigr)_{j \ne A} -
    i\, \mu_{j}^{\rm (an)}\, \displaystyle\frac{m_{Aj}}{m_{A}}\, \sigmabf \cdot \Bigl[ \displaystyle\sum_{k=1, k \ne j}^{A-1} \vb{\tilde{p}}_{Ak} \times \vb{e}^{(\alpha)} \Bigr]
  \biggr\}
\end{array}
\label{eq.app.1.8}
\end{equation}
%
and solutions for operators $\hat{H}_{P}$, $\hat{H}_{p}$, $\hat{H}_{k}$ in relative coordinates are given by Eqs.~(16)--(18) in Appendix~B in Ref.~\cite{Liu_Maydanyuk_Zhang_Liu.2019.PRC.hypernuclei}.
But, operator $\hat{H}_{\gamma}$ was not presented in that paper in a final form. So, we calculate it in this Appendix.

The formula~(\ref{eq.app.1.8}) includes only internal moments of nucleons, so it can be separated on two groups of terms, including and excluding spin.
On such a basis, rewrite Eq.~(\ref{eq.app.1.8}) as
\begin{equation}
  \Delta \hat{H}_{\gamma} = \Delta \hat{H}_{\gamma E} + \Delta \hat{H}_{\gamma M},
\label{eq.app.1.2.1}
\end{equation}
where
\begin{equation}
\begin{array}{lll}
\vspace{1.1mm}
  \Delta \hat{H}_{\gamma E} & = &
  -\, \sqrt{\displaystyle\frac{2\pi c^{2}}{\hbar w_{\rm ph}}}\: \mu_{N}\, \displaystyle\frac{2 m_{\rm p}}{m_{\alpha}}\,
  \displaystyle\sum\limits_{\alpha=1,2} \vb{e}^{(\alpha)}
  \displaystyle\sum_{i=1}^{n} z_{i}\, e^{-i\, \vb{k_{\rm ph}r}_{i}}\,
  \biggl\{
    \Bigl( \displaystyle\frac{m_{\alpha}}{m_{\alpha i}}\, \vb{\tilde{p}}_{\alpha i} \Bigr)_{i \ne n} -
    \displaystyle\sum_{k=1}^{n-1} \vb{\tilde{p}}_{\alpha k}
  \biggr\}\; - \\
\vspace{2.5mm}
  & - &
  \sqrt{\displaystyle\frac{2\pi c^{2}}{\hbar w_{\rm ph}}}\: \mu_{N}\, \displaystyle\frac{2 m_{\rm p}}{m_{A}}\,
  \displaystyle\sum\limits_{\alpha=1,2} \vb{e}^{(\alpha)}
  \displaystyle\sum_{j=1}^{A} z_{j}\, e^{-i\, \vb{k_{\rm ph}r}_{j}}\,
  \biggl\{
    \Bigl( \displaystyle\frac{m_{A}}{m_{Aj}}\, \vb{\tilde{p}}_{Aj} \Bigr)_{j \ne A} -
    \displaystyle\sum_{k=1}^{A-1} \vb{\tilde{p}}_{Ak}
  \biggr\}, \\

\vspace{1.1mm}
  \Delta \hat{H}_{\gamma M} & = &
  -\, i\, \sqrt{\displaystyle\frac{2\pi c^{2}}{\hbar w_{\rm ph}}}\;
  \displaystyle\sum\limits_{\alpha=1,2}
  \displaystyle\sum_{i=1}^{n}
    \mu_{i}^{\rm (an)}\, e^{-i\, \vb{k_{\rm ph}r}_{i}}\,
  \biggl\{
    \Bigl( \sigmabf \cdot \bigl[ \vb{\tilde{p}}_{\alpha i} \times \vb{e}^{(\alpha)} \bigr] \Bigr)_{i \ne n} -
    \displaystyle\frac{m_{\alpha i}}{m_{\alpha}}\, \displaystyle\sum_{k=1}^{n-1} \sigmabf \cdot \bigl[ \vb{\tilde{p}}_{\alpha k} \times \vb{e}^{(\alpha)} \bigr]
  \biggr\}\; - \\
  & - &
  i\, \sqrt{\displaystyle\frac{2\pi c^{2}}{\hbar w_{\rm ph}}}\;
  \displaystyle\sum\limits_{\alpha=1,2}
  \displaystyle\sum_{j=1}^{A}
    \mu_{j}^{\rm (an)}\, e^{-i\, \vb{k_{\rm ph}r}_{j}}\,
  \biggl\{
    \Bigl( \sigmabf \cdot \bigl[ \vb{\tilde{p}}_{Aj} \times \vb{e}^{(\alpha)} \bigr] \Bigr)_{j \ne A} -
    \displaystyle\frac{m_{Aj}}{m_{A}}\, \displaystyle\sum_{k=1}^{A-1} \sigmabf \cdot \bigl[ \vb{\tilde{p}}_{Ak} \times \vb{e}^{(\alpha)} \bigr]
  \biggr\},
\end{array}
\label{eq.app.1.2.9}
\end{equation}
where we take into account property:
\begin{equation}
\begin{array}{lll}
  \displaystyle\sum_{i=1}^{n}
    a_{i}\,
    \biggl\{
      \bigl( b_{i}\, \vb{\tilde{p}}_{\alpha i} \bigr)_{i \ne n} +
      b_{i}\, \displaystyle\sum_{k=1, k \ne i}^{n-1} \vb{\tilde{p}}_{\alpha k}
    \biggr\} =

  \bigl( a_{1} b_{1} + a_{2} b_{2} + a_{3} b_{3} + a_{4} b_{4} \bigr)\, \Bigl\{ \vb{\tilde{p}}_{\alpha 1} + \vb{\tilde{p}}_{\alpha 2} + \vb{\tilde{p}}_{\alpha 3} \Bigr\} =
  \displaystyle\sum\limits_{i=1}^{n} a_{i} b_{i} \cdot \displaystyle\sum\limits_{k=1}^{n-1} \vb{\tilde{p}}_{\alpha k}.
\end{array}
\label{eq.app.1.2.5}
\end{equation}


We rewrite these solutions in relative coordinates.
Using Eqs.~(\ref{eq.2.4.1.6}):
\[
\begin{array}{ll}
  \vb{r}_{\alpha i} = \rhobf_{\alpha i} + \vb{R} + c_{A}\, \vb{r}, &
  \vb{r}_{Aj} = \rhobf_{A j} + \vb{R} - c_{\alpha}\, \vb{r}, \\
  \vb{r}_{\alpha n} = \vb{R} + c_{A} \vb{r} - \displaystyle\frac{1}{m_{n}} \displaystyle\sum_{k=1}^{n-1} m_{k}\, \rhobf_{\alpha k}, &
  \vb{r}_{AA} = \vb{R} - c_{\alpha} \vb{r} - \displaystyle\frac{1}{m_{AA}} \displaystyle\sum_{k=1}^{A-1} m_{k}\, \rhobf_{A k},
\end{array}
\]
we obtain:
\begin{equation}
\begin{array}{lcl}
\vspace{0.4mm}
  \Delta \hat{H}_{\gamma E} & = &
  -\, \sqrt{\displaystyle\frac{2\pi c^{2}}{\hbar w_{\rm ph}}}\:
    2\, \mu_{N}\, e^{-i\, \vb{k_{\rm ph}}\vb{R}}\,
    \displaystyle\sum\limits_{\alpha=1,2} \vb{e}^{(\alpha)}\; \times \\
\vspace{0.4mm}
  & \times &
  \biggl\{
  \biggl[
    e^{-i\, c_{A}\, \vb{k_{\rm ph}} \vb{r}}\,
    \displaystyle\sum_{i=1}^{n-1}
      \displaystyle\frac{z_{i}\,m_{\rm p}}{m_{\alpha i}}\, e^{-i\, \vb{k_{\rm ph}} \rhobf_{\alpha i}}\, \vb{\tilde{p}}_{\alpha i} +
    e^{i\, c_{\alpha}\, \vb{k_{\rm ph}} \vb{r}}\,
    \displaystyle\sum_{j=1}^{A-1}
      \displaystyle\frac{z_{j}\,m_{\rm p}}{m_{Aj}}\, e^{-i\, \vb{k_{\rm ph}} \rhobf_{Aj}}\, \vb{\tilde{p}}_{Aj}
  \biggr]\; - \\
  & - &
  \biggl[
    \displaystyle\frac{m_{\rm p}}{m_{\alpha}}\, e^{-i\, c_{A}\, \vb{k_{\rm ph}} \vb{r}}\,
      \displaystyle\sum_{i=1}^{n} z_{i}\, e^{-i\, \vb{k_{\rm ph}} \rhobf_{\alpha i}}\, \displaystyle\sum_{k=1}^{n-1} \vb{\tilde{p}}_{\alpha k} +
    \displaystyle\frac{m_{\rm p}}{m_{A}}\, e^{i\, c_{\alpha}\, \vb{k_{\rm ph}} \vb{r}}\,
      \displaystyle\sum_{j=1}^{A} z_{j}\, e^{-i\, \vb{k_{\rm ph}} \rhobf_{Aj}}\, \displaystyle\sum_{k=1}^{A-1} \vb{\tilde{p}}_{Ak}
  \biggr]
  \biggr\},
\end{array}
\label{eq.app.1.3.3}
\end{equation}
\begin{equation}
\begin{array}{lll}
\vspace{0.4mm}
  & \Delta \hat{H}_{\gamma M} =
  -\, i\, \sqrt{\displaystyle\frac{2\pi c^{2}}{\hbar w_{\rm ph}}}\;
    e^{-i\, \vb{k_{\rm ph}} \vb{R}}\,
    \displaystyle\sum\limits_{\alpha=1,2} \; \times \\

\vspace{0.4mm}
  \times &
  \biggl\{
  \biggl[
    e^{-i\, \vb{k_{\rm ph}} c_{A}\, \vb{r}}\,
    \displaystyle\sum_{i=1}^{n-1}
      \mu_{i}^{\rm (an)}\, e^{-i\, \vb{k_{\rm ph}} \rhobf_{\alpha i}}\, \sigmabf \cdot \bigl[ \vb{\tilde{p}}_{\alpha i} \times \vb{e}^{(\alpha)} \bigr] +

    e^{i\, \vb{k_{\rm ph}} c_{\alpha}\, \vb{r}}\,
    \displaystyle\sum_{j=1}^{A-1}
      \mu_{j}^{\rm (an)}\, e^{-i\, \vb{k_{\rm ph}} \rhobf_{Aj}}\, \sigmabf \cdot \bigl[ \vb{\tilde{p}}_{Aj} \times \vb{e}^{(\alpha)} \bigr]
  \biggr]\; - \\

  \times &
  \biggl[
    e^{-i\, \vb{k_{\rm ph}} c_{A}\, \vb{r}}\,
    \displaystyle\sum_{i=1}^{n}
      \mu_{i}^{\rm (an)}\, \displaystyle\frac{m_{\alpha i}}{m_{\alpha}}\,
      e^{-i\, \vb{k_{\rm ph}} \rhobf_{\alpha i}}\,
      \displaystyle\sum_{k=1}^{n-1} \sigmabf \cdot \bigl[ \vb{\tilde{p}}_{\alpha k} \times \vb{e}^{(\alpha)} \bigr] +

    e^{i\, \vb{k_{\rm ph}} c_{\alpha}\, \vb{r}}\,
    \displaystyle\sum_{j=1}^{A}
      \mu_{j}^{\rm (an)}\, \displaystyle\frac{m_{Aj}}{m_{A}}\,
      e^{-i\, \vb{k_{\rm ph}} \rhobf_{Aj}}\,
      \displaystyle\sum_{k=1}^{A-1} \sigmabf \cdot \bigl[ \vb{\tilde{p}}_{Ak} \times \vb{e}^{(\alpha)} \bigr]
  \biggr\}.
\end{array}
\label{eq.app.1.3.4}
\end{equation}

\section{Integration over space variable $\vb{R}$
\label{sec.2.8}}

In Eq.~(\ref{eq.2.6.1}) we defined the wave function of the full nuclear system, which after ignoring correction $\Delta \Psi$ obtains form:
\begin{equation}
\begin{array}{lcl}
  \Psi = \Phi (\vb{R}) \cdot F (\vb{r}, \beta_{A}, \beta_{\alpha}), &
  F (\vb{r}, \beta_{A}, \beta_{\alpha}) =
  \Phi_{\rm \alpha - nucl} (\vb{r}) \cdot
  \psi_{\rm nucl} (\beta_{A}) \cdot
  \psi_{\alpha} (\beta_{\alpha}),
\end{array}
\label{eq.2.8.1}
\end{equation}
Here, $\Phi (\vb{R})$ is wave function describing evolution of center of mass of the full nuclear system.
We shall assume approximated form for the function $\Phi_{\bar{s}}$ before and after emission of photons as
\begin{equation}
  \Phi_{\bar{s}} (\vb{R}) =  e^{-i\,\vb{K}_{\bar{s}}\cdot\vb{R}},
\label{eq.2.8.2}
\end{equation}
where $\bar{s} = i$ or $f$ (indexes $i$ and $f$ denote the initial state,
i.e. the state before emission of photon,
and the final state, i.e. the state after emission of photon),
$\vb{K}_{s}$ is momentum of the total system~\cite{Kopitin.1997.YF}.
Previously, in $\alpha$ decay study we assumed $\vb{K}_{i} = 0$, as we considered the $\alpha$-decaying nuclear system before emission of photons as not moving in the laboratory frame.
However, for $\alpha$-nucleus scattering $\vb{K}_{i} \ne 0$ is more reasonable
(it seems for $\alpha$ decay we should also use $\vb{K}_{i} \ne 0$, that should be taken into account in further study).

Let us calculate the contribution $M_{p}$ [starting from Eq.~(\ref{eq.2.7.3.b})]:
\begin{equation}
\begin{array}{lll}
\vspace{-0.1mm}
  & & M_{p} =
  -\, \displaystyle\int_{-\infty}^{+\infty}  e^{i\, (\vb{K}_{f} - \vb{K}_{i} - \vb{k}_{\rm ph}) \cdot\vb{R}}\;  \vb{dR}\; \times \\
\vspace{-0.1mm}
  & \times &
  \displaystyle\sum\limits_{\alpha=1,2}
  \biggl\langle F_{f}\, \biggl|\,
  2\, \mu_{N}\,  m_{\rm p}\,
  \Bigl\{
    e^{-i\, c_{A} \vb{k_{\rm ph}} \vb{r}}\, \displaystyle\frac{1}{m_{\alpha}}\,
      \displaystyle\sum_{i=1}^{4} z_{i}\, e^{-i\, \vb{k_{\rm ph}} \rhobf_{\alpha i}} -
    e^{i\, c_{\alpha} \vb{k_{\rm ph}} \vb{r}}\,  \displaystyle\frac{1}{m_{A}}\,
      \displaystyle\sum_{j=1}^{A} z_{j}\, e^{-i\, \vb{k_{\rm ph}} \rhobf_{Aj}}
  \Bigr\}\; \vb{e}^{(\alpha)} \cdot \vu{p}\; + \\
  & + &
  i\,
  \Bigl\{
    e^{-i\, c_{A} \vb{k_{\rm ph}} \vb{r}} \displaystyle\frac{1}{m_{\alpha}}\,
    \displaystyle\sum_{i=1}^{4}
      \mu_{i}^{\rm (an)}\, m_{\alpha i}\;
      e^{-i\, \vb{k_{\rm ph}} \rhobf_{\alpha i}}\, \sigmabf -
    e^{i\, c_{\alpha} \vb{k_{\rm ph}} \vb{r}} \displaystyle\frac{1}{m_{A}}
    \displaystyle\sum_{j=1}^{A}
      \mu_{j}^{\rm (an)}\, m_{Aj}\;
      e^{-i\, \vb{k_{\rm ph}} \rhobf_{Aj}}\, \sigmabf \Bigr\}
    \cdot \bigl[ \vu{p} \times \vb{e}^{(\alpha)} \bigr]\,
  \biggr|\, F_{i}\, \biggr\rangle.
\end{array}
\label{eq.2.8.5}
\end{equation}
$\delta$-Function is defined as
\begin{equation}
  \displaystyle\int_{-\infty}^{+\infty}  e^{i\, (\vb{K}_{f} - \vb{K}_{i} - \vb{k}) \cdot\vb{R}}\;  \vb{dR}\; =
  (2\pi)^{3} \delta (\vb{K}_{f} - \vb{K}_{i} - \vb{k}).
\label{eq.2.8.6}
\end{equation}
And from (\ref{eq.2.8.5}) we obtain:
\begin{equation}
\begin{array}{lll}
\vspace{-0.1mm}
  & & M_{p} = -\, (2\pi)^{3} \delta (\vb{K}_{f} - \vb{K}_{i} - \vb{k}_{\rm ph})\; \times \\
\vspace{-0.1mm}
  & \times &
  \displaystyle\sum\limits_{\alpha=1,2}
  \biggl\langle F_{f}\, \biggl|\,
  2\, \mu_{N}\,  m_{\rm p}\,
  \Bigl\{
    e^{-i\, c_{A} \vb{k_{\rm ph}} \vb{r}}\, \displaystyle\frac{1}{m_{\alpha}}\,
      \displaystyle\sum_{i=1}^{4} z_{i}\, e^{-i\, \vb{k_{\rm ph}} \rhobf_{\alpha i}} -
    e^{i\, c_{\alpha} \vb{k_{\rm ph}} \vb{r}}\,  \displaystyle\frac{1}{m_{A}}\,
      \displaystyle\sum_{j=1}^{A} z_{j}\, e^{-i\, \vb{k_{\rm ph}} \rhobf_{Aj}}
  \Bigr\}\; \vb{e}^{(\alpha)} \cdot \vu{p}\; + \\
  & + &
  i\,
  \Bigl\{
    e^{-i\, c_{A} \vb{k_{\rm ph}} \vb{r}} \displaystyle\frac{1}{m_{\alpha}}\,
    \displaystyle\sum_{i=1}^{4}
      \mu_{i}^{\rm (an)}\, m_{\alpha i}\;
      e^{-i\, \vb{k_{\rm ph}} \rhobf_{\alpha i}}\, \sigmabf -
    e^{i\, c_{\alpha} \vb{k_{\rm ph}} \vb{r}} \displaystyle\frac{1}{m_{A}}
    \displaystyle\sum_{j=1}^{A}
      \mu_{j}^{\rm (an)}\, m_{Aj}\;
      e^{-i\, \vb{k_{\rm ph}} \rhobf_{Aj}}\, \sigmabf \Bigr\}
    \cdot \bigl[ \vu{p} \times \vb{e}^{(\alpha)} \bigr]\,
  \biggr|\, F_{i}\, \biggr\rangle.
\end{array}
\label{eq.2.8.7}
\end{equation}
In this formula we have integration over space variables $\vb{r}$, $\rhobf_{\alpha i}$, $\rhobf_{Aj}$ ($i = 1 \ldots n-1$, $j = 1 \ldots A$).

\vspace{1.5mm}
For other matrix elements we obtain:
\begin{equation}
\begin{array}{lcl}
\vspace{-0.1mm}
  M_{k} & = & (2\pi)^{3} \delta (\vb{K}_{f} - \vb{K}_{i} - \vb{k}_{\rm ph})\; \times \\
\vspace{-0.1mm}
  & \times &
  i\, \hbar\,
  \displaystyle\sum\limits_{\alpha=1,2}
  \biggl\langle F_{f}\, \biggl|\,
  \biggl\{
    e^{-i\, c_{A}\, \vb{k_{\rm ph}} \vb{r}}\,
    \displaystyle\sum_{i=1}^{4}
      \mu_{i}^{\rm (an)}\, e^{-i\, \vb{k_{\rm ph}} \rhobf_{\alpha i}}\, \sigmabf +
    e^{i\, c_{\alpha}\, \vb{k_{\rm ph}} \vb{r}}\,
    \displaystyle\sum_{j=1}^{A}
      \mu_{j}^{\rm (an)}\, e^{-i\, \vb{k_{\rm ph}} \rhobf_{Aj}}\, \sigmabf
  \biggr\}
  \cdot \bigl[ \vb{k_{\rm ph}} \cp \vb{e}^{(\alpha)} \bigr]
  \biggr|\, F_{i} \biggr\rangle,
\end{array}
\label{eq.2.8.8}
\end{equation}
\begin{equation}
\begin{array}{lcl}
\vspace{-0.1mm}
  M_{\Delta E} & = & -\, (2\pi)^{3} \delta (\vb{K}_{f} - \vb{K}_{i} - \vb{k}_{\rm ph})\; \times \\
\vspace{-0.1mm}
  & \times &
  2\, \mu_{N}
  \displaystyle\sum\limits_{\alpha=1,2} \vb{e}^{(\alpha)}
  \biggl\langle F_{f}\, \biggl|\,
  \biggl[
    e^{-i\, c_{A}\, \vb{k_{\rm ph}} \vb{r}}\,
    \displaystyle\sum_{i=1}^{3}
      \displaystyle\frac{z_{i} m_{\rm p}}{m_{\alpha i}}\,
      e^{-i\, \vb{k_{\rm ph}} \rhobf_{\alpha i}}\,
      \vb{\tilde{p}}_{\alpha i} +
    e^{i\, c_{\alpha}\, \vb{k_{\rm ph}} \vb{r}}\,
    \displaystyle\sum_{j=1}^{A-1}
      \displaystyle\frac{z_{j} m_{\rm p}}{m_{Aj}}\,
      e^{-i\, \vb{k_{\rm ph}} \rhobf_{Aj}}\,
      \vb{\tilde{p}}_{Aj}
  \biggr]\; - \\

  & - &
  \biggl[
    \displaystyle\frac{m_{\rm p}}{m_{\alpha}}\,
    e^{-i\, c_{A}\, \vb{k_{\rm ph}} \vb{r}}\,
    \displaystyle\sum_{i=1}^{4}
      z_{i}\, e^{-i\, \vb{k_{\rm ph}} \rhobf_{\alpha i}}\,
    \Bigl( \displaystyle\sum_{k=1}^{n-1} \vb{\tilde{p}}_{\alpha k} \Bigr) +

    \displaystyle\frac{m_{\rm p}}{m_{A}}\,
    e^{i\, c_{\alpha}\, \vb{k_{\rm ph}} \vb{r}}\,
    \displaystyle\sum_{j=1}^{A}
      z_{j}\, e^{-i\, \vb{k_{\rm ph}} \rhobf_{Aj}}
    \Bigl( \displaystyle\sum_{k=1}^{A-1} \vb{\tilde{p}}_{Ak} \Bigr)\,
  \biggr]
  \biggr|\, F_{i} \biggr\rangle,
\end{array}
\label{eq.2.8.9}
\end{equation}
\begin{equation}
\begin{array}{lll}
\vspace{0.2mm}
  & M_{\Delta M} = -\, i\, (2\pi)^{3} \delta (\vb{K}_{f} - \vb{K}_{i} - \vb{k}_{\rm ph})\; \times \\

\vspace{0.2mm}
  \times &
    \displaystyle\sum\limits_{\alpha=1,2} \biggl\langle F_{f}\, \biggl|\;
  \biggl[
    e^{-i\, \vb{k_{\rm ph}} c_{A}\, \vb{r}}\,
    \displaystyle\sum_{i=1}^{n-1}
      \mu_{i}^{\rm (an)}\, e^{-i\, \vb{k_{\rm ph}} \rhobf_{\alpha i}}\, \sigmabf \cdot \bigl[ \vb{\tilde{p}}_{\alpha i} \times \vb{e}^{(\alpha)} \bigr] +
    e^{i\, \vb{k_{\rm ph}} c_{\alpha}\, \vb{r}}\,
    \displaystyle\sum_{j=1}^{A-1}
      \mu_{j}^{\rm (an)}\, e^{-i\, \vb{k_{\rm ph}} \rhobf_{Aj}}\, \sigmabf \cdot \bigl[ \vb{\tilde{p}}_{Aj} \times \vb{e}^{(\alpha)} \bigr]
  \biggr]\; - \\

  - &
  \Bigl[
    e^{-i\, \vb{k_{\rm ph}} c_{A}\, \vb{r}}\,
    \displaystyle\sum_{i=1}^{n}
      \mu_{i}^{\rm (an)}\, \displaystyle\frac{m_{\alpha i}}{m_{\alpha}}\,
      e^{-i\, \vb{k_{\rm ph}} \rhobf_{\alpha i}}\,
      \displaystyle\sum_{k=1}^{n-1} \sigmabf \cdot \bigl[ \vb{\tilde{p}}_{\alpha k} \times \vb{e}^{(\alpha)} \bigr] +

    e^{i\, \vb{k_{\rm ph}} c_{\alpha}\, \vb{r}}\,
    \displaystyle\sum_{j=1}^{A}
      \mu_{j}^{\rm (an)}\, \displaystyle\frac{m_{Aj}}{m_{A}}\,
      e^{-i\, \vb{k_{\rm ph}} \rhobf_{Aj}}\,
      \displaystyle\sum_{k=1}^{A-1} \sigmabf \cdot \bigl[ \vb{\tilde{p}}_{Ak} \times \vb{e}^{(\alpha)} \bigr]
  \Bigr]
  \biggr|\, F_{i} \biggr\rangle.
\end{array}
\label{eq.2.8.10}
\end{equation}

For $\alpha$-nucleus scattering for $M_{P}$ we have $\vb{K}_{i} \ne 0$ and obtain property:
\begin{equation}
\begin{array}{lcl}
  \vu{P} \Psi_{i} = \vu{P} \Phi_{i} (\vb{R}) F_{i} = -\,\hbar\, \vb{K}_{i}\, \Phi_{\bar{s}} (\vb{R}) F_{i}, &
  \vb{K}_{i} \ne 0.
\end{array}
\label{eq.2.8.11}
\end{equation}
We calculate the last matrix element:
\begin{equation}
\begin{array}{lcl}
\vspace{-0.1mm}
  M_{P} & = & \hbar\, (2\pi)^{3}\, \delta (\vb{K}_{f} - \vb{K}_{i} - \vb{k}_{\rm ph})\; \times \\
\vspace{-0.1mm}
  & \times &
  \displaystyle\frac{1}{m_{A} + m_{\alpha}}\,
  \displaystyle\sum\limits_{\alpha=1,2}
  \biggl\langle F_{f}\, \biggl|\,
    2\, \mu_{N}\, m_{\rm p}\;
    \biggl\{
      e^{-i\, c_{A}\, \vb{k_{\rm ph}} \vb{r}}
        \displaystyle\sum_{i=1}^{4} z_{i}\, e^{-i\, \vb{k_{\rm ph}} \rhobf_{\alpha i}} +
      e^{i\, c_{\alpha}\, \vb{k_{\rm ph}} \vb{r} }
        \displaystyle\sum_{j=1}^{A} z_{j}\, e^{-i\, \vb{k_{\rm ph}} \rhobf_{Aj}}
    \biggr\}\, \vb{e}^{(\alpha)} \cdot \vb{K}_{i}\; + \\

  & + &
    i\,
    \biggl\{
      e^{-i\, c_{A}\, \vb{k_{\rm ph}} \vb{r}}\,
        \displaystyle\sum_{i=1}^{4} \mu_{i}^{\rm (an)}\, m_{\alpha i}\, e^{-i\, \vb{k_{\rm ph}} \rhobf_{\alpha i}}\, \sigmabf  +
      e^{i\, c_{\alpha}\, \vb{k_{\rm ph}} \vb{r}}\,
        \displaystyle\sum_{j=1}^{A} \mu_{j}^{\rm (an)}\, m_{Aj}\, e^{-i\, \vb{k_{\rm ph}} \rhobf_{Aj}}\, \sigmabf
  \biggr\}\, \cdot \bigl[ \vb{K}_{i} \cp \vb{e}^{(\alpha)} \bigr]\,
  \biggr|\, F_{i} \biggr\rangle.
\end{array}
\label{eq.2.8.12}
\end{equation}

\section{Electric and magnetic form factors
\label{sec.2.9}}

In calculation of the matrix element $M_{p}$, we follow to formalism of Ref.~\cite{Liu_Maydanyuk_Zhang_Liu.2019.PRC.hypernuclei}
[see Eqs.~(C1)--(C10), in Appendix~C in that paper] and obtain:
\begin{equation}
\begin{array}{lll}
\vspace{-0.1mm}
  M_{p} & = &
  i \hbar\, (2\pi)^{3} \delta (\vb{K}_{f} - \vb{K}_{i} - \vb{k}_{\rm ph}) \cdot
  \displaystyle\sum\limits_{\alpha=1,2}
  \displaystyle\int\limits_{}^{}
    \Phi_{\rm \alpha - nucl, f}^{*} (\vb{r})\;
    e^{i\, \vb{k}_{\rm ph} \vb{r}}\; \times \\
\vspace{0.5mm}
  & \times &
  \biggl\{
  2\, \mu_{N}\,  m_{\rm p}\,
  \Bigl[
    e^{-i\, c_{A} \vb{k_{\rm ph}} \vb{r}}\, \displaystyle\frac{1}{m_{\alpha}}\, F_{\alpha,\, {\rm el}} -
    e^{i\, c_{\alpha} \vb{k_{\rm ph}} \vb{r}}\,  \displaystyle\frac{1}{m_{A}}\, F_{A,\, {\rm el}}
  \Bigr]\,
  e^{- i\, \vb{k}_{\rm ph} \vb{r}} \cdot
  \vb{e}^{(\alpha)}\, \vb{\displaystyle\frac{d}{dr}}\; + \\
  & + &
  i\,
  \Bigl[
    e^{-i\, c_{A} \vb{k_{\rm ph}} \vb{r}} \displaystyle\frac{1}{m_{\alpha}}\, \vb{F}_{\alpha,\, {\rm mag}} -
    e^{i\, c_{\alpha} \vb{k_{\rm ph}} \vb{r}} \displaystyle\frac{1}{m_{A}}\, \vb{F}_{A,\, {\rm mag}}
  \Bigr]\,
  e^{- i\, \vb{k}_{\rm ph} \vb{r}} \cdot
  \Bigl[ \vb{\displaystyle\frac{d}{dr}} \times \vb{e}^{(\alpha)} \Bigr]\,
  \biggr\} \cdot
  \Phi_{\rm \alpha - nucl, i} (\vb{r})\; \vb{dr},
\end{array}
\label{eq.2.9.10}
\end{equation}
where we introduced \emph{electric and magnetic form factors} of the $\alpha$-particle and nucleus as
\begin{equation}
\begin{array}{llllll}
\vspace{1mm}
  F_{\alpha,\, {\rm el}} (\vb{k}_{\rm ph}) & = &
    \displaystyle\sum\limits_{n=1}^{4}
    \Bigl\langle \psi_{\alpha, f} (\beta_{\alpha})\, \Bigl|\,
      z_{n}\, e^{-i \vb{k}_{\rm ph} \rhobf_{\alpha n} }
    \Bigr|\,  \psi_{\alpha, i} (\beta_{\alpha}) \Bigr\rangle , \\

\vspace{1mm}
  F_{A,\, {\rm el}} (\vb{k}_{\rm ph}) & = &
    \displaystyle\sum\limits_{m=1}^{A}
    \Bigl\langle \psi_{\rm nucl, f} (\beta_{A}) \Bigl|\,
      z_{m}\, e^{-i \vb{k}_{\rm ph} \rhobf_{A m} }
    \Bigr|\, \psi_{\rm nucl, i} (\beta_{A}) \Bigr\rangle , \\

\vspace{1mm}
  \vb{F}_{\alpha,\, {\rm mag}} (\vb{k}_{\rm ph}) & = &
    \displaystyle\sum_{i=1}^{4}
    \Bigl\langle \psi_{\alpha, f} (\beta_{\alpha})\, \Bigl|\,
      \mu_{i}^{\rm (an)}\, m_{\alpha i}\; e^{-i\, \vb{k_{\rm ph}} \rhobf_{\alpha i}}\, \sigmabf
    \Bigr| \psi_{\alpha, i} (\beta_{\alpha}) \Bigr\rangle, \\

  \vb{F}_{A,\, {\rm mag}} (\vb{k}_{\rm ph}) & = &
    \displaystyle\sum_{j=1}^{A}
    \Bigl\langle \psi_{\rm nucl, f} (\beta_{A})\, \Bigl|\,
        \mu_{j}^{\rm (an)}\, m_{Aj}\; e^{-i\, \vb{k_{\rm ph}} \rhobf_{Aj}}\, \sigmabf
    \Bigr| \psi_{\rm nucl, i} (\beta_{A}) \Bigr\rangle.
\end{array}
\label{eq.2.9.7}
\end{equation}
Matrix elements $M_{\Delta E}$, $M_{\Delta M}$, $M_{k}$ and $M_{P}$ were not found in that paper in final form.
So, we will obtain them here.

\subsection{Calculations for the matrix elements $M_{\Delta E}$ and $M_{\Delta M}$
\label{sec.2.9.b}}

We substitute explicit representation (\ref{eq.2.8.1}) for wave function $F (\vb{r}, \beta_{A}, \beta_{\alpha})$ to the matrix elements
$M_{\Delta E}$ and $M_{\Delta M}$ in form (\ref{eq.2.8.9}) and (\ref{eq.2.8.10}) and
rewrite integration over variable $\vb{r}$ explicitly.
We take into account that function $\psi_{\alpha, s} (\beta_{\alpha})$ is dependent of variables $\rhobf_{\alpha n}$ (i.e. it is not dependent on variables $\rhobf_{A m}$),
as the function $\psi_{\rm nucl, s} (\beta_{A})$ is dependent on variables $\rhobf_{A m}$ (i.e. it is not dependent on variables $\rhobf_{\alpha n}$).
We take into account normalization condition for wave functions as
\begin{equation}
\begin{array}{lll}
  \Bigl\langle \psi_{\rm nucl, f} (\beta_{A}) \Bigr|\, \psi_{\rm nucl, i} (\beta_{A}) \Bigr\rangle = 1, &
  \Bigl\langle \psi_{\alpha, f} (\beta_{\alpha})\, \Bigl|\, \psi_{\alpha, i} (\beta_{\alpha}) \Bigr\rangle = 1,
\end{array}
\label{eq.2.9.5}
\end{equation}
and we take into account definition for relative momentum as
\begin{equation}
  \vb{p_{\rm small}} = - i \hbar\, \vb{\displaystyle\frac{d}{dr}}.
\label{eq.2.9.8}
\end{equation}
And we obtain the following solutions:
\begin{equation}
\begin{array}{lll}
\vspace{-0.1mm}
  & M_{\Delta E} =
  -\, (2\pi)^{3} \delta (\vb{K}_{f} - \vb{K}_{i} - \vb{k}_{\rm ph}) \cdot
  2\, \mu_{N}
  \displaystyle\sum\limits_{\alpha=1,2} \vb{e}^{(\alpha)}
  \displaystyle\int\limits_{}^{}
    \Phi_{\rm \alpha - nucl, f}^{*} (\vb{r})\; \times \\
  \times &
  \biggl\{
    \Bigl[
      e^{-i\, c_{A}\, \vb{k_{\rm ph}} \vb{r}}\, \vb{D}_{\alpha 1,\, {\rm el}} +
      e^{i\, c_{\alpha}\, \vb{k_{\rm ph}} \vb{r}}\, \vb{D}_{A 1,\, {\rm el}}
    \Bigr] -

  \Bigl[
    \displaystyle\frac{m_{\rm p}}{m_{\alpha}}\, e^{-i\, c_{A}\, \vb{k_{\rm ph}} \vb{r}}\, \vb{D}_{\alpha 2,\, {\rm el}} +
    \displaystyle\frac{m_{\rm p}}{m_{A}}\, e^{i\, c_{\alpha}\, \vb{k_{\rm ph}} \vb{r}}\, \vb{D}_{A 2,\, {\rm el}}
  \Bigr]
  \biggr\} \cdot
  \Phi_{\rm \alpha - nucl, i} (\vb{r})\; \vb{dr},
\end{array}
\label{eq.2.9.b.1}
\end{equation}
\begin{equation}
\begin{array}{lll}
\vspace{-0.1mm}
  M_{\Delta M} & = &
  -\, i\, (2\pi)^{3} \delta (\vb{K}_{f} - \vb{K}_{i} - \vb{k}_{\rm ph}) \cdot
  \displaystyle\sum\limits_{\alpha=1,2}
  \displaystyle\int\limits_{}^{}
    \Phi_{\rm \alpha - nucl, f}^{*} (\vb{r})\; \times \\

\vspace{-0.1mm}
  & \times &
  \biggl\{
    \Bigl[
      e^{-i\, c_{A}\, \vb{k_{\rm ph}} \vb{r}}\; D_{\alpha 1,\, {\rm mag}} (\vb{e}^{(\alpha)}) +
      e^{i\, c_{\alpha}\, \vb{k_{\rm ph}} \vb{r}}\; D_{A 1,\, {\rm mag}} (\vb{e}^{(\alpha)})
    \Bigr]\; - \\

  & - &
  \Bigl[
    e^{-i\, c_{A}\, \vb{k_{\rm ph}} \vb{r}}\; D_{\alpha 2,\, {\rm mag}} (\vb{e}^{(\alpha)}) +
    e^{i\, c_{\alpha}\, \vb{k_{\rm ph}} \vb{r}}\; D_{A 2,\, {\rm mag}} (\vb{e}^{(\alpha)})
  \Bigr]
  \biggr\} \cdot
  \Phi_{\rm \alpha - nucl, i} (\vb{r})\; \vb{dr},
\end{array}
\label{eq.2.9.b.2}
\end{equation}
where
\begin{equation}
\begin{array}{lll}
\vspace{1mm}
  \vb{D}_{\alpha 1,\, {\rm el}} = &
    \displaystyle\sum\limits_{i=1}^{3}
      \displaystyle\frac{z_{i} m_{\rm p}}{m_{\alpha i}}\,
      \Bigl\langle \psi_{\alpha, f} (\beta_{\alpha})\, \Bigl|\,
        e^{-i \vb{k}_{\rm ph} \rhobf_{\alpha i}} \vb{\tilde{p}}_{\alpha i}
      \Bigr|\,  \psi_{\alpha, i} (\beta_{\alpha}) \Bigr\rangle , \\

\vspace{1mm}
  \vb{D}_{A 1,\, {\rm el}} = &
    \displaystyle\sum\limits_{i=1}^{A-1}
      \displaystyle\frac{z_{j} m_{\rm p}}{m_{Aj}}\,
      \Bigl\langle \psi_{A, f} (\beta_{A})\, \Bigl|\,
        e^{-i \vb{k}_{\rm ph} \rhobf_{Aj}} \vb{\tilde{p}}_{Aj}
      \Bigr|\,  \psi_{Aj} (\beta_{A}) \Bigr\rangle , \\

\vspace{1mm}
  \vb{D}_{\alpha 2,\, {\rm el}} = &
    \displaystyle\sum\limits_{i=1}^{4}
      z_{i}\,
      \Bigl\langle \psi_{\alpha, f} (\beta_{\alpha})\, \Bigl|\,
        e^{-i \vb{k}_{\rm ph} \rhobf_{\alpha i}}\,
        \Bigl( \displaystyle\sum_{k=1}^{n-1} \vb{\tilde{p}}_{\alpha k} \Bigr)
      \Bigr|\,  \psi_{\alpha, i} (\beta_{\alpha}) \Bigr\rangle, \\

  \vb{D}_{A 2,\, {\rm el}} = &
    \displaystyle\sum\limits_{i=1}^{A}
      z_{j}\,
      \Bigl\langle \psi_{A, f} (\beta_{A})\, \Bigl|\,
        e^{-i \vb{k}_{\rm ph} \rhobf_{Aj}}
        \Bigl( \displaystyle\sum_{k=1}^{A-1} \vb{\tilde{p}}_{Ak} \Bigr)
      \Bigr|\,  \psi_{Aj} (\beta_{A}) \Bigr\rangle,
\end{array}
\label{eq.2.9.b.3}
\end{equation}
\begin{equation}
\begin{array}{lll}
\vspace{1mm}
  D_{\alpha 1,\, {\rm mag}} (\vb{e}^{(\alpha)}) = &
  \displaystyle\sum\limits_{i=1}^{3}
    \mu_{i}^{\rm (an)}\,
    \Bigl\langle \psi_{\alpha, f} (\beta_{\alpha})\, \Bigl|\,
      e^{-i\, \vb{k_{\rm ph}} \rhobf_{\alpha i}}\; \sigmabf \cdot
      \bigl[ \vb{\tilde{p}}_{\alpha i} \times \vb{e}^{(\alpha)} \bigr]
    \Bigr|\,  \psi_{\alpha, i} (\beta_{\alpha}) \Bigr\rangle, \\

\vspace{1mm}
  D_{A 1,\, {\rm mag}} (\vb{e}^{(\alpha)}) = &
    \displaystyle\sum\limits_{j=1}^{A-1}
    \mu_{j}^{\rm (an)}\,
    \Bigl\langle \psi_{A, f} (\beta_{A})\, \Bigl|\,
      e^{-i\, \vb{k_{\rm ph}} \rhobf_{Aj}}\; \sigmabf \cdot
      \bigl[ \vb{\tilde{p}}_{Aj} \times \vb{e}^{(\alpha)} \bigr]
    \Bigr|\,  \psi_{Aj} (\beta_{A}) \Bigr\rangle, \\

\vspace{1mm}
  D_{\alpha 2,\, {\rm mag}} (\vb{e}^{(\alpha)}) = &
    \displaystyle\sum\limits_{i=1}^{4}
      \mu_{i}^{\rm (an)}\,
      \displaystyle\frac{m_{\alpha i}}{m_{\alpha}}\,
      \Bigl\langle \psi_{\alpha, f} (\beta_{\alpha})\, \Bigl|\,
      e^{-i\, \vb{k_{\rm ph}} \rhobf_{\alpha i}}\; \sigmabf \cdot
      \Bigl[ \Bigl( \displaystyle\sum_{k=1}^{n-1} \vb{\tilde{p}}_{\alpha k} \Bigr) \times \vb{e}^{(\alpha)} \Bigr]
      \Bigr|\,  \psi_{\alpha, i} (\beta_{\alpha}) \Bigr\rangle, \\

  D_{A 2,\, {\rm mag}} (\vb{e}^{(\alpha)}) = &
    \displaystyle\sum\limits_{j=1}^{A}
      \mu_{j}^{\rm (an)}\,
      \displaystyle\frac{m_{Aj}}{m_{A}}\,
      \Bigl\langle \psi_{A, f} (\beta_{A})\, \Bigl|\,
      e^{-i\, \vb{k_{\rm ph}} \rhobf_{Aj}}\; \sigmabf \cdot
      \Bigl[ \Bigl( \displaystyle\sum_{k=1}^{A-1} \vb{\tilde{p}}_{Ak} \Bigr) \times \vb{e}^{(\alpha)} \Bigr]
      \Bigr|\,  \psi_{A, i} (\beta_{A}) \Bigr\rangle.
\end{array}
\label{eq.2.9.b.4}
\end{equation}

\subsection{Calculations for the matrix elements $M_{k}$
\label{sec.2.9.c}}

For $M_{k}$ we obtained solution (\ref{eq.2.8.8}):
\begin{equation}
\begin{array}{lcl}
\vspace{-0.1mm}
  M_{k} & = &
  i\, \hbar\, (2\pi)^{3} \delta (\vb{K}_{f} - \vb{K}_{i} - \vb{k}_{\rm ph}) \cdot
  \displaystyle\sum\limits_{\alpha=1,2}
    \bigl[ \vb{k_{\rm ph}} \cp \vb{e}^{(\alpha)} \bigr]
  \displaystyle\int\limits_{}^{}
    \Phi_{\rm \alpha - nucl, f}^{*} (\vb{r})\; \times \\
\vspace{-0.1mm}
  & \times &
  \biggl\{
    e^{-i\, c_{A}\, \vb{k_{\rm ph}} \vb{r}}\, \vb{D}_{\alpha,\, {\rm k}} +
    e^{i\, c_{\alpha}\, \vb{k_{\rm ph}} \vb{r}}\, \vb{D}_{A,\, {\rm k}}
  \biggr\} \cdot
  \Phi_{\rm \alpha - nucl, i} (\vb{r})\; \vb{dr},
\end{array}
\label{eq.2.9.c.1}
\end{equation}
where
\begin{equation}
\begin{array}{lll}
\vspace{1mm}
  \vb{D}_{\alpha,\, {\rm k}} = &
  \displaystyle\sum\limits_{i=1}^{4}
    \mu_{i}^{\rm (an)}\,
    \Bigl\langle \psi_{\alpha, f} (\beta_{\alpha})\, \Bigl|\,
      e^{-i\, \vb{k_{\rm ph}} \rhobf_{\alpha i}}\, \sigmabf
    \Bigr|\,  \psi_{\alpha, i} (\beta_{\alpha}) \Bigr\rangle , \\

  \vb{D}_{A,\, {\rm k}} = &
  \displaystyle\sum\limits_{j=1}^{A}
    \mu_{j}^{\rm (an)}\,
    \Bigl\langle \psi_{A, f} (\beta_{A})\, \Bigl|\,
      e^{-i\, \vb{k_{\rm ph}} \rhobf_{Aj}}\, \sigmabf
    \Bigr|\,  \psi_{A, j} (\beta_{A}) \Bigr\rangle.
\end{array}
\label{eq.2.9.c.2}
\end{equation}

\subsection{Calculations for the matrix elements $M_{P}$
\label{sec.2.9.d}}

For $M_{P}$ we obtained solution (\ref{eq.2.8.12}):
\begin{equation}
\begin{array}{lcl}
\vspace{-0.1mm}
  M_{P} & = &
  \hbar\, (2\pi)^{3} \delta (\vb{K}_{f} - \vb{K}_{i} - \vb{k}_{\rm ph}) \cdot
  \displaystyle\frac{1}{m_{A} + m_{\alpha}}\,
  \displaystyle\sum\limits_{\alpha=1,2}
  \displaystyle\int\limits_{}^{}
    \Phi_{\rm \alpha - nucl, f}^{*} (\vb{r})\; \times \\

\vspace{-0.1mm}
  & \times &
  \biggl\{
    2\, \mu_{N}\, m_{\rm p}\;
    \Bigl[
      e^{-i\, c_{A}\, \vb{k_{\rm ph}} \vb{r}} D_{\alpha,P\, {\rm el}} +
      e^{i\, c_{\alpha}\, \vb{k_{\rm ph}} \vb{r} } D_{A,P\, {\rm el}}
    \Bigr]\, \vb{e}^{(\alpha)} \cdot \vb{K}_{i} + \\

  & + &
  i\,
  \Bigl[
    e^{-i\, c_{A}\, \vb{k_{\rm ph}} \vb{r}}\,
      \vb{D}_{\alpha,P\, {\rm mag}} +
    e^{i\, c_{\alpha}\, \vb{k_{\rm ph}} \vb{r}}\,
      \vb{D}_{A,P\, {\rm mag}}
  \Bigr]\, \cdot
    \bigl[ \vb{K}_{i} \cp \vb{e}^{(\alpha)} \bigr]
  \biggr\} \cdot
  \Phi_{\rm \alpha - nucl, i} (\vb{r})\; \vb{dr},
\end{array}
\label{eq.2.9.b.1}
\end{equation}
where
\begin{equation}
\begin{array}{lll}
\vspace{1mm}
  D_{\alpha,P\, {\rm el}} = &
  \displaystyle\sum\limits_{i=1}^{4}
    z_{i}\,
    \Bigl\langle \psi_{\alpha, f} (\beta_{\alpha})\, \Bigl|\,
      e^{-i\, \vb{k_{\rm ph}} \rhobf_{\alpha i}}
    \Bigr|\,  \psi_{\alpha, i} (\beta_{\alpha}) \Bigr\rangle = F_{\alpha,\, {\rm el}}, \\

\vspace{1mm}
  D_{A,P\, {\rm el}} = &
  \displaystyle\sum\limits_{j=1}^{A}
    z_{j}\,
    \Bigl\langle \psi_{A, f} (\beta_{A})\, \Bigl|\,
      e^{-i\, \vb{k_{\rm ph}} \rhobf_{Aj}}
    \Bigr|\,  \psi_{A, j} (\beta_{A}) \Bigr\rangle = F_{A,\, {\rm el}}, \\

\vspace{1mm}
  \vb{D}_{\alpha,P\, {\rm mag}} = &
  \displaystyle\sum\limits_{i=1}^{4}
    \mu_{i}^{\rm (an)}\, m_{\alpha i}\,
    \Bigl\langle \psi_{\alpha, f} (\beta_{\alpha})\, \Bigl|\,
      e^{-i\, \vb{k_{\rm ph}} \rhobf_{\alpha i}}\, \sigmabf
    \Bigr|\,  \psi_{\alpha, i} (\beta_{\alpha}) \Bigr\rangle = \vb{F}_{\alpha,\, {\rm mag}}, \\

  \vb{D}_{A,P\, {\rm mag}} = &
  \displaystyle\sum\limits_{j=1}^{A}
    \mu_{j}^{\rm (an)}\, m_{Aj}\,
    \Bigl\langle \psi_{A, f} (\beta_{A})\, \Bigl|\,
      e^{-i\, \vb{k_{\rm ph}} \rhobf_{Aj}}\, \sigmabf
    \Bigr|\,  \psi_{A, j} (\beta_{A}) \Bigr\rangle = \vb{F}_{A,\, {\rm mag}}.
\end{array}
\label{eq.2.9.b.2}
\end{equation}

\subsection{Effective electric charge and effective magnetic moment of the full system
\label{sec.2.10}}

Following to logics in Ref.~\cite{Liu_Maydanyuk_Zhang_Liu.2019.PRC.hypernuclei},
we introduce \emph{effective electric charge} and
\emph{effective magnetic moment} of the full $\alpha$-nucleus system [see Eqs.~(30)--(35) in that paper] as
\begin{equation}
\begin{array}{lll}
\vspace{1mm}
  Z_{\rm eff} (\vb{k}_{\rm ph}, \vb{r}) =
  e^{i\, \vb{k_{\rm ph}} \vb{r}}\,
  \Bigl[
    e^{-i\, c_{A} \vb{k_{\rm ph}} \vb{r}}\, \displaystyle\frac{m_{A}}{m_{\alpha} + m_{A}}\, F_{\alpha,\, {\rm el}} -
    e^{i\, c_{\alpha} \vb{k_{\rm ph}} \vb{r}}\,  \displaystyle\frac{m_{\alpha}}{m_{\alpha} + m_{A}}\, F_{A,\, {\rm el}}
  \Bigr], \\

  \textbf{M}_{\rm eff} (\vb{k}_{\rm ph}, \vb{r}) =
  e^{i\, \vb{k_{\rm ph}} \vb{r}}\,
  \Bigl[
    e^{-i\, c_{A} \vb{k_{\rm ph}} \vb{r}}\, \displaystyle\frac{m_{A}}{m_{\alpha} + m_{A}}\, \textbf{F}_{\alpha,\, {\rm mag}} -
    e^{i\, c_{\alpha} \vb{k_{\rm ph}} \vb{r}}\,  \displaystyle\frac{m_{\alpha}}{m_{\alpha} + m_{A}}\, \textbf{F}_{A,\, {\rm mag}}
  \Bigr].
\end{array}
\label{eq.2.10.3}
\end{equation}
%
%
%
On such a basis, expression (\ref{eq.2.9.10}) for $M_{p}$ can be rewritten in a compact form as%
\footnote{The formula (\ref{eq.2.10.6}) is different from Eq.~(35) in Ref.~\cite{Liu_Maydanyuk_Zhang_Liu.2019.PRC.hypernuclei},
as includes additional momentum $\vb{K}_{i}$ in the initial state in $\delta$-function, that corresponds to momentum of the $\alpha$-particles in beam in the scattering process.}
\begin{equation}
\begin{array}{lll}
\vspace{-0.2mm}
  M_{p} & = &
  i \hbar\, (2\pi)^{3} \displaystyle\frac{1}{\mu}\: \delta (\vb{K}_{f} - \vb{K}_{i} - \vb{k}_{\rm ph}) \cdot
  \displaystyle\sum\limits_{\alpha=1,2}
  \displaystyle\int\limits_{}^{}
    \Phi_{\rm \alpha - nucl, f}^{*} (\vb{r})\;
    e^{-i\, \vb{k}_{\rm ph} \vb{r}}\; \times \\
  & \times &
  \biggl\{
  2\, \mu_{N}\,  m_{\rm p} \cdot
  Z_{\rm eff} (\vb{k}_{\rm ph}, \vb{r}) \cdot
  \vb{e}^{(\alpha)}\, \vb{\displaystyle\frac{d}{dr}} +

  i\,
  \vb{M}_{\rm eff} (\vb{k}_{\rm ph}, \vb{r}) \cdot
  \Bigl[ \vb{\displaystyle\frac{d}{dr}} \times \vb{e}^{(\alpha)} \Bigr]
  \biggr\} \cdot
  \Phi_{\rm \alpha - nucl, i} (\vb{r})\; \vb{dr},
\end{array}
\label{eq.2.10.6}
\end{equation}
where
\begin{equation}
  \mu = \displaystyle\frac{m_{\alpha}\, m_{A}}{m_{\alpha} + m_{A}}
\label{eq.2.10.2}
\end{equation}
is \emph{reduced mass of system of $\alpha$-particle and the nucleus}.

\section{Integration over momentum $K_{f}$
\label{sec.2.12}}

We will calculate cross-section of emission of photons, not dependent on vector $\vb{K}_{f}$ (i.e., momentum of the full nuclear system after emission of photon in laboratory frame). Therefore, we have to average matrix element over all degrees of freedom related with $\mathbf{K}_{f}$, i.e. we integrate these matrix elements over $\mathbf{K}_{f}$.
Using property:
\begin{equation}
  \displaystyle\int \delta (\mathbf{K}_{i} - \mathbf{K}_{f} - \mathbf{k})\; \mathbf{dK}_{f} = 1,
\label{eq.app.2.12.1}
\end{equation}
we integrate each matrix element as
\begin{equation}
  M_{s} \Rightarrow \displaystyle\int M_{s} (\vb{K}_{f})\; \vb{dK}_{f},
\label{eq.app.2.12.2}
\end{equation}
where $s$ is index indicating the matrix elements $M_{p}$, $M_{P}$, $M_{k}$, $M_{\Delta E}$ and $M_{\Delta M}$.
In particular, from (\ref{eq.2.10.6}), (\ref{eq.2.9.b.3}), (\ref{eq.2.9.c.1}), (\ref{eq.2.9.b.1}), (\ref{eq.2.9.b.2}) we obtain:
\begin{equation}
\begin{array}{lll}
\vspace{-0.2mm}
  M_{p} & = &
  i \hbar\, (2\pi)^{3} \displaystyle\frac{1}{\mu}\;
  \displaystyle\sum\limits_{\alpha=1,2}
  \displaystyle\int\limits_{}^{}
    \Phi_{\rm \alpha - nucl, f}^{*} (\vb{r})\; e^{-i\, \vb{k}_{\rm ph} \vb{r}}\; \times \\
  & \times &
  \biggl\{
  2\, \mu_{N}\,  m_{\rm p} \cdot Z_{\rm eff} (\vb{k}_{\rm ph}, \vb{r}) \cdot \vb{e}^{(\alpha)}\, \vb{\displaystyle\frac{d}{dr}} +

  i\,
  \vb{M}_{\rm eff} (\vb{k}_{\rm ph}, \vb{r}) \cdot \Bigl[ \vb{\displaystyle\frac{d}{dr}} \times \vb{e}^{(\alpha)} \Bigr]
  \biggr\} \cdot
  \Phi_{\rm \alpha - nucl, i} (\vb{r})\; \vb{dr}.
\end{array}
\label{eq.app.2.12.3}
\end{equation}
\begin{equation}
\begin{array}{lcl}
\vspace{-0.1mm}
  M_{P} & = &
  \displaystyle\frac{\hbar\, (2\pi)^{3}}{m_{A} + m_{\alpha}}\,
  \displaystyle\sum\limits_{\alpha=1,2}
  \displaystyle\int\limits_{}^{}
    \Phi_{\rm \alpha - nucl, f}^{*} (\vb{r})\:

  \biggl\{
    2\, \mu_{N}\, m_{\rm p}\;
    \Bigl[
      e^{-i\, c_{A}\, \vb{k_{\rm ph}} \vb{r}} F_{\alpha,\, {\rm el}} +
      e^{i\, c_{\alpha}\, \vb{k_{\rm ph}} \vb{r}} F_{A,\, {\rm el}}
    \Bigr]\, \vb{e}^{(\alpha)} \cdot \vb{K}_{i} + \\

  & + &
  i\,
  \Bigl[
    e^{-i\, c_{A}\, \vb{k_{\rm ph}} \vb{r}}\, \vb{F}_{\alpha,\, {\rm mag}} +
    e^{i\, c_{\alpha}\, \vb{k_{\rm ph}} \vb{r}}\, \vb{F}_{A,\, {\rm mag}}
  \Bigr] \cdot
    \bigl[ \vb{K}_{i} \cp \vb{e}^{(\alpha)} \bigr]
  \biggr\} \cdot
  \Phi_{\rm \alpha - nucl, i} (\vb{r})\; \vb{dr}.
\end{array}
\label{eq.app.2.12.4}
\end{equation}
\begin{equation}
\begin{array}{lcl}
  M_{k} & = &
  i\, \hbar\, (2\pi)^{3}\,
  \displaystyle\sum\limits_{\alpha=1,2}
    \bigl[ \vb{k_{\rm ph}} \cp \vb{e}^{(\alpha)} \bigr]
  \displaystyle\int\limits_{}^{}
    \Phi_{\rm \alpha - nucl, f}^{*} (\vb{r})\;
  \biggl\{
    e^{-i\, c_{A}\, \vb{k_{\rm ph}} \vb{r}}\, \vb{D}_{\alpha,\, {\rm k}} +
    e^{i\, c_{\alpha}\, \vb{k_{\rm ph}} \vb{r}}\, \vb{D}_{A,\, {\rm k}}
  \biggr\} \cdot
  \Phi_{\rm \alpha - nucl, i} (\vb{r})\; \vb{dr},
\end{array}
\label{eq.app.2.12.5}
\end{equation}
\begin{equation}
\begin{array}{lll}
\vspace{-0.1mm}
  M_{\Delta E} & = &
  -\, (2\pi)^{3}\; 2\, \mu_{N}
  \displaystyle\sum\limits_{\alpha=1,2} \vb{e}^{(\alpha)}
  \displaystyle\int\limits_{}^{}
    \Phi_{\rm \alpha - nucl, f}^{*} (\vb{r})\;
  \biggl\{
    \Bigl[
      e^{-i\, c_{A}\, \vb{k_{\rm ph}} \vb{r}}\, \vb{D}_{\alpha 1,\, {\rm el}} +
      e^{i\, c_{\alpha}\, \vb{k_{\rm ph}} \vb{r}}\, \vb{D}_{A 1,\, {\rm el}}
    \Bigr]\; - \\

  & - &
  \Bigl[
    \displaystyle\frac{m_{\rm p}}{m_{\alpha}}\, e^{-i\, c_{A}\, \vb{k_{\rm ph}} \vb{r}}\, \vb{D}_{\alpha 2,\, {\rm el}} +
    \displaystyle\frac{m_{\rm p}}{m_{A}}\, e^{i\, c_{\alpha}\, \vb{k_{\rm ph}} \vb{r}}\, \vb{D}_{A 2,\, {\rm el}}
  \Bigr]
  \biggr\} \cdot
  \Phi_{\rm \alpha - nucl, i} (\vb{r})\; \vb{dr},
\end{array}
\label{eq.app.2.12.6}
\end{equation}
\begin{equation}
\begin{array}{lll}
\vspace{-0.1mm}
  M_{\Delta M} & = &
  -\, i\, (2\pi)^{3}\,
  \displaystyle\sum\limits_{\alpha=1,2}
  \displaystyle\int\limits_{}^{}
    \Phi_{\rm \alpha - nucl, f}^{*} (\vb{r})\;
  \biggl\{
    \Bigl[
      e^{-i\, c_{A}\, \vb{k_{\rm ph}} \vb{r}}\; D_{\alpha 1,\, {\rm mag}} (\vb{e}^{(\alpha)}) +
      e^{i\, c_{\alpha}\, \vb{k_{\rm ph}} \vb{r}}\; D_{A 1,\, {\rm mag}} (\vb{e}^{(\alpha)})
    \Bigr]\; - \\
  & - &
  \Bigl[
    e^{-i\, c_{A}\, \vb{k_{\rm ph}} \vb{r}}\; D_{\alpha 2,\, {\rm mag}} (\vb{e}^{(\alpha)}) +
    e^{i\, c_{\alpha}\, \vb{k_{\rm ph}} \vb{r}}\; D_{A 2,\, {\rm mag}} (\vb{e}^{(\alpha)})
  \Bigr]
  \biggr\} \cdot
  \Phi_{\rm \alpha - nucl, i} (\vb{r})\; \vb{dr},
\end{array}
\label{eq.app.2.12.7}
\end{equation}
Also we have
%
\begin{equation}
  \mathbf{K}_{i} = \mathbf{K}_{f} + \mathbf{k}.
\label{eq.app.2.12.8}
\end{equation}



\section{Electric form factor of the nucleus
\label{sec.app.2}}

In this Appendix we describe many nucleon formalism of nucleus needed in calculations of form factors and matrix elements of emission.
We follow to formalism given in Appendix~A in Ref.~\cite{Maydanyuk_Zhang.2015.PRC}.
We define the space wave function of one nucleon in the gaussian form as [see Eqs.~(A6)--(A8) in that paper]:
\begin{equation}
  \varphi_{i} (\mathbf{r}) =
  N_{x}\,N_{y}\,N_{z}\,
  \exp[-\,\displaystyle\frac{1}{2}\,\Bigl(\displaystyle\frac{x^{2}}{a^{2}} + \displaystyle\frac{y^{2}}{b^{2}} + \displaystyle\frac{z^{2}}{c^{2}}\Bigr)] \cdot
  H_{n_{x}} \Bigl(\displaystyle\frac{x}{a} \Bigr)\,
  H_{n_{y}} \Bigl(\displaystyle\frac{y}{b} \Bigr)\,
  H_{n_{z}} \Bigl(\displaystyle\frac{z}{c} \Bigr),
\label{eq.app.2.2.1}
\end{equation}
where $H_{n_{x}}$, $H_{n_{y}}$ and $H_{n_{z}}$ are the Hermitian polynomials,
$a$, $b$ and $c$ are parameters of model of deformed oscillator shells (these parameters are not free, they are obtained from calculations of binding energy for the studied nucleus),
$N_{x}$, $N_{y}$, $N_{z}$ are the normalized coefficients in form:
\begin{equation}
\begin{array}{ccc}
  N_{x} = \displaystyle\frac{1}{\pi^{1/4} \sqrt{ a\, 2^{n_{x}}\, n_{x}! }}, &
  N_{y} = \displaystyle\frac{1}{\pi^{1/4} \sqrt{ b\, 2^{n_{y}}\, n_{y}! }}, &
  N_{z} = \displaystyle\frac{1}{\pi^{1/4} \sqrt{ c\, 2^{n_{z}}\, n_{z}! }}.
\end{array}
\label{eq.app.2.2.4}
\end{equation}
%
We obtain the form factor of nucleus as [see Eq.~(16)--(18) in Ref.~\cite{Maydanyuk_Zhang.2015.PRC}]
\begin{equation}
\begin{array}{lcl}
  Z_{\rm A} (\vb{k}_{\rm ph}) & = &
  2\, e^{-\, (a^{2} k_{x}^{2} + b^{2} k_{y}^{2} + c^{2} k_{z}^{2})\,/4}\; \cdot
  f_{1}\, (\mathbf{k}, n_{1} \ldots n_{\rm A}),
\end{array}
\label{eq.app.2.3.17}
\end{equation}
where
\begin{equation}
\begin{array}{lcl}
  f_{1}\, (\vb{k}_{\rm ph}, n_{1} \ldots n_{\rm A}) & = &
  \displaystyle\sum\limits_{n_{x}, n_{y},n_{z} = 0}^{n_{x} + n_{y} + n_{z} \le B}
    L_{n_{x}} \Bigl[a^{2} k_{x}^{2}/2\Bigr]\:
    L_{n_{y}} \Bigl[b^{2} k_{y}^{2}/2\Bigr]\:
    L_{n_{z}} \Bigl[c^{2} k_{z}^{2}/2\Bigr].
\end{array}
\label{eq.app.2.3.18}
\end{equation}
Here, function $f_{1}$ is summation over all states of one-nucleon space wave functions for protons of nucleus,
$L_{n} = L_{n}^{0}$ is Rodrigues polynomial,
$B$ is number of states of the space wave function of protons (in nucleus)%
\footnote{We have corrected previous formula (A17) in Ref.~\cite{Maydanyuk_Zhang.2015.PRC}.},
For isotopes of \isotope[4,5,6,7,8...]{He} we obtain $B = 1$.
If energy of photon tends to zero, electric form factor of nucleus tends to its electric charge:
\begin{equation}
\begin{array}{lllll}
  Z_{\rm A} (\vb{k}_{\rm ph}) \to Z_{\rm A}, &
  Z_{\alpha} (\vb{k}_{\rm ph}) \to Z_{\alpha}, &
  f_{1}\, (\vb{k}_{\rm ph}, n_{1} \ldots n_{\rm A}) \to B &
  \mbox{\rm при } \vb{k}_{\rm ph} \to 0.
\end{array}
\label{eq.app.2.3.19}
\end{equation}
%



\section{Matrix elements on moments of nucleons of nucleus. I
\label{sec.app.3}}

Let us find the first two matrix elements $\vb{D}_{\alpha 1,\, {\rm el}}$ and $\vb{D}_{A 1,\, {\rm el}}$ в (\ref{eq.2.9.b.3}), which include moments of nucleons of nuclear fragment:
%
\begin{equation}
\begin{array}{lll}
\vspace{1mm}
  \vb{D}_{\alpha 1,\, {\rm el}} (\vb{k}_{\rm ph}) = &
    \displaystyle\sum\limits_{i=1}^{3}
      \displaystyle\frac{z_{i} m_{\rm p}}{m_{\alpha i}}\,
      \Bigl\langle \psi_{\alpha, f} (\beta_{\alpha})\, \Bigl|\,
        e^{-i \vb{k}_{\rm ph} \rhobf_{\alpha i}} \vb{\tilde{p}}_{\alpha i}
      \Bigr|\,  \psi_{\alpha, i} (\beta_{\alpha}) \Bigr\rangle , \\

  \vb{D}_{A 1,\, {\rm el}} (\vb{k}_{\rm ph}) = &
    \displaystyle\sum\limits_{i=1}^{A-1}
      \displaystyle\frac{z_{j} m_{\rm p}}{m_{Aj}}\,
      \Bigl\langle \psi_{A, f} (\beta_{A})\, \Bigl|\,
        e^{-i \vb{k}_{\rm ph} \rhobf_{Aj}} \vb{\tilde{p}}_{Aj}
      \Bigr|\,  \psi_{Aj} (\beta_{A}) \Bigr\rangle , \\
\end{array}
\label{eq.app.3.1.1}
\end{equation}
We will calculate the second matrix element (as calculations for the first one are the similar).
Substituting to here many-nucleon wave function in form~(\ref{eq.2.6.1})
[where one-nucleon wave functions is $\psi_{\lambda_{s}} (s) = \varphi_{n_{s}} (\vb{r}_{s})\, \bigl|\, \sigma^{(s)} \tau^{(s)} \bigr\rangle$],
we obtain:
\begin{equation}
\begin{array}{lcl}
  \vb{D}_{A1,\, {\rm el}} (\vb{k}_{\rm ph}) = &
  \displaystyle\frac{1}{A}
  \displaystyle\sum\limits_{i=1}^{A-1}
  \displaystyle\sum\limits_{k=1}^{A}\,
    \Bigl\langle \psi_{k}(i)\, \Bigl|\,
      \displaystyle\frac{z_{k} m_{\rm p}}{m_{Ak}}\,
      e^{-i \mathbf{k}_{\rm ph} \rhobfsm_{Ai}}\,
      \vb{\tilde{p}}_{A i}
    \Bigl|\, \psi_{k}(i)\, \Bigr\rangle.
\end{array}
\label{eq.app.3.1.2}
\end{equation}

\subsubsection{Summation over spin-isospin states
\label{sec.app.3.1.2}}

Taking into account zero charge of neutron, we sum Eq.~(\ref{eq.app.3.1.2}) over spin-isospin states.
For nuclei with even number of protons (and arbitrary number of neutrons!), we obtain:
\begin{equation}
\begin{array}{lcl}
  \vb{D}_{A1,\, {\rm el}} (\vb{k}_{\rm ph}) = &
  \displaystyle\frac{1}{A}
  \displaystyle\sum\limits_{i=1}^{A-1}
  \displaystyle\sum\limits_{k=1}^{B}\,
    \displaystyle\frac{z_{k} m_{\rm p}}{m_{Ak}}\,
    \Bigl\langle \varphi_{k}(\rhobf_{i})\, \Bigl|\,
      e^{-i \mathbf{k}_{\rm ph} \rhobfsm_{Ai}}\,
      \vb{\tilde{p}}_{A i}
    \Bigl|\, \varphi_{k}(\rhobf_{i})\, \Bigr\rangle =

  \displaystyle\frac{2}{A}
  \displaystyle\sum\limits_{i=1}^{A-1}
  \displaystyle\sum\limits_{k=1}^{B}\,
    z_{k}\,
    \Bigl\langle \varphi_{k}(\rhobf_{i})\, \Bigl|\,
      e^{-i \mathbf{k}_{\rm ph} \rhobfsm_{Ai}}\,
      \vb{\tilde{p}}_{A i}
    \Bigl|\, \varphi_{k}(\rhobf_{i})\, \Bigr\rangle,
\end{array}
\label{eq.app.3.1.2.1}
\end{equation}
where $B$ is number of states of the space wave function of protons (in nucleus), and $z_{k}=1$ for protons.

\subsubsection{Calculation of one-nucleon matrix element
\label{sec.app.3.1.3}}

We substitute the one-nucleon space wave function (\ref{eq.app.2.2.1}) into matrix element:
\begin{equation}
\begin{array}{lll}
  & \vb{D}_{A1,\, {\rm el}} (\vb{k}_{\rm ph}) =


  - i\, \hbar \cdot
  \displaystyle\frac{2}{A}\;
  \displaystyle\sum\limits_{i=1}^{A-1}
  \displaystyle\sum\limits_{n_{x}, n_{y}, n_{z}}^{B}
  N_{x}^{2}\, N_{y}^{2}\, N_{z}^{2} \cdot
  \displaystyle\int
    \exp[-\,\displaystyle\frac{(x_{i})^{2}}{2a^{2}} - \displaystyle\frac{(y_{i})^{2}}{2b^{2}} - \displaystyle\frac{(z_{i})^{2}}{2c^{2}} ] \cdot
    H_{n_{x}} \Bigl(\displaystyle\frac{x_{i}}{a} \Bigr)\,
    H_{n_{y}} \Bigl(\displaystyle\frac{y_{i}}{b} \Bigr)\,
    H_{n_{z}} \Bigl(\displaystyle\frac{z_{i}}{c} \Bigr)\; \times \\

  \times &
    e^{-i \mathbf{k}_{\rm ph} \rhobfsm_{i}}\;
    \Bigl( \vb{e}_{x}\, \displaystyle\frac{d}{d x_{i}} + \vb{e}_{y}\, \displaystyle\frac{d}{dy_{i}} + \vb{e}_{z}\, \displaystyle\frac{d}{dz_{i}} \Bigr)\;

    \exp[-\,\displaystyle\frac{(x_{i})^{2}}{2a^{2}} - \displaystyle\frac{(y_{i})^{2}}{2b^{2}} - \displaystyle\frac{(z_{i})^{2}}{2c^{2}}]\;
    H_{n_{x}} \Bigl(\displaystyle\frac{x_{i}}{a} \Bigr)\,
    H_{n_{y}} \Bigl(\displaystyle\frac{y_{i}}{b} \Bigr)\,
    H_{n_{z}} \Bigl(\displaystyle\frac{z_{i}}{c} \Bigr)\;
    dx_{i}\,dy_{i}\,dz_{i}\; = \\

  = &
  \displaystyle\frac{2}{A}\;
  \displaystyle\sum\limits_{i=1}^{A-1}
  \displaystyle\sum\limits_{n_{x}, n_{y}, n_{z}}^{B}
    \Bigl( \vb{e}_{x} J_{x}(n_{x}) + \vb{e}_{y} J_{y}(n_{y}) + \vb{e}_{z} J_{z}(n_{z}) \Bigr).
\end{array}
\label{eq.app.3.1.3.1}
\end{equation}
We have separated formula on different coordinate components $J_{x}(n_{x})$, $J_{y}(n_{y})$, $J_{z}(n_{z})$.
Let us analyze one integral:
%
\begin{equation}
\begin{array}{lcl}
\vspace{1mm}
  J_{x}(n_{x}) & = &
  - i\, \hbar\; N_{x}^{2}\;
  \displaystyle\int
    \exp[-\,\displaystyle\frac{(x_{i})^{2}}{2a^{2}}] \cdot H_{n_{x}} \Bigl(\displaystyle\frac{x_{i}}{a} \Bigr)\;
    e^{-i k_{x} x_{i}}\;
    \Bigl( \mathbf{e}_{x}\, \displaystyle\frac{d}{d x_{i}} \Bigr)\:
    \Bigl\{ \exp[ -\,\displaystyle\frac{(x_{i})^{2}}{2a^{2}}]\; H_{n_{x}} \Bigl(\displaystyle\frac{x_{i}}{a} \Bigr) \Bigr\}\;
  dx_{i}\; \times \\

& \times &
  N_{y}^{2}\, \displaystyle\int \exp[-\,\displaystyle\frac{(y_{i})^{2}}{b^{2}}]\; H_{n_{y}}^{2} \Bigl(\displaystyle\frac{y_{i}}{b} \Bigr)\; e^{-i k_{y} y_{i}}\; dy_{i} \times
  N_{z}^{2}\, \displaystyle\int \exp[-\,\displaystyle\frac{(z_{i})^{2}}{c^{2}}]\; H_{n_{z}}^{2} \Bigl(\displaystyle\frac{z_{i}}{c} \Bigr)\; e^{-i k_{z} z_{i}}\; dz_{i}.
\end{array}
\label{eq.app.3.1.3.2}
\end{equation}
One can see that last two integrals (in last line) represent functions $I_{y} (n_{y}, b)$ and $I_{z} (n_{z}, c)$ in
Eq.~(A10), (A15) in Appendix~A in Ref.~\cite{Maydanyuk_Zhang.2015.PRC}:
\begin{equation}
\begin{array}{lcl}
  I_{y}
  & = &
  N_{y}^{2} \cdot
  \displaystyle\int
    \exp[-\,\displaystyle\frac{(y_{i})^{2}}{b^{2}} ] \cdot \exp{-i\, k_{y} y_{i}}\, H_{n_{y}}^{2} \Bigl(\displaystyle\frac{y_{i}}{b} \Bigr)\; dy_{i} =

  L_{n_{y}} \Bigl[b^{2} k_{y}^{2}/2\Bigr] \cdot \exp[-\, b^{2} k_{y}^{2}/4].
\end{array}
\label{eq.app.2.3.3}
\end{equation}
%
We integrate in Eq.~(\ref{eq.app.2.3.3}) over variable $x$ by parts:
\begin{equation}
\begin{array}{lcl}
\vspace{0.8mm}
  J_{x}(n_{x}) & = &
  \mathbf{e}_{x}\;
  i\, \hbar\;
  N_{x}^{2}
  \displaystyle\int
    \displaystyle\frac{d}{d x_{i}}\,
    \Bigl\{ \exp[-\,\displaystyle\frac{(x_{i})^{2}}{2a^{2}}]\, H_{n_{x}} \Bigl(\displaystyle\frac{x_{i}}{a} \Bigr) \Bigr\}\;
    e^{-i\, k_{\rm x} x_{i}}\;
    \exp[ -\,\displaystyle\frac{(x_{i})^{2}}{2a^{2}} ]\: H_{n_{x}} \Bigl(\displaystyle\frac{x_{i}}{a} \Bigr)\; dx_{i}\;
    \cdot I_{y} (n_{y}, b) \cdot I_{z} (n_{z}, c)\; + \\

& + &
  \mathbf{e}_{x}\;
  (-i\, k_{\rm x})\:
  i\, \hbar\;
  N_{x}^{2}
  \displaystyle\int
    \exp[-\,\displaystyle\frac{(x_{i})^{2}}{a^{2}} ]\, e^{-i\, k_{\rm x} x_{i}}\,
    H_{n_{x}}^{2} \Bigl(\displaystyle\frac{x_{i}}{a} \Bigr)\; dx_{i}
    \cdot I_{y} (n_{y}, b) \cdot I_{z} (n_{z}, c).
\end{array}
\label{eq.app.3.1.3.3}
\end{equation}
One can see that integral over $x$ in the first term in the obtained solution is expressed via definition for $J_{x}(n_{x})$,
and integral over $x$ in the second term --- via $I_{x} (n_{x}, a)$:
%
\begin{equation}
\begin{array}{lcl}
  J_{x}(n_{x}) & = &
  -\; J_{x}(n_{x})\; + \;
  \mathbf{e}_{x}\: \hbar\, k_{\rm x} \cdot
  I_{x} (n_{x}, a) \cdot I_{y} (n_{y}, b) \cdot I_{z} (n_{z}, c).
\end{array}
\label{eq.app.3.1.3.4}
\end{equation}
Transfering integral $J_{x}(n_{x})$ from the right part to left one, we obtain:
%
\begin{equation}
\begin{array}{lcl}
  J_{x}(n_{x}) & = &
  \vb{e}_{x}\: \displaystyle\frac{\hbar\, k_{\rm x}}{2} \cdot I_{x} (n_{x}, a)\; I_{y}\; (n_{y}, b)\; I_{z} (n_{z}, c).
\end{array}
\label{eq.app.3.1.3.5}
\end{equation}
We substitute this solution to Eq.~(\ref{eq.app.3.1.3.1}) and calculate function $\vb{D}_{A1,\, {\rm el}} (\vb{k}_{\rm ph})$:
\begin{equation}
\begin{array}{lll}
  \vb{D}_{A1,\, {\rm el}} (\vb{k}_{\rm ph}) & = &

  \displaystyle\frac{2}{A}\;
  \displaystyle\sum\limits_{i=1}^{A-1}
  \displaystyle\sum\limits_{n_{x}, n_{y}, n_{z}}^{B}
    \Bigl(
      \vb{e}_{x} \displaystyle\frac{\hbar\, k_{\rm x}}{2}\: +\:
      \vb{e}_{y} \displaystyle\frac{\hbar\, k_{\rm y}}{2}\: +\:
      \vb{e}_{z} \displaystyle\frac{\hbar\, k_{\rm z}}{2}
    \Bigr) \cdot
    I_{x} (n_{x}, a)\; I_{y}\; (n_{y}, b)\; I_{z} (n_{z}, c)\; = \\

  & = &
  \displaystyle\frac{\hbar}{A}\;
  \displaystyle\sum\limits_{i=1}^{A-1}
  \displaystyle\sum\limits_{n_{x}, n_{y}, n_{z}}^{B}
    \Bigl( \vb{e}_{x}\, k_{\rm x} + \vb{e}_{y}\, k_{\rm y} + \vb{e}_{z}\, k_{\rm z} \Bigr) \cdot
    I_{x} (n_{x}, a)\; I_{y}\; (n_{y}, b)\; I_{z} (n_{z}, c)\; = \\

  & = &
  \vb{k}_{\rm ph}\;
  \displaystyle\frac{\hbar}{A}\;
  \displaystyle\sum\limits_{i=1}^{A-1}
  \displaystyle\sum\limits_{n_{x}, n_{y}, n_{z}}^{B}
    I_{x} (n_{x}, a)\; I_{y}\; (n_{y}, b)\; I_{z} (n_{z}, c).
\end{array}
\label{eq.app.3.1.3.6}
\end{equation}
Taking into account representation~(A9)--(A10) in Appendix~A in Ref.~\cite{Maydanyuk_Zhang.2015.PRC} for form factors,
one can rewrite the found solution (\ref{eq.app.3.1.3.6}) via these form factors as
\begin{equation}
\begin{array}{lll}
  \vb{D}_{\alpha 1,\, {\rm el}} = \displaystyle\frac{3\, \hbar}{8}\; \vb{k}_{\rm ph}\; Z_{\rm \alpha} (\vb{k}_{\rm ph}), &
  \vb{D}_{A 1,\, {\rm el}} = \displaystyle\frac{\hbar}{2}\; \displaystyle\frac{A-1}{A}\; \vb{k}_{\rm ph}\; Z_{\rm A} (\vb{k}_{\rm ph}).
\end{array}
\label{eq.app.3.1.3.7}
\end{equation}

We have properties:

\vspace{0.5mm}
\noindent
1) For such solutions and taking into account $\vb{e}^{(\alpha)} \cdot \vb{k}_{\rm ph} = 0$ (as ortogonality of vectors $\vb{e}^{(\alpha)}$ and $\vb{k}_{\rm ph}$), we obtain:
\begin{equation}
\begin{array}{llllllll}
  \vb{e}^{(\alpha)} \cdot \vb{D}_{\alpha 1,\, {\rm el}} = 0, &
  \vb{e}^{(\alpha)} \cdot \vb{D}_{A 1,\, {\rm el}} = 0.
\end{array}
\label{eq.app.3.1.3.8}
\end{equation}
2) At tending energy of photon to zero $\vb{D}_{A 1,\, {\rm el}}$ tends to zero (in contrast to $Z_{\rm A} (\vb{k}_{\rm ph}=0)$):
%
\begin{equation}
\begin{array}{llll}
  \vb{D}_{\alpha 1,\, {\rm el}} \to 0, &
  \vb{D}_{A 1,\, {\rm el}} \to 0 &
  \mbox{\rm at } k \to 0.
\end{array}
\label{eq.app.3.1.3.9}
\end{equation}

\section{Matrix elements on moments of nucleons of nucleus. II
\label{sec.app.3.2}}

Finally, we will find the second two matrix elements $\vb{D}_{\alpha 2,\, {\rm el}}$ and $\vb{D}_{A 2,\, {\rm el}}$ в (\ref{eq.2.9.b.3}), which include momenta of nucleons of the $\alpha$-particle and nucleus:
%
\begin{equation}
\begin{array}{lll}
\vspace{1mm}
  \vb{D}_{\alpha 2,\, {\rm el}} = &
    \displaystyle\sum\limits_{i=1}^{4}
      z_{i}\,
      \Bigl\langle \psi_{\alpha, f} (\beta_{\alpha})\, \Bigl|\,
        e^{-i \vb{k}_{\rm ph} \rhobf_{\alpha i}}\,
        \Bigl( \displaystyle\sum_{k=1}^{n-1} \vb{\tilde{p}}_{\alpha k} \Bigr)
      \Bigr|\,  \psi_{\alpha, i} (\beta_{\alpha}) \Bigr\rangle, \\

  \vb{D}_{A 2,\, {\rm el}} = &
    \displaystyle\sum\limits_{i=1}^{A}
      z_{j}\,
      \Bigl\langle \psi_{A, f} (\beta_{A})\, \Bigl|\,
        e^{-i \vb{k}_{\rm ph} \rhobf_{Aj}}
        \Bigl( \displaystyle\sum_{k=1}^{A-1} \vb{\tilde{p}}_{Ak} \Bigr)
      \Bigr|\,  \psi_{Aj} (\beta_{A}) \Bigr\rangle,
\end{array}
\label{eq.app.3.2.1}
\end{equation}
We will calculate the second matrix element (as calculations for the first one are the similar).
Substituting to here many-nucleon wave function from Eq.~(\ref{eq.2.6.1}), we obtain:
\begin{equation}
\begin{array}{lll}
\vspace{1mm}
  &
  \displaystyle\sum\limits_{j=1}^{A}
    z_{Aj}\; e^{-i \mathbf{k}\, \rhobfsm_{Aj}}\;
    \displaystyle\sum_{k=1}^{A-1} \mathbf{\tilde{p}}_{Ak} =

  \Bigl\{
    \displaystyle\sum\limits_{j=1}^{A-1}
      z_{Aj}\; e^{-i \mathbf{k}\, \rhobfsm_{Aj}} +
    z_{AA}\; e^{-i \mathbf{k}\, \rhobfsm_{AA}}\;
  \Bigr\}\,
    \displaystyle\sum_{k=1}^{A-1} \mathbf{\tilde{p}}_{Ak}\; = \\

  = &
  \Bigl\{
    \displaystyle\sum\limits_{j=1}^{A-1}
      z_{Aj}\; e^{-i \mathbf{k}\, \rhobfsm_{Aj}} +
      z_{AA}\; e^{-i \mathbf{k}\,
        \bigl[-\, \frac{1}{m_{A}} \sum_{k=1}^{A-1} m_{k}\, \rhobfsm_{A k} \bigr]}\;
  \Bigr\}\,
    \displaystyle\sum_{k=1}^{A-1} \mathbf{\tilde{p}}_{Ak}.
\end{array}
\label{eq.app.3.2.2}
\end{equation}
Let us use the following property of summation:
%
\begin{equation}
\begin{array}{lcl}
  & & \vb{D}_{A 2,\, {\rm el}} (\vb{k}_{\rm ph}) =
  \displaystyle\frac{1}{A\,(A-1)}
  \displaystyle\sum\limits_{i=1}^{A}
  \displaystyle\sum\limits_{k=1}^{A}
  \displaystyle\sum\limits_{m=1, m \ne k}^{A}
  \biggl\{
    \Bigl\langle \psi_{k}(i)\,\psi_{m}(j)\, \Bigl|\,
      z_{i}\; e^{-i \vb{k}_{\rm ph} \rhobf_{i}}
      \Bigl( \displaystyle\sum_{j=1}^{A-1} \vb{\tilde{p}}_{j} \Bigr)
    \Bigl|\, \psi_{k}(i)\,\psi_{m}(j)\, \Bigr\rangle\; - \\
  & - &
    \Bigl\langle \psi_{k}(i)\,\psi_{m}(j)\, \Bigl|\,
      z_{i}\; e^{-i \vb{k}_{\rm ph} \rhobf_{i}}
      \Bigl( \displaystyle\sum_{j=1}^{A-1} \vb{\tilde{p}}_{j} \Bigr)
    \Bigl|\, \psi_{m}(i)\,\psi_{k}(j)\, \Bigr\rangle
  \biggr\}.
\end{array}
\label{eq.app.3.2.3}
\end{equation}
One can see that these matrix elements are proportional to wave vector of photon:
\begin{equation}
\begin{array}{lll}
  \vb{D}_{\alpha 2,\, {\rm el}} \sim \vb{k}_{\rm ph}, &
  \vb{D}_{A 2,\, {\rm el}} \sim \vb{k}_{\rm ph}.
\end{array}
\label{eq.app.3.2.4}
\end{equation}
%
%
For such solutions and taking into account $\vb{e}^{(\alpha)} \cdot \vb{k}_{\rm ph} = 0$ (as ortogonality of vectors $\vb{e}^{(\alpha)}$ and $\vb{k}_{\rm ph}$), we obtain:
\begin{equation}
\begin{array}{llllllll}
  \vb{e}^{(\alpha)} \cdot \vb{D}_{\alpha 2,\, {\rm el}} = 0, &
  \vb{e}^{(\alpha)} \cdot \vb{D}_{A 2,\, {\rm el}} = 0.
\end{array}
\label{eq.app.3.2.5}
\end{equation}




\begin{thebibliography}{99}
\bibitem{Maydanyuk_Zhang_Zou.2019.PRC.microscopy}
S.~P.~Maydanyuk, P.-M.~Zhang, and L.-P.~Zou,
\newblock
  \emph{Nucleon microscopy in proton-nucleus scattering via analysis of bremsstrahlung emission},
\newblock
  Phys. Rev. \textbf{C 99}, 064602 (2019);
\newblock
  arXiv:1812.07180.

\bibitem{Liu_Maydanyuk_Zhang_Liu.2019.PRC.hypernuclei}
X.~Liu, S.~P.~Maydanyuk, P.-M.~Zhang, and L.~Liu,
\newblock
  \emph{First investigation of hypernuclei in reactions via analysis of emitted bremsstrahlung photons},
\newblock
  Phys. Rev. \textbf{C 99}, 064614 (2019);
\newblock
  arXiv:1810.11942.
\bibitem{Glassel.1989.NPA}
  P.~Gl\"{a}ssel, R.~Schmidt-Fabian, D.~Schwalm, D.~Habs, and H.~V.~Helmdt,
\newblock
  Nucl. Phys. \textbf{A502}, 315c (1989).

\bibitem{Luke.1991.PRC}
  S.~J.~Luke, C.~A.~Gossett, and R.~Vandenbosch,
\newblock
  \emph{Search for high energy $\gamma$ rays from spontaneos fission of \isotope[252]{Cf}},
\newblock
  Phys. Rev. \textbf{C44} (4), 1548 (1991).

\bibitem{Ploeg.1992.PRL}
  H.~van~der Ploeg, R.~Postma, J.~C.~Bacelar, T.~van den Berg, V.~E.~Iacob, J.~R.~Jongman, and A.~van~der~Woude,
\newblock
  \emph{Large gamma anisotropy observed in the \isotope[252]{Cf} spontaneous-fission process},
\newblock
  Phys.~Rev.~Lett. \textbf{68} (21), 3145 (1992).

\bibitem{Hofman.1993.PRC}
  D.~J.~Hofman, B.~B.~Back, C.~P.~Montoya, S.~Schadmand, R.~Varma, and P.~Paul,
\newblock
  \emph{High energy $\gamma$ rays from \isotope[252]{Cm} spontaneos fission},
\newblock
  Phys.~Rev. \textbf{C47} (3), 1103 (1993).

\bibitem{Ploeg.1995.PRC}
  H.~van~der Ploeg, J.~C.~S.~Bacelar, A.~Buda, C.~R.~Laurens, and A.~van~der~Woude,
\newblock
  \emph{Emission of photons in spontaneous fission of \isotope[252]{Cf}},
\newblock
  Phys.~Rev. \textbf{C52} (4), 1915 (1995).

\bibitem{Pandit.2009.DAESNP}
  D.~Pandit, S.~Mukhopadhyay, S.~Pal, S.~Bhattacharya, A.~De, T.~K.~Rana, K.~Banerjee, J.~K.~Meena, and S.~R.~Banerjee,
\newblock
  \emph{Nucleus-nucleus coherent bremsstrahlung in \isotope[252]{Cf} spontaneous fission},
\newblock
  DAE Symp. Nucl.Phys.  \textbf{54}, 360--361 (2009).

\bibitem{Pandit.2010.PLB}
  D.~Pandit, S.~Mukhopadhyay, S.~Bhattacharya, S.~Pal, A.~De, and S.~R.~Banerjee,
\newblock
  \emph{Coherent bremsstrahlung and GDR width from \isotope[252]{Cf} cold fission},
\newblock
  Phys.~Lett. \textbf{B 690}, 473--476 (2010);
\newblock
  arXiv:0912.4578.





\bibitem{Kasagi.1989.JPSJ}
  J.~Kasagi, H.~Hama, K.~Yoschida et al.,
\newblock
  Journ. Phys. Soc. Jpn. \textbf{58}, 620 (1989).

\bibitem{Varlachev.2007.BRASP}
  V.~A.~Varlachev, G.~N.~Dudkin, V.~N.~Padalko,
\newblock
  Bull. Russ. Acad. Sci.: Phys. \textbf{71} (11), 1635--1639 (2007).

\bibitem{Eremin.2010.IJMPE}
  N.~V.~Eremin, et al.,
\newblock
  Int. J. Mod. Phys. E \textbf{19}, 1183 (2010).






\bibitem{Edington.1966.NP}
  J.~Edington, and B.~Rose,
\newblock
  Nucl. Phys. \textbf{89}, 523 (1966).

\bibitem{Koehler.1967.PRL}
  P.~F.~M.~Koehler, K.~W.~Rothe, and E.~H.~Thorndike,
\newblock
  Phys. Rev. Lett. \textbf{18}, 933 (1967).

\bibitem{Kwato_Njock.1988.PLB}
  M.~Kwato~Njock, M.~Maurel, H.~Nifenecker, J.~Pinston, F.~Schussler, D.~Barneoud, S.~Drissi,
  J.~Kern, and J.~P.~Vorlet,
\newblock
  Phys. Lett. \textbf{B207}, 269 (1988).

\bibitem{Pinston.1989.PLB}
  J.~A.~Pinston, D.~Barneoud, V.~Bellini, S.~Drissi, J.~Guillot, J.~Julien, M.~Kwato~Njock,
  H.~Nifenecker, M.~Maurel, F.~Schussler, and J.~P.~Vorlet,
\newblock
  Phys. Lett. \textbf{B 218}, 128 (1989).

\bibitem{Pinston.1990.PLB}
  J.~A.~Pinston, D.~Barneoud, V.~Bellini, S.~Drissi, J.~Guillot, J.~Julien,
  H.~Nifenecker, and F.~Schussler,
\newblock
  Phys. Lett. \textbf{B 249}, 402 (1990).

\bibitem{Clayton.1992.PRC}
  J.~Clayton, W.~Benenson, M.~Cronqvist, R.~Fox, D.~Krofcheck, R.~Pfaff,
  M.~F.~Mohar,
  C.~Bloch, D.~E.~Fields,
\newblock
  Phys. Rev. \textbf{C45}, 1815 (1992).

\bibitem{Pluiko.1987.PEPAN}
  V.~A.~Pluyko and V.~A.~Poyarkov,
\newblock
  \emph{Bremsstrahlung in reactions induced by protons},
\newblock
  Phys. Elem. Part. At. Nucl. \textbf{18} (2), 374--418 (1987).

\bibitem{Kamanin.1989.PEPAN}
  V.~V.~Kamanin, A.~Kugler, Yu.~E.~Penionzhkevich, I.~S.~Batkin, and I.~V.~Kopytin,
\newblock
  \emph{High-energy gamma-ray emission in heavy-ion reactions at nonrelativistic energies},
\newblock
  Phys. Elem. Part. At. Nucl. \textbf{20} (4), 741--829 (1989).

\bibitem{Clayton.1991.PhD}
  J.~E.~Clayton,
\newblock
  \emph{High energy gamma ray production in proton induced reactions
  at energies of 104, 145, and 195 MeV},
\newblock
  PhD thesis (Michigan State University, 1991).

\bibitem{Chakrabarty.1999.PRC}
  D.~R.~Chakrabarty, V.~M.~Datar, Y.~K.~Agarwal, C.~V.~K.~Baba, M.~S.~Samant,
  I.~Mazumdar, A.~K.~Sinha, and P.~Sugathan,
\newblock
  Phys. Rev. \textbf{C60}, 024606 (1999).

\bibitem{Goethem.2002.PRL}
  M.~J.~van~Goethem, L.~Aphecetche, J.~C.~S.~Bacelar, H.~Delagrange, J.~Diaz, D.~d'Enterria, M.~Hoefman,
  R.~Holzmann, H.~Huisman, N.~Kalantar-Nayestanaki, A.~Kugler, H.~L\"{o}hner, G.~Martinez,
  J.~G.~Messchendorp, R.~W.~Ostendorf, S.~Schadmand, R.~H.~Siemssen, R.~S.~Simon,
  Y.~Schutz, R.~Turrisi, M.~Volkerts, V.~Wagner, and H.~W.~Wilschut,
\newblock
  \emph{Suppresion of soft nuclear bremsstrahlung in proton-nucleus collisions},
\newblock
  Phys. Rev. Lett. \textbf{88} (12), 122302 (2002).

\bibitem{Nakayama.1986.PRC}
  K.~Nakayama, G.~Bertsch,
\newblock
  \emph{High energy photon production in nuclear collisions},
\newblock
  Phys. Rev. \textbf{C34} (6), 2190 (1986).

\bibitem{Nakayama.1989.PRC}
  K.~Nakayama,
\newblock
  \emph{High-energy photons in neutron-proton and proton-nucleus collisions},
\newblock
  Phys. Rev. \textbf{C39} (4), 1475--1487 (1989).

\bibitem{Nakayama.1989.PRCv40}
  K.~Nakayama, G.~F.~Bertsch,
\newblock
  Phys. Rev. \textbf{C40}, 685 (1989).

\bibitem{Knoll.1989.NPA}
  J.~Knoll, C.~Guet,
\newblock
  Nucl. Phys. \textbf{A494}, 334--348 (1989).

\bibitem{Herrmann.1991.PRC}
  V.~Herrmann, J.~Speth, K.~Nakayama,
\newblock
  \emph{Nucleon-nucleon bremsstrahlung at intermediate energies},
\newblock
  Phys. Rev. \textbf{C43} (2), 394--415 (1991).

\bibitem{Liou.1987.PRC}
  M.~K.~Liou, and Z.~M.~Ding,
\newblock
  Phys. Rev. \textbf{C35} (2), 651 (1987).

\bibitem{Liou.1993.PRC}
  M.~K.~Liou, D.~Lin, and B.~F.~Gibson,
\newblock
  Phys. Rev. \textbf{C47}, 973 (1993).

\bibitem{Liou.1995.PLB.v345}
  M.~K.~Liou, R.~Timmermans, and B.~F.~Gibson,
\newblock
  Phys. Lett. \textbf{B345}, 372 (1995).

\bibitem{Liou.1995.PLB.v355}
  M.~K.~Liou, R.~Timmermans, and B.~F.~Gibson,
\newblock
  Phys. Lett. \textbf{B355}, 606(E) (1995).

\bibitem{Liou.1996.PRC}
  M.~K.~Liou, R.~Timmermans, and B.~F.~Gibson,
\newblock
  Phys. Rev. \textbf{C54} (4), 1574--1584 (1996).

\bibitem{Li.1998.PRC.v57}
  Yi~Li, M.~K.~Liou, and W.~M.~Schreiber,
\newblock
  Phys. Rev. \textbf{C57}, 507 (1998).

\bibitem{Li.1998.PRC.v58}
  Yi~Li, M.~K.~Liou, R.~Timmermans, and B.~F.~Gibson,
\newblock
  Phys. Rev. \textbf{C58}, R1880 (1998).

\bibitem{Timmermans.2001.PRC}
  R.~G.~E.~Timmermans, B.~F.~Gibson, Yi~Li, and M.~K.~Liou,
\newblock
  Phys. Rev. \textbf{C65}, 014001 (2001) [15~p.].

\bibitem{Liou.2004.PRC}
  M.~K.~Liou, T.~D.~Penninga, R.~G.~E.~Timmermans, and B.~F.~Gibson,
\newblock
  Phys. Rev. \textbf{C69}, 011001 (2004).

\bibitem{Li.2005.PRC}
  Y.~Li, M.~K.~Liou, and W.~M.~Schreiber,
\newblock
  Phys. Rev. \textbf{C72}, 024005 (2005).

\bibitem{Timmermans.2006.PRC}
  R.~G.~E.~Timmermans, T.~D.~Penninga, B.~F.~Gibson, and M.~K.~Liou,
\newblock
  Phys. Rev. \textbf{C73}, 034006 (2006).

\bibitem{Li.2011.PRC}
  Yi~Li, M.~K.~Liou, W.~M.~Schreiber, and B.~F.~Gibson,
\newblock
  Phys. Rev. \textbf{C84}, 034007 (2011) [10~p.].

\bibitem{Kurgalin.2001.IRAN}
  S.~D.~Kurgalin, Yu.~M.~Chuvilskiy, and T.~A.~Churakova,
\newblock
  Izv. Acad. Nauk: Ser. Fiz. \textbf{65}, 672 (2001) [in Russian].

\bibitem{Buck.1993.ADNDT}
  B.~Buck, A.~C.~Merchant, and S.~M.~Perez,
\newblock
  \emph{Half-lives of favored alpha decays from nuclear ground states},
\newblock
  Atom. Data Nucl. Data Tables \textbf{54} (1), 53--74 (1993).

\bibitem{Akovali.1998.NDS}
  Y.~A.~Akovali,
\newblock
  \emph{Review of alpha-decay data from doubly-even nuclei},
\newblock
  Nucl. Data Sheets \textbf{84} (1), 1--114 (1998).

\bibitem{Duarte.2002.ADNDT}
  S.~B.~Duarte, O.~A.~P.~Tavares, F.~Guzman, A.~Dimarco, et al.,
\newblock
  \emph{Half-lives of favored alpha decays from nuclear ground states},
\newblock
  Atom. Data Nucl. Data Tables \textbf{80} (2), 235--299 (2002).

\bibitem{Audi.2003.NPA}
  G.~Audi, O.~Bersillon, J.~Blachot, and A.~H.~Wapstra,
\newblock
  \emph{The NUBASE evaluation of nuclear and decay properties},
\newblock
  Nucl. Phys. \textbf{A 729} (1), 3--128 (2003).

\bibitem{Dasgupta-Schubert.2007.ADNDT}
  N.~Dasgupta-Schubert and M.~A.~Reyes,
\newblock
  \emph{The generalized liquid drop model alpha-decay formula: Predictability analysis and superheavy element alpha half-lives},
\newblock
  Atom. Data Nucl. Data Tables \textbf{93} (6), 907--930 (2007).

\bibitem{Silisteanu.2012.ADNDT}
  I.~Silisteanu and A.~I.~Budaca,
\newblock
  \emph{Structure and alpha-decay properties of the heaviest nuclei},
\newblock
  Atom. Data Nucl. Data Tables \textbf{98} (6), 1096--1108 (2012).

\bibitem{Lovas.1998.PRep}
  R.~G.~Lovas, R.~J.~Liotta, A.~Insolia, K.~Varga, and D.~S.~Delion,
\newblock
  \emph{Microscopic theory of cluster radioactivity},
\newblock
  Phys. Rep. \textbf{294} (5), 265--362 (1998).

\bibitem{Sobiczewski.2007.PPNP}
  A.~Sobiczewski and K.~Pomorski,
\newblock
  \emph{Description of structure and properties of superheavy nuclei},
\newblock
  Prog. Part. Nucl. Phys. \textbf{58}, 292--349 (2007).


\bibitem{Denisov.2009.ADNDT}
  V.~Yu.~Denisov and A.~A.~Khudenko,
\newblock
  \emph{$\alpha$ decay half-lives, $\alpha$ capture, and $\alpha$ nucleus potential},
\newblock
  Atom. Data Nucl. Data Tables \textbf{95} (6), 815--835 (2009).




\bibitem{Denisov.2015.PRC}
  V.~Yu.~Denisov, O.~I.~Davidovskaya, and I.~Yu.~Sedykh,
\newblock
  \emph{Improved parametrization of the unified model for $\alpha$ decay and $\alpha$ capture},
\newblock
  Phys. Rev. \textbf{C 92}, 014602 (2015).


\bibitem{Stewart.1996.NPA}
  T.~L.~Stewart, M.~W.~Kermode, D.~J.~Beachey, N.~Rowley, I.~S.~Grant, A.~T.~Kruppa,
\newblock
  \emph{Alpha-particle decay through a deformed barrier},
\newblock
  Nucl. Phys. \textbf{A 611} (2--3), 332--354 (1996).

\bibitem{Xu.2006.PRC}
  C.~Xu, Z.~Ren,
\newblock
  \emph{New deformed model of alpha-decay half-lives with a microscopic potential},
\newblock
  Phys. Rev. \textbf{C 73} (4), 041301(R) (2006).

\bibitem{Nazarewicz.2012.PRC}
  R.~Id~Betan and W.~Nazarewicz,
\newblock
  \emph{$\alpha$ Decay in the complex-energy shell model},
\newblock
  Phys. Rev. \textbf{C 86}, 034338 (2012).

\bibitem{Delion.2013.PRC}
  D.~S.~Delion and R.~J.~Liotta,
\newblock
  \emph{Shell-model representation to describe alpha emission},
\newblock
  Phys. Rev. \textbf{C 87} (4), 041302(R) (2013).

\bibitem{Silisteanu.2015.RJP}
  I.~Silisteanu and C.~I.~Anghel,
\newblock
  \emph{Alpha-decay and spontaneous fission half-lives of super-heavy nuclei around the double magic nucleus $^{270}{\rm Hs}$},
\newblock
  Rom. J. Phys. \textbf{60}, 444--451 (2015).

\bibitem{Silisteanu.2017.PRC}
   C.~I.~Anghel and I.~Silisteanu,
\newblock
  \emph{$\alpha$ decay and spontaneous fission half-lives of nuclei around $^{270}{\rm Hs}$},
\newblock
  Phys. Rev. \textbf{C95} (3), 034611 (2017).


\bibitem{www_library}
  http://www.nndc.bnl.gov,
  http://www-nds.iaea.org.











\bibitem{Ahiezer.1981}
  A.~I.~Ahiezer and V.~B.~Berestetskii,
\newblock
  \emph{Kvantovaya Elektrodinamika}
\newblock
  ({Nauka}, {Mockva}, 1981) p.~432 --- [in Russian].

\bibitem{Maydanyuk.2003.PTP}
S.~P.~Maydanyuk and V.~S.~Olkhovsky,
\newblock
  \textit{Does sub-barrier bremsstrahlung in $\alpha$-decay of \isotope[210]{Po} exist?}
\newblock
  Prog. Theor. Phys. \textbf{109} (2), 203--211 (2003);
\newblock
  arXiv: nucl-th/0404090.

\bibitem{Maydanyuk.2006.EPJA}
S.~P.~Maydanyuk and V.~S.~Olkhovsky,
\newblock
  \emph{Angular analysis of bremsstrahlung in $\alpha$-decay},
\newblock
  Eur. Phys. Journ. \textbf{A 28} (3), 283--294 (2006);
\newblock
  arXiv:nucl-th/0408022.

\bibitem{Maydanyuk.2008.EPJA}
  G.~Giardina, G.~Fazio, G.~Mandaglio, M.~Manganaro, C.~Sacc\'{a}, N.~V.~Eremin, A.~A.~Paskhalov, D.~A.~Smirnov, S.~P.~Maydanyuk, and V.~S.~Olkhovsky,
\newblock
  \emph{Bremsstrahlung emission accompanying the $\alpha$-decay of \isotope[214]{Po}},
\newblock
  Europ. Phys. Journ. \textbf{A36} (1), 31--36 (2008).

\bibitem{Maydanyuk.2008.MPLA}
  G.~Giardina, G.~Fazio, G.~Mandaglio, M.~Manganaro, S.~P.~Maydanyuk, V.~S.~Olkhovsky, N.~V.~Eremin, A.~A.~Paskhalov, D.~A.~Smirnov, and C.~Sacc\'{a},
\newblock
  \emph{Bremsstrahlung emission during $\alpha$-decay of \isotope[226]{Ra}},
\newblock
  Mod. Phys. Lett. \textbf{A 23} (31), 2651--2663 (2008);
\newblock
  arxiv:~0804.2640.

\bibitem{Maydanyuk.2009.JPS}
S.~P.~Maydanyuk,
\newblock
  \emph{A multipolar approach for the description of bremsstrahlung during $\alpha$-decay},
\newblock
  Jour. Phys. Study. \textbf{13} (3), 3201 (2009).

\bibitem{Maydanyuk.2009.TONPPJ}
  S.~P.~Maydanyuk,
\newblock
  \textit{Multipolar approach for description of bremsstrahlung during $\alpha$-decay and unified formula of the bremsstrahlung probability},
\newblock
  Open Nucl. Part. Phys. J. \textbf{2}, 17--33 (2009).

\bibitem{Maydanyuk.2009.NPA}
  S.~P.~Maydanyuk, V.~S.~Olkhovsky, G.~Giardina, G.~Fazio, G.~Mandaglio, and M.~Manganaro,
\newblock
  \emph{Bremsstrahlung emission accompanying $\alpha$-decay of deformed nuclei},
\newblock
  Nucl.~Phys. A \textbf{823}, 38--46 (2009).

\bibitem{Maydanyuk.2010.PRC}
  S.~P.~Maydanyuk, V.~S.~Olkhovsky, G.~Mandaglio, M.~Manganaro, G.~Fazio, and G.~Giardina,
\newblock
   \emph{Bremsstrahlung emission of high energy accompanying spontaneous fission of \isotope[252]{Cf}},
\newblock
  Phys. Rev. \textbf{C82}, 014602 (2010).

\bibitem{Maydanyuk.2011.JPG}
  S.~P.~Maydanyuk,
\newblock
  \emph{Multipolar model of bremsstrahlung accompanying proton decay of nuclei},
\newblock
  J. Phys. \textbf{G 38} (8), 085106 (2011);
\newblock
  arXiv:1102.2067.

\bibitem{Maydanyuk.2011.JPCS}
  S.~P.~Maydanyuk, V.~S.~Olkhovsky, G.~Mandaglio, M.~Manganaro, G.~Fazio, G.~Giardina, and C.~Sacc\'{a},
\newblock
  \emph{Bremsstrahlung emission of photons accompanying ternary fission of \isotope[252]{Cf}},
\newblock
\newblock
  Journ. Phys.: Conf. Ser. \textbf{282}, 012016 (2011).

\bibitem{Maydanyuk.2012.PRC}
S.~P.~Maydanyuk,
\newblock
  \emph{Model for bremsstrahlung emission accompanying interactions between protons and nuclei from low energies up to intermediate energies: Role of magnetic emission},
\newblock
  Phys. Rev. \textbf{C86}, 014618 (2012),
\newblock
  arXiv:1203.1498.

\bibitem{Maydanyuk_Zhang.2015.PRC}
S.~P.~Maydanyuk and P.-M.~Zhang,
\newblock
  \emph{New approach to determine proton-nucleus interactions from experimental bremsstrahlung data},
\newblock
  Phys. Rev. C \textbf{91}, 024605 (2015);
\newblock
  arXiv:1309.2784.

\bibitem{Maydanyuk_Zhang_Zou.2016.PRC}
S.~P.~Maydanyuk, P.-M.~Zhang, and L.-P.~Zou,
\newblock
  \emph{New approach for obtaining information on the many-nucleon structure in $\alpha$ decay from accompanying bremsstrahlung emission},
\newblock
  Phys. Rev. \textbf{C93}, 014617 (2016);
\newblock
  arXiv:1505.01029.

\bibitem{Maydanyuk_Zhang_Zou.2018.PRC}
S.~P.~Maydanyuk, P.-M.~Zhang, and L.-P.~Zou,
\newblock
  \emph{Manifestation of the important role of nuclear forces in the emission of photons in pion scattering off nuclei},
\newblock
  Phys. Rev. \textbf{C98}, 054613 (2018);
\newblock
  arXiv:1809.10403.





\bibitem{Back.2014.RMP}
  B.~B.~Back, H.~Esbensen, C.~L.~Jiang, and K.~E.~Rehm,
\newblock
  \emph{Recent developments in heavy-ion fusion reactions},
\newblock
  Rev. Mod. Phys. \textbf{86}, 317 (2014).

\bibitem{Birkelund.1979.PRep}
  J.~R.~Birkelund, L.~E.~Tubbs, J.~R.~Huizenga, J.~N.~De, and D.~Sperber,
\newblock
  \emph{Heavy-ion fusion: Comparison of experimental data with classical trajectory models},
\newblock
  Phys. Rep. \textbf{56}, 107 (1979).

\bibitem{Vaz.1981.PRep}
  L.~C.~Vaz, J.~M.~Alexander, and G.~R.~Satchler,
\newblock
  Phys. Rep. \textbf{69}, 373 (1981).

\bibitem{Birkelund.1983.ARNPS}
  J.~R.~Birkelund, and J.~R.~Huizenga,
\newblock
  \emph{Fusion Reactions Between Heavy Nuclei},
\newblock
  Annu. Rev. Nucl. Part. Sci. \textbf{33}, 265 (1983).

\bibitem{Beckerman.1985.PRep}
  M.~Beckerman,
\newblock
  Phys. Rep. \textbf{129}, 145 (1985).

\bibitem{Steadman.1986.ARNPS}
  S.~G.~Steadman, and M.~J.~Rhoades-Brown,
\newblock
  \emph{Sub-barrier fusion reactions},
\newblock
  Annu. Rev. Nucl. Part. Sci. \textbf{36}, 649 (1986).

\bibitem{Beckerman.1988.RPP}
  M.~Beckerman,
\newblock
  Rep. Prog. Phys. \textbf{51}, 1047 (1988).

\bibitem{Rowley.1991.PLB}
  N.~Rowley, G. R. Satchler, and P. H. Stelson,
\newblock
  \emph{On the “distribution of barriers” interpretation of heavy-ion fusion},
\newblock
  Phys. Lett. B \textbf{254}, 25 (1991).

\bibitem{Vandenbosch.1992.ARNPS}
  R.~Vandenbosch,
\newblock
  \emph{Angular momentum distributions in subbarrier fusion reactions},
\newblock
  Annu. Rev. Nucl. Part. Sci. \textbf{42}, 447 (1992).

\bibitem{Reisdorf.1994.JPG}
  W.~Reisdorf,
\newblock
  \emph{Heavy-ion reactions close to the Coulomb barrier},
\newblock
  J.~Phys.~G \textbf{20}, 1297 (1994).

\bibitem{Dasgupta.1998.ARNPS}
  M.~Dasgupta, D.~J.~Hinde, N.~Rowley, and A.~M.~Stefanini,
\newblock
  \emph{Measuring barriers to fusion},
\newblock
  Annu. Rev. Nucl. Part. Sci. \textbf{48}, 401 (1998).

\bibitem{Balantekin.1998.RMP}
  A.~B.~Balantekin, N.~Takigawa,
\newblock
  \emph{Quantum tunneling in nuclear fusion},
\newblock
  Rev. Mod. Phys. \textbf{70}, 77 (1998).

\bibitem{Liang.2005.IJMPE}
  J.~F.~Liang, and C.~Signorini,
\newblock
  Int. J. Mod. Phys. E \textbf{14}, 1121 (2005).

\bibitem{Canto.2006.PRep}
  L.~F.~Canto, P.~R.~S.~Gomes, R.~Donangelo, and M.~S.~Hussein,
\newblock
  \emph{Fusion and breakup of weakly bound nuclei},
\newblock
  Phys. Rep. \textbf{424}, 1 (2006).

\bibitem{Keeley.2007.PPNP}
  N.~Keeley, R.~Raabe, N.~Alamanos, J.~L.~Sida,
\newblock
  \emph{Fusion and direct reactions of halo nuclei at energies around the Coulomb barrier},
\newblock
  Prog. Part. Nucl. Phys. \textbf{59}, 579 (2007).

\bibitem{Hagino.2012.PTP}
  K.~Hagino, and N.~Takigawa,
\newblock
  \emph{Subbarrier fusion reactions and many-particle quantum tunneling},
\newblock
  Prog. Theor. Phys. \textbf{128}, 1001 (2012).

\bibitem{Glas.1975.NPA}
  D.~Glas and U.~Mosel,
\newblock
  Nucl.~Phys. \textbf{A237}, 429 (1975).

\bibitem{Glas.1974.PRC}
  D.~Glas and U.~Mosel,
\newblock
  \emph{Limitation on complete fusion during heavy-ion collisions},
\newblock
  Phys.~Rev. \textbf{C10}, 2620 (1974).

\bibitem{Maydanyuk_Zhang_Zou.2017.PRC}
  S.~P.~Maydanyuk, P.-M.~Zhang, and L.-P.~Zou,
\newblock
  \emph{New quasibound states of the compound nucleus in $\alpha$-particle capture by the nucleus},
\newblock
  Phys. Rev. \textbf{C 96}, 014602 (2017);
\newblock
  arXiv:1711.07012.

\bibitem{Maydanyuk.2015.NPA}
  S.~P.~Maydanyuk, P.-M.~Zhang, and S.~V.~Belchikov,
\newblock
  \emph{Quantum design using a multiple internal reflections method in a study of fusion processes in the capture of alpha-particles by nuclei},
\newblock
  Nucl. Phys. A \textbf{940}, 89--118 (2015);
\newblock
  arXiv:1504.00567.



\bibitem{Eberhard.1979.PRL} 
  K.~A.~Eberhard, Ch.~Appel, R.~Bangert, L.~Cleemann, J.~Eberth, and V.~Zobel,
\newblock
  \emph{Fusion cross sections for $\alpha + \isotope[40, 44]{Ca}$ and the problem of anomalous large-angle scattering},
\newblock
  Phys. Rev. Lett. \textbf{43} (2), 107--110 (1979).

\bibitem{DAuria.1968.PR}
  J.~M.~D'Auria, M.~J.~Fluss, L.~Kowalski, and J.~M.~Miller,
\newblock
  \emph{Reaction cross section for low-energy alpha particles on \isotope[59]{Co}},
\newblock
  Phys. Rev. \textbf{168}, 1224 (1968).

\bibitem{Barnett.2000.PRC}
  A.~R.~Barnett and J.~S.~Lilley,
\newblock
  \emph{Interaction of alpha particles in the lead region near the Coulomb barrier},
\newblock
  Phys. Rev. \textbf{C 9}, 2010 (1974).

\bibitem{Karpeshin.2007.JPG}
  F.~F.~Karpeshin, G.~LaRana, E.~Vardaci, A.~Brondi, R.~Moro, S.~N.~Abramovich, and V.~I.~Serov,
\newblock
  \emph{Resonances in alpha-nuclei interactions},
\newblock
  J. Phys. G: Nucl. Part. Phys. \textbf{34}, 587--595 (2007).

\bibitem{Gaul.1969.NPA}
  G.~Gaul, H.~Ludecke, R.~Santo, and H.~Schmeing,
\newblock
  \emph{Effects of particle correlations in elastic $\alpha$-scattering},
\newblock
  Nucl. Phys. \textbf{A137}, 177--192 (1969).







\bibitem{Sharan.1993.PRC}
  M.~K.~Sharan, Y.~K.~Agarwal, C.~V.~K.~Baba, D.~R.~Chakrabarty, and V.~M.Datar,
\newblock
  \emph{Ultradipole photon production in 40 and 50 MeV $\alpha$-nucleus collisions},
\newblock
  Phys. Rev. \textbf{C48} (6), 2845 (1993).




\bibitem{D'Arrigo.1994.PHLTA}
  A.~D'Arrigo, N.~V.~Eremin, G.~Fazio, G.~Giardina, M.~G.~Glotova, T.~V.~Klochko, M.~Sacchi, and A.~Taccone,
\newblock
  \emph{Investigation of bremsstrahlung emission in $\alpha$-decay of heavy nuclei},
\newblock
   Phys. Lett. \textbf{B332} (1--2), 25--30 (1994).

\bibitem{Kasagi.1997.JPHGB}
  J.~Kasagi, H.~Yamazaki, N.~Kasajima, T.~Ohtsuki, and H.~Yuki,
\newblock
  \emph{Bremsstrahlung emission in $\alpha$ decay and tunneling motion of $\alpha$ particle},
\newblock
  Journ. Phys. \textbf{G 23}, 1451--1457 (1997).

\bibitem{Kasagi.1997.PRLTA}
  J.~Kasagi, H.~Yamazaki, N. Kasajima, T.~Ohtsuki, and H.~Yuki,
\newblock
  \emph{Bremsstrahlung in $\alpha$ decay of \isotope[210]{Po}: do $\alpha$ particles emit photons in tunneling?}
\newblock
  Phys. Rev. Lett. \textbf{79} (3), 371--374 (1997).

\bibitem{Boie.2007.PRL} 
  H.~Boie, H.~Scheit, U.~D.~Jentschura, F.~K\"{o}ck, M.~Lauer, A.~I.~Milstein, I.~S.~Terekhov, and D.~Schwalm,
\newblock
  \emph{Bremsstrahlung in $\alpha$ decay reexamined},
\newblock
  Phys. Rev. Lett. \textbf{99}, 022505 (2007);
\newblock
  arXiv:0706.2109.

\bibitem{Boie.2009.PhD}
  H.~Boie,
\newblock
  \emph{Bremsstrahlung emission probability in the $\alpha$ decay of \isotope[210]{Po}},
\newblock
  Ph.D. thesis, Ruperto-Carola University of Heidelberg, Germany, 2009, p.~193.

\bibitem{RewPartPhys_PDG.2018}
  M.~Tanabashi et al. (Particle Data Group),
\newblock
  Phys. Rev. \textbf{D 98}, 030001 (2018).

\bibitem{Landau.v3.1989}
  L.~D.~Landau and E.~M.~Lifshitz,
\newblock
  \emph{Kvantovaya Mehanika, kurs Teoreticheskoi Fiziki}
  (Quantum mechanics, course of Theoretical Physics),
\newblock
  Vol.~3 ({Nauka}, {Moskva}, 1989) p.~768 ---
  [in Russian; eng. variant: Oxford, Uk, Pergamon, 1982].


\bibitem{Papenbrock.1998.PRLTA}
  T.~Papenbrock and G.~F.~Bertsch,
\newblock
  \emph{Bremsstrahlung in $\alpha$ decay},
\newblock
  Phys. Rev. Lett. \textbf{80} (19), 4141--4144 (1998),
\newblock
 nucl-th/9801044.

\bibitem{Tkalya.1999.PHRVA}
  E.~V.~Tkalya,
\newblock
  \emph{Bremsstrahlung in $\alpha$ decay and ``interference of space regions''},
\newblock
  Phys.~Rev. \textbf{C60}, 054612 (1999).

\end{thebibliography}
\end{document}